\documentclass[12pt,a4paper,dvips]{article}
\usepackage{graphicx,epsfig}
\usepackage{hhline}

\usepackage{amsmath,amssymb}
\usepackage{times}
\DeclareMathAlphabet{\mathbold}{OML}{txr}{b}{it}

\usepackage{array,multirow,dcolumn}
\usepackage[mathlines,displaymath]{lineno}
\usepackage{rotating}

\usepackage[numbers,square,comma,sort&compress]{natbib}
\usepackage{hypernat}
\usepackage{textcomp}
\newcommand{\tablecaption}{%
\caption}

\newcolumntype{.}{D{.}{.}{-1}}
\newcolumntype{-}{D{-}{-}{-1}}

\usepackage[dvipsnames]{color}
\definecolor{rltred}{rgb}{0.75,0,0}
\definecolor{rltgreen}{rgb}{0,0.5,0}
\definecolor{rltblue}{rgb}{0,0,0.5}

\newcounter{pdfadd}    % need for correct PDF hyperlinks and bookmarks :-(
\usepackage[hyperindex,bookmarks,bookmarksnumbered,breaklinks,a4paper,unicode]{hyperref}
\hypersetup{%
  pdftitle        = {Measurement of the inclusive ep Scattering Cross Section
 at low Q2 and x at HERA},
  urlcolor        = rltblue,       % \href{...}{...} external (URL)
  urlbordercolor  = 0 0 0.5,
  filecolor       = rltblue,       % \href{...} local file
  filebordercolor = 0 0 0.5,
  linkcolor       = rltred,        % \ref{...}
  linkbordercolor = 0.75 0 0,
  citecolor       = rltgreen,      % \cite{...}
  citebordercolor = 0 0.5 0,
  pagecolor       = rltgreen,      % \pageref{...}
  pagebordercolor = 0 0.5 0,
  menucolor       = rltgreen,      % Acrobat menu items
  menubordercolor = 0 0.5 0,
  colorlinks    = true,
  pdfauthor     = {H1 Collaboration},
  pdfsubject    = { },
  pdfkeywords   = {High-Energy Physics, Particle Physics, Proton Structure, DIS}
}

\newlength{\dinwidth}
\newlength{\dinmargin}
\newcommand{\empz}{\mbox{$E$$-$$P_z$}}

\newcommand{\ee}{\mbox{$E_e^{\prime}$}}
\newcommand{\thetae}{\mbox{$\theta_e$}}

\newcommand{\Fig}{\mbox{figure}}
\newcommand{\Tab}{\mbox{table}}
\newcommand{\Eq}{\mbox{equation}}
\newcommand{\Sec}{\mbox{section}}

\newcommand{\FFig}{\mbox{Figure}}
\newcommand{\TTab}{\mbox{Table}}
\newcommand{\EEq}{\mbox{Equation}}
\newcommand{\SSec}{\mbox{Section}}

\def\ytrans{0.56}

\newcommand{\Figs}{\mbox{figures}}
\newcommand{\Tabs}{\mbox{tables}}

\newcommand{\Secs}{\mbox{sections}}

\def\dof{\mathop{n_{\rm dof}}\nolimits}

\newcommand\hqcjc{medium $Q^2$ CJC $E_p=920$~GeV}
\newcommand\lqbst{low $Q^2$ BST $E_p=920$~GeV}
\newcommand\ler{\mbox{$E_p=460$~GeV}}
\newcommand\mer{\mbox{ $E_p=575$~GeV}}
\newcommand\lermer{$E_p=460$~GeV and $E_p=575$~GeV}
\newcommand\ner{$E_p=920$~GeV}

\newcommand\Hqcjc{Medium $Q^2$ CJC $E_p=920$~GeV}
\newcommand\Lqbst{Low $Q^2$ BST $E_p=920$~GeV}
\newcommand\Ler{\mbox{$E_p=460$~GeV}}
\newcommand\Mer{\mbox{$E_p=575$~GeV}}

\newcommand\lumicjcplus{\mbox{$53.2$}}
\newcommand\lumicjcminus{\mbox{$44.4$}}
\newcommand\lumibstplus{\mbox{$3.4$}}
\newcommand\lumibstminus{\mbox{$2.5$}}

\newcommand\lunit{\mbox{pb$^{-1}$}}
\newcommand\lumiler{\mbox{$12.2$}}
\newcommand\lumimer{\mbox{$5.9$}}
\newcommand\lumicjc{\mbox{$97.6$}}
\newcommand\lumibst{\mbox{$5.9$}}
\newcommand\gbw{GBW}
\newcommand\gbwdglap{GBW+DGLAP$_{\rm valence}$}
\newcommand\iim{IIM}
\newcommand\iimdglap{IIM+DGLAP$_{\rm valence}$}
\newcommand\bsat{B-SAT}
\newcommand\bsatdglap{B-SAT+DGLAP$_{\rm valence}$}

\newcommand\ndfdefone{782}
\newcommand\chacotone{722.7}
\newcommand\chrtone{773.2}
\newcommand\ndfdef{781}
\newcommand\chacot{715.2}
\newcommand\chrt{764.5}

\newcommand\chrtqcut{288.8}
\newcommand\chacotqcut{248.3}
\newcommand\ndfdglapqcut{249}

\newcommand\bsatchsq{$424.9/352$} 
\newcommand\bsatqchsq{$261.7/252$} 
\newcommand\bsatqdvchsq{$371.4/252$} 
\newcommand\gbwqdvchsq{$739.5/252$} 
\newcommand\iimqdvchsq{$287.6/252$} 
\newcommand\gbwchsq{$718.8/352$} 
\newcommand\gbwqchsq{$559.7/252$} 
\newcommand\iimchsq{$397.6/352$} 
\newcommand\iimqchsq{$259.4/252$} 

\begin{document}

% change natbib spacing between numbers in citations
\makeatletter \def\NAT@space{} \makeatother

\begin{titlepage}
 
\noindent
DESY 10-228 \hfill ISSN 0418-9833 \\
 December 2010 \\

\vspace*{4.5cm}

\begin{center}
\begin{Large}

{\bfseries Measurement of the Inclusive $\boldsymbol{e^{\pm}p}$ Scattering
 Cross Section 
at High Inelasticity $\boldsymbol{y}$
and 
%Determination  
of 
the Structure Function $\boldsymbol{F_L}$
% by the H1
%Experiment.
}

\vspace*{2cm}

H1 Collaboration

\end{Large}
\end{center}

\vspace*{2cm}

\begin{abstract} \noindent
A  measurement is presented of the inclusive neutral current
$e^\pm p$ scattering cross section 
using data collected by the H1 experiment at HERA
 during the years  2003 to 2007 
with proton beam energies $E_p$ of $920$, $575$, and $460$~GeV.
The kinematic range of the measurement covers low absolute
four-momentum transfers squared, $1.5$~GeV$^2<Q^2<120$~GeV$^2$,  small 
values of Bjorken $x$, $2.9 \cdot 10^{-5}<x<0.01$,
and extends to high inelasticity up to $y=0.85$.
The structure function $F_L$
is measured by combining the new results with previously published H1 data 
 at $E_p=920$~GeV and $E_p=820$~GeV.
The new measurements are used to test several phenomenological and
QCD models applicable in this low $Q^2$ and low $x$ kinematic domain.
\end{abstract}

\vspace*{1.5cm}

\begin{center}
{\slshape Submitted to EPJC}
\end{center}

\end{titlepage}

\begin{flushleft}
 %\input{h1auts}
%-- H1AUTS Author list by names 
%-- Status: Wed Dec  1 16:54:27 CET 2010  Number of authors = 212 

F.D.~Aaron$^{5,49}$,           %BUCH-PD        11/06           Aaron               
C.~Alexa$^{5}$,                %BUCH-PD        06/06           Alexa               
V.~Andreev$^{25}$,             %LPI -PD        8/88            Andreev             
S.~Backovic$^{30}$,            %PODG-PD        03/02           Backovic            
A.~Baghdasaryan$^{38}$,        %YERE-PD        09/03           Baghdasaryana       
S.~Baghdasaryan$^{38}$,        %YERE-ST        02/10           Baghdasaryans       
E.~Barrelet$^{29}$,            %PARI-PD        11/99           Barrelet            
W.~Bartel$^{11}$,              %DESY-PD        8/88            Bartel  
O.~Behrend$^{8,54}$            %DORT-PD      --- exceptionally for FL paper 
P.~Belov$^{11}$
% ...

%-- H1 Institutes 
\bigskip{\it
% ...
$ ^{2}$ Vinca Institute of Nuclear Sciences, University of Belgrade,
          1100 Belgrade, Serbia \\
% ...
$ ^{5}$ National Institute for Physics and Nuclear Engineering (NIPNE) ,
          Bucharest, Romania$^{ m}$ \\
% ...
$ ^{8}$ Institut f\"ur Physik, TU Dortmund, Dortmund, Germany$^{ a}$ \\
% ...
$ ^{11}$ DESY, Hamburg, Germany \\
% ...
$ ^{25}$ Lebedev Physical Institute, Moscow, Russia$^{ e}$ \\
% ...
$ ^{29}$ LPNHE, Universit\'e Pierre et Marie Curie Paris 6,
          Universit\'e Denis Diderot Paris 7, CNRS/IN2P3, Paris, France \\
 $ ^{30}$ Faculty of Science, University of Montenegro,
          Podgorica, Montenegro$^{ e}$ \\
% ...
$ ^{38}$ Yerevan Physics Institute, Yerevan, Armenia \\
% ...
$ ^{49}$ Also at Faculty of Physics, University of Bucharest,
          Bucharest, Romania \\
% ...
$ ^{54}$ Now at Siemens, N\"urnberg, Germany \\

\smallskip
% ...
$ ^a$ Supported by the Bundesministerium f\"ur Bildung und Forschung, FRG,
      under contract numbers 05H09GUF, 05H09VHC, 05H09VHF,  05H16PEA \\
% ...
$ ^e$ Supported by the Deutsche Forschungsgemeinschaft \\
% ...
$ ^m$ Supported by the Romanian National Authority for Scientific Research
      under the contract PN 09370101 \\
}
\end{flushleft}
 
%\newpage
%\tableofcontents
\newpage

\section{Introduction}

\label{sec:intro}
Deep inelastic lepton-nucleon scattering (DIS) plays a pivotal role in
determining the structure of the proton. 
The electron-proton collider HERA
covers a wide range of absolute four-momentum transfer squared, $Q^2$, and of
Bjorken $x$.
Previous measurements
of the DIS cross section, performed 
by the H1~\cite{H1:2009bp,Aaron:2009kv,h1alphas,Adloff:1999ah,Adloff:2000qj,Adloff:2003uh} 
and ZEUS~\cite{Breitweg:1997hz,Breitweg:2000yn,Breitweg:1998dz,Chekanov:2001qu,zeuscc97,
Chekanov:2002ej,Chekanov:2002zs,Chekanov:2003yv,Chekanov:2003vw}
 collaborations, using data at proton beam energies of
$E_p=820$~GeV and $E_p=920$~GeV and a lepton beam energy of 
$E_e=27.6$~GeV,
as well as their combination~\cite{h1zeus:2009wt},
have enabled studies of 
perturbative Quantum 
Chromodynamics (QCD) with unprecedented precision. 
These measurements are complemented here with new data including the 
data taken at $E_p=460$~GeV and $E_p=575$~GeV.

At low $Q^2$, the scattering cross section is defined by the two structure
functions, $F_2$ and $F_L$.
In a reduced form, the double differential cross section is given by 
\begin{equation}
  \sigma_r(x,Q^2)  \equiv \frac{\textstyle Q^4 x}
{\textstyle 2\pi \alpha^2 \left[1+(1-y)^2\right] }\cdot 
\frac{\textstyle {\rm d}^2 \sigma}
{\textstyle {\rm d}x\,{\rm d}Q^2} 
= F_2(x,Q^2) - \frac{y^2}{1+(1-y)^2} F_L(x,Q^2)\,. \label{eq:one}
\end{equation}
Here $\alpha$ is the fine structure constant and $0\le y \le 1$ 
is the process inelasticity which is related
to $Q^2,~x$ and the centre-of-mass energy squared 
$s=4 E_e E_p$ by $y = Q^2/sx$.
The two structure functions are defined by 
the cross sections for the scattering of the longitudinally and
transversely polarised photons off protons $\sigma_L$ and $\sigma_T$ as 
\begin{eqnarray}
  F_L(x,Q^2) &=& \frac  {Q^2}  { 4 \pi^2 \alpha}  (1-x)\cdot \sigma_L\,, \\
  F_2(x,Q^2) &=& \frac {Q^2 }  { 4 \pi^2 \alpha} (1-x) \cdot ( \sigma_L + \sigma_T)\,.
\end{eqnarray}
These relations  are valid to  good approximation at low $x$ and imply that
$0\le F_L \le F_2$.
To disentangle the two structure functions in a model-independent way, 
measurements at different values of $s$ are required. 

Using the ratio $R(x,Q^2)$, defined as
\begin{equation}
R = {\sigma_L \over \sigma_T} = {F_L \over {F_2 - F_L}}, 
\label{ratio}
\end{equation}
the reduced cross section in equation~\ref{eq:one} can also be written as
\begin{equation} 
\sigma_r = F_2(x,Q^2) \cdot \left [1 - f(y) \cdot \frac{R}{1 + R} \right ],
 \label{eq:rrr}
\end{equation}
where $f(y)=y^2/(1+(1-y)^2)$.

In the quark-parton model, $F_2$ is given  
by the charge squared weighted sum of the quark densities while $F_L$
is zero because of helicity conservation. 
In QCD, the gluon emission gives rise to a non-vanishing $F_L$. 
Measuring the structure function $F_L$ %in the region of DIS
therefore provides a way of studying the gluon density
and a test of perturbative QCD.

The contribution of the term containing $F_L$ to the scattering cross 
section  can %only 
be sizeable only at large values of $y$.
For low values  of $y$, the 
reduced DIS neutral current (NC) scattering cross section
is well approximated by the structure function $F_2$.
Kinematically, for low $Q^2$, 
large values of $y$ correspond to low energies of the scattered lepton.
Selecting high $y$ events is thus complicated due to a
possibly large  background from hadronic final state particles.
%
%Since $F_L$ gives sizable contribution only for large values of
%inelasticity $y$,
%For most of the kinematic domain 
% 
% To determine the structure function $F_L$, measurements at the same $Q^2$ and $x$, but different $y$ are needed.
%
%To disentangle $F_2$ and $F_L$ in a model
%independent way, measurements at different values of $s$ are required.

This paper reports  new measurements of the DIS cross 
section at low $Q^2$ and high $y$ values, 
using data collected by the H1 collaboration in the years 
$2003$ to $2007$.
% as listed in  \Tab~\ref{tab:data}.
%
The data samples are taken with dedicated  high $y$ and low $Q^2$ 
triggers. Methods relying on data are used to determine the hadronic background
contribution.
The first data sample consists of a new cross-section measurement
for the nominal proton beam energy of
$E_p=920$~GeV for $8.5\le Q^2\le 90$\,GeV$^2$ with an increased integrated 
luminosity of \lumicjc~\lunit\
compared to~\cite{Aaron:2009kv}.
The second new data sample at $E_p=920$~GeV
covers the region $ 2.5 \le Q^2 \le 12$\,GeV$^2$ using
a dedicated silicon tracker  
for the measurement of the charge of backward scattered particles.
This analysis is based on an integrated luminosity of \lumibst~\lunit. 
Two further  data  samples correspond to the
measurements at reduced proton beam energy,
$E_p=575$~GeV and $E_p=460$~GeV, covering 
the kinematic domain of $1.5\le Q^2\le90$~GeV$^2$ 
with integrated luminosities of
\lumimer~\lunit\ and \lumiler~\lunit, respectively. 
Combined with the previously published H1 
measurements~\cite{H1:2009bp,Aaron:2009kv},  the data are used to measure
the structure function $F_L$. This new measurement supersedes the previous H1
result~\cite{h1fl}.

The measurements are used to test several phenomenological and QCD models
describing the low $x$ behaviour of the DIS cross section.
The phenomenological models  
include the power-law dependence 
of $F_2$~\cite{adloff-2001-520} and several dipole 
models~\cite{Nikolaev:1990ja,Golec-Biernat:1998js,Iancu:2003ge,Kowalski:2006hc,Albacete:2009fh}
applicable at low $x<0.01$.
For the first time, dipole model analyses are 
extended to account for 
the non-negligible valence-quark contributions at small $x$.
Fits using the DGLAP evolution  
equations~\cite{Gribov:1972ri,Gribov:1972rt,Lipatov:1974qm,Dokshitzer:1977sg,Altarelli:1977zs}
at NLO  \cite{Curci:1980uw,Furmanski:1980cm} are applied for $Q^2\ge3.5$~GeV$^2$.
For the DGLAP fits, different treatments of the heavy quark contributions are 
compared~\cite{Thorne:1997ga,Thorne:2006qt,Kramer:2000hn}. 
A study of possible non-DGLAP contributions at low $x$ and low 
$Q^2$ is performed by varying kinematic cuts applied to  the data. 
The dipole and DGLAP models are compared 
to with other by performing fits in a common
kinematic domain.

This paper is organised as follows. 
The measurement technique is presented in \Sec~\ref{sec:measurement}.
The data analysis and event selection are described in \Sec~\ref{sec:dataanalysis}. 
The cross section and $F_L$ measurement
procedures are explained in \Secs~\ref{sec:xs} and \ref{sec:fl}.
\SSec~\ref{sec:theory} contains the phenomenological analysis of the
measurement. The results presented in the paper are summarised in \Sec~\ref{sec:conclusion}. 
\section{Measurement Technique}
\label{sec:measurement}
\subsection{H1 Detector}

\label{sec:detector}
\begin{figure}[b]
\begin{center}
\includegraphics[angle =90, width=0.6\linewidth]{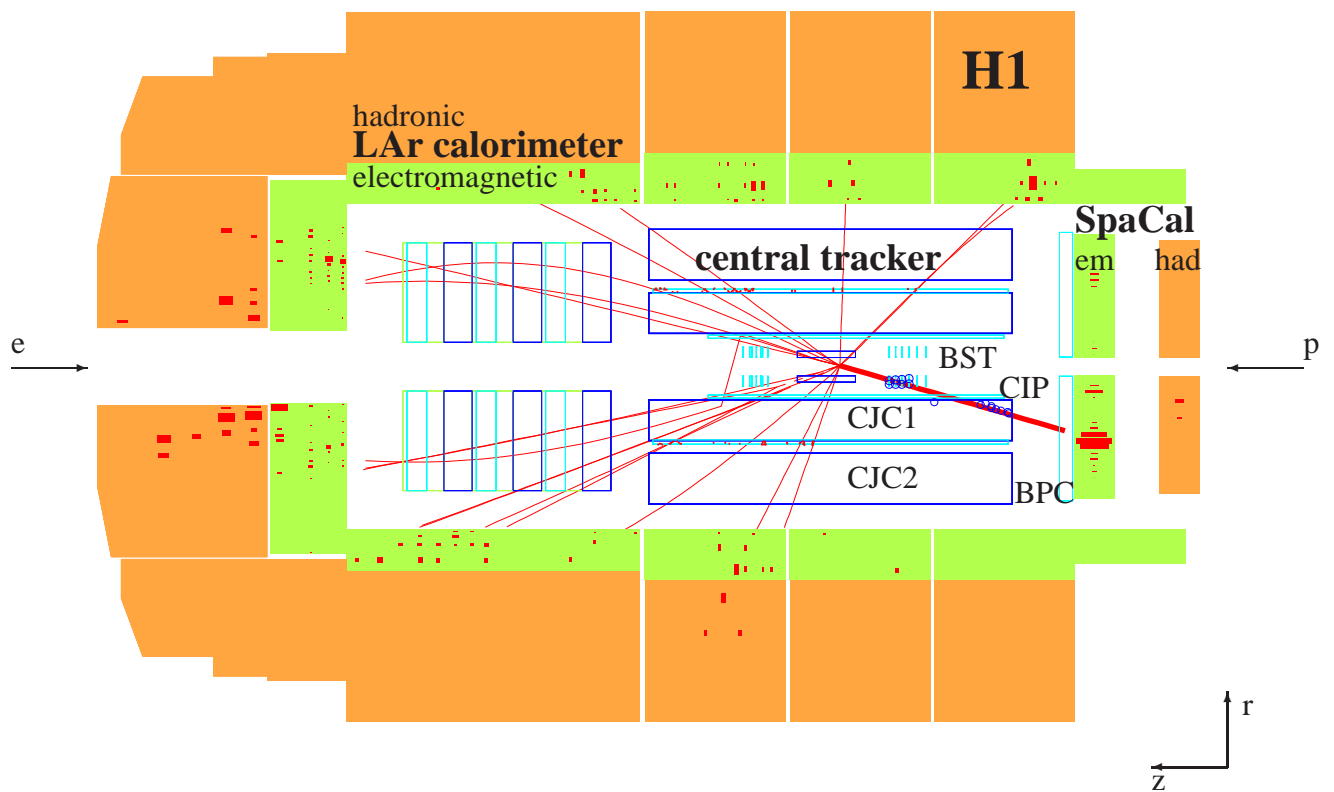}
%\epsfig{file=figs/event.eps, bburx=500, bbury=543,bbllx=69,bblly=0, angle=90,width=0.6\linewidth}
%\end{center}
\put(-130,53){\vector(1,0){10}}
\put(-130,55){e}
\put(40,53){\vector(-1,0){10}}
\put(40,55){p}
\put(30,0){\vector(-1,0){10}}
\put(30,0){\vector(0,1){10}}
\put(20,-3){z}
\put(32,7){r}
\put(-5,90){\LARGE\bf H1}
\put(-85,81){\large\bf LAr calorimeter}
\put(-85,85){hadronic}
\put(-85,77){electromagnetic}
\put(-40,66){\large\bf central tracker}
\put(-20,45){CJC1}
\put(-20,37){CJC2}
\put(-8,53){BST}
\put(2,35){BPC}
\put(10,71){\large\bf SpaCal}
\put(10,66){em~~~~~had}
\put(0,49){CIP}
\end{center}
\caption{A 
 high $y$ event as reconstructed in the H1 detector.
The scattered electron is measured in the SpaCal calorimeter.
The electron trajectory, shown by a thick line, is reconstructed
in the backward silicon tracker (BST)  and in the inner
central jet chamber (CJC1). The trajectory crosses the central 
inner proportional chamber (CIP)
which is used for triggering. 
The backward proportional chamber (BPC) may assists
the measurement of the scattering angle.
The hadronic final state particles are
detected in the central tracker, liquid argon calorimeter (LAr) and
in the SpaCal electromagnetic and hadronic sections.
\label{fig:event}}
\end{figure}

A detailed description of the H1 detector can be found in~\cite{h1det,h1det2}.
A view of a high $y$, low $Q^2$ event reconstructed in the H1 detector is shown in
\Fig~\ref{fig:event}.
The origin of the H1 coordinate system is the nominal $ep$ interaction point.
The direction of the proton beam defines the positive $z$-axis (forward direction). Transverse momenta are measured in the $x-y$ plane. Polar ($\theta$) and
azimuthal ($\phi$) angles are measured with respect to the reference system.

The most relevant
detector components for this analysis are the central tracker (CT), 
the backward
lead-scintillator calorimeter (SpaCal)~\cite{spacalc} and the liquid argon calorimeter (LAr)~\cite{Andrieu:1993kh}.
The central tracker
consists of the central jet drift chambers (CJC1 and CJC2), the $z$ drift
chamber~\cite{Barwolff:1989fc}, the central inner
proportional chamber (CIP)~\cite{Becker:2007ms},
 the central silicon tracker
(CST)~\cite{cst}. 
The detector operates in a $1.16$~T solenoidal magnetic field.
The drift chambers and the CST are used for the
measurement of tracks from the hadronic final state and to determine the 
interaction vertex. 
The energy of the scattered lepton $\ee$ is measured by the SpaCal.
The polar angle of the scattered lepton $\thetae$ is determined by the SpaCal
and the vertex position. The
tracking information obtained from the backward silicon tracker
(BST)~\cite{Eick:1996gv}, partially in combination
with the CJC, determines the charge of the scattered
electron candidate using the measured curvature. 
The BST is equipped with silicon pad sensors to provide a fast trigger signal.
The SpaCal contains electromagnetic and
hadronic sections. Its energy resolution for electromagnetic energy
depositions is $\delta E/E \approx 0.07/\sqrt{E/\rm{GeV}} \oplus
0.01$. It also provides a trigger based on the scattered lepton
energy, time and location inside the calorimeter. The LAr allows the hadronic final state to be reconstructed.
Its energy resolution was determined to be $\delta E/E \approx
0.50/\sqrt{E/\rm{GeV}} \oplus 0.02$ with pion test beam
data~\cite{Andrieu:1993tz}.
Two electromagnetic calorimeters, a tungsten/quartz-fibre sampling
calorimeter 
(``photon tagger'') and a compact lead/scintillator %spaghetti
calorimeter  
(``electron tagger''), 
are located close to the beam pipe 
at $z = -103.1$\,m and $z = -6$\,m,
respectively. 
The photon tagger is used for monitoring the luminosity via the measurement
of the Bethe-Heitler process $ep\to \gamma ep$.
The electron tagger is  used to select pure samples of photoproduction events
used for background estimation.

The H1 data collection employs a four level trigger system. The first level 
trigger (L1) is based on various sub-detector components, 
which are combined at the second level (L2); the  decision is
refined at the third level (L3). The fully reconstructed events are 
subject to an additional selection at the software filter farm (L4). 
%%\subsection{Definitions}
%%\input{defs}
\subsection{Reconstruction}
% of Event Kinematics and Cross Section Extraction}
%\input{kine}

\label{sec:kine}
At HERA, the DIS kinematics can be reconstructed using the 
scattered lepton, the hadronic final state or a combination of both.
For the measurement at high inelasticity $y$, however, 
the reconstruction of kinematics
using the scattered lepton, the so called electron method, 
has superior resolution and  is used here.

The kinematic variables in the electron method are determined by
\begin{equation}
y_e = 1 - \frac{\ee(1-\cos \thetae)}{2 E_e},~~~~~~~~
Q^2_e = \frac{{\ee}^2\sin^2\theta_e}{1-y_e},~~~~~~~~
x_e = \frac{Q^2_e}{4E_p E_e y_e}\,.
\end{equation}
%where $E'_e$ and $\theta_e$ are the energy and the polar angle of the 
%scattered electron.
%Measurement of $\ee$ is based solely on the SpaCal cluster energy 
%while  $\thetae$ is determined using the event vertex and the SpaCal cluster position.
Energy-momentum conservation implies that
\begin{equation}
2E_e \approx (E - P_z)_{\rm in} = (E - P_z)_{\rm out} 
\approx E'_e(1-\cos \thetae) + \sum_i \left( E_z - P_{z,i}\right) \equiv \empz 
\,,
\end{equation}
where the subscripts ``in'' and ``out'' denote the total {\empz} before and after
the interaction, $E_i$ ($P_{z,i} $) is the reconstructed energy 
(longitudinal component of the momentum) of a particle $i$ from the
hadronic final state and the sum runs over all measured
hadronic final state particles. For events with hard QED initial state radiation (ISR), the radiated
photon escapes in the beam pipe and $\empz$ is reduced.
Therefore measurement of $\empz$ 
allows
control of the effective beam energy and the reduction
of  contamination from ISR
events.
Requiring $\empz$ to be close to the nominal value also suppresses 
photoproduction background events, in which the scattered lepton escapes 
in the beam pipe.
\subsection{Monte Carlo Simulation}

\label{sec:mc}
The Monte Carlo (MC) simulation is used to correct for detector
acceptance and resolution effects.
The inclusive 
DIS signal events are generated using the DJANGOH~\cite{Schuler:1991yg}
event generator, which also contains a simplified simulation of  difractive
processes.
Elastic QED Compton events are generated using the COMPTON event
generator~\cite{Courau:1992ht}. The cross section measurement is
corrected for QED radiation up to order $\alpha$ using
HERACLES~\cite{heracles} which is included in DJANGOH. The radiative corrections are cross checked
with HECTOR~\cite{hector}. 
%An agreement to better than $0.3\%$ is found in the kinematic range of this measurement.

All generated events are passed through the full
GEANT~\cite{geant} based simulation of the H1 apparatus and are
reconstructed using the same program chain as  the data. 
Shower models are used to speed up the simulation in 
the LAr~\cite{gflash} and SpaCal~\cite{Glazov:2010zz} calorimeters.
The calibrations of the SpaCal and the LAr, as well as
the alignment, are performed for the reconstructed MC events in the
same way as for the data. 
%The simulated events are reweighted to the cross section derived
%from the QCD fit which is described in section~\ref{sec:qcdfit}.
%
%
\section{Data Analysis}
\label{sec:dataanalysis}
\subsection{Data Sets and Event Selection}
The analysis is based on several data samples 
which are listed in \Tab~\ref{tab:data}.
These samples are distinguished based on the kinematic region,
online trigger condition, offline selection criteria and proton beam energy.

\subsubsection{Online Event Selection}

\label{sec:data}

\renewcommand{\arraystretch}{1.}
\begin{table}
\begin{center}
\begin{tabular}{l|ccccc}
\hline\\[-11pt]
{\bf Sample}          & {\bf Years} & {\bf $\boldsymbol{Q^2}$ range} 
&{\bf $\boldsymbol{{\cal L}_{e^-p}}$  }& {\bf $\boldsymbol{{\cal L}_{e^+p}}$ }& {\bf Total $\boldsymbol{\cal L}$ }\\
                 &       & {\bf GeV$^2$} & {\bf \lunit}          &  {\bf \lunit}          &  {\bf \lunit} \\
\hline
\\[-11pt]
 \Hqcjc & $2003$-$2005$  & $8.5-90$  &$\lumicjcminus$ & $\lumicjcplus$ & $\lumicjc$ \\
 \Lqbst & $2006$-$2007$       & $2.5-12$ &$\lumibstminus$ & $\lumibstplus$ & $\lumibst$  \\
 \Mer     & $2007$ & $1.5-90$ &---  & $\lumimer$ & $\lumimer$ \\
 \Ler     & $2007$ & $1.5-90$ &---  & $\lumiler$ & $\lumiler$ \\
\hline
\end{tabular}
\end{center}
\tablecaption{\label{tab:data}Data samples used in the analysis with their 
 $Q^2$ coverage and integrated luminosities.}
\end{table}
\renewcommand{\arraystretch}{1.}

The two  high $y$ data samples for  $E_p=920$~GeV (`{\hqcjc}' and 
'{\lqbst}')
are collected with dedicated low energy SpaCal triggers which require at L1
a compact energy deposit in the SpaCal with energy above $2$~GeV. In addition,
to suppress non-$ep$ background, a CIP track segment pointing
to the nominal interaction vertex position is required. Several additional
veto conditions, which are based on scintillator counters 
positioned up- and downstream the 
nominal interaction region and on the 
hadronic section of the SpaCal are  used 
to further suppress non-$ep$ background.

The \hqcjc\ analysis uses an L2 trigger condition,
which requires that the energy deposit
is reconstructed in the outer SpaCal region,
at  distances from the beam line
of $R_{sp}\ge 38$~cm,
corresponding approximately to the inner CJC acceptance.  
The \lqbst\ analysis uses another L2 condition, which requires
$R_{sp}\ge 17$~cm corresponding to the acceptance of the BST. 
Both triggers are further filtered at L4 using fully reconstructed
events to validate low level trigger conditions.
  
For the reduced proton beam energy data, 
the main trigger is the SpaCal 
trigger % at low energy  
%with energy above 
with an energy threshold of $2$~GeV. 
In addition a track segment  has to be reconstructed
in the BST pad detector or the CIP. This trigger is complemented with 
a SpaCal trigger at a higher energy threshold of
$6.5$~GeV and no tracking condition. No L4 filtering is imposed for
the low $E_p$ runs.

The data are subject to offline cuts which are listed in \Tab~\ref{tab:selection}
and discussed below.
Some of the cuts are common to all data samples, others differ, primarily
because of the different tracking conditions
used for the scattered lepton validation.

In order to ensure an
accurate kinematic reconstruction and to suppress 
non-$ep$ background, 
the $z$ coordinate of the interaction vertex, $z_{\rm vtx}$, is required to be 
reconstructed close to the nominal position and with sufficient accuracy,
$\sigma(z_{\rm vtx})$. 

The scattered lepton is identified with the localised energy deposition 
(cluster) reconstructed in the SpaCal
calorimeter, which has the highest transverse energy $E_T$.
Here $E_T$ is calculated using the cluster energy and position, the event
interaction vertex position and the  beam line.
If the highest $E_T$ cluster does not satisfy all of the
identification cuts mentioned below, 
the cluster with the second highest $E_t$ is tried. 
The procedure 
is repeated for up to three clusters. If none of the three clusters 
satisfies the selection cuts, the event is rejected.
\subsubsection{Offline Event Selection} 
\renewcommand{\arraystretch}{1.2}
\begin{table}
\begin{center}
\begin{footnotesize}
\begin{tabular}{l|c|c|c}
\hline%\\[-10pt]
{\bf Selection criteria}  & {\bf Medium $\boldsymbol{Q^2}$ CJC} & {\bf Low 
$\boldsymbol{Q^2}$ BST} & {\bf %Reduced 
$\boldsymbol{E_p=460}$~GeV } \\
                    &$\boldsymbol{E_p=920}$~{\bf GeV}       & 
$\boldsymbol{E_p=920}$~{\bf GeV}& {\bf and }$\boldsymbol{E_p=575}$~{\bf GeV}  \\
\hline
 Vertex $z$ position & \multicolumn{3}{c}{$|z_{\rm vtx}|<35$~cm} \\
 Vertex $z$ precision  & \multicolumn{3}{c}{$\sigma (z_{\rm vtx})<8$~cm} \\
 Scattered lepton energy &  \multicolumn{3}{c}{$\ee > 3.4$~GeV} \\
 Radial cluster position & $40<R_{sp}<74$~cm & $18<R_{sp}<74$~cm & $18<R_{sp}<74$~cm \\
 Cluster transverse shape & \multicolumn{3}{c}{$R_{4}>0.8$} \\
                          & \multicolumn{3}{c}{$R_{\rm log}<4.5$~cm for $R_{sp}>60$~cm} \\
%$R_{4}>0.8$~cm & $R_{\log}<5$~cm & $R_4>0.8$ \\
                          & ${\tt ECRA}<4.5$~cm & ${\tt ECRA}<4.5$~cm  &  ---\\                          
 Energy in hadronic section & \multicolumn{3}{c}{$E_h/\ee<0.15$} \\
 Tracker validation        &   $D_{\rm CJC}<6$~cm   &  \multicolumn{2}{c}{$D_{\rm BC}<3$~cm}   \\
 Lepton charge             &  \multicolumn{3}{c}{ Agree with beam charge for $y\ge \ytrans$} \\     
 Energy-momentum match     &  ---             &   \multicolumn{2}{c}{$|\ee/P|>0.5$ for $\ee <7$~GeV}   \\
 Longitudinal momentum balance & \multicolumn{3}{c}{$\empz>35$~GeV} \\
 Total energy in hadronic SpaCal & --- & --- & $E_{h,{\rm tot}}<16$~GeV \\ 
 QED Compton rejection     & \multicolumn{3}{c}{Topological veto} \\
 Kinematic range           & $Q^2>7.5$~GeV$^2$  & $Q^2>2.37$~GeV$^2$   & $Q^2>1.33$~GeV$^2$ \\
\hline
\end{tabular}
\end{footnotesize}
\end{center}
\caption{\label{tab:selection}Selection criteria used for the analyses.}
\end{table}
\renewcommand{\arraystretch}{1.}
   
The energy of the scattered lepton is required to exceed $\ee>3.4$~GeV
to ensure a high trigger efficiency. 
The radial position of the scattered lepton is required to be well inside
the SpaCal acceptance and within the active trigger region.

Several cuts are applied to suppress photoproduction background. 
In photoproduction events, hadronic final state particles may scatter in
the SpaCal calorimeter and mimic the electron signal. The main sources of such
background are charged hadrons 
(pions, kaons and (anti-)protons) as well as $\pi^0\to \gamma\gamma$
decays for which one of the photons converts into 
an $e^+e^-$ pair prior to entering the 
tracking devices.  
The selection against photoproduction events
includes cuts on the transverse shower radius, estimated using logarithmic ($R_{\log}$) 
and square root ({\tt ECRA}) energy weighting \cite{Glazov:2010zz}, 
as well as the fraction of energy of the cluster  contained in the four highest energy cells, $R_4$.
The cut  $R_4>0.8$ is found to be more efficient than the cluster radius estimators,
but an L4 condition requires  the {\tt ECRA} cut 
to be used for the $E_p=920$~GeV analyses.
The transverse shape requirements are efficient against hadronic background
as well as background from $\pi^0 \to \gamma\gamma$
where the two photon clusters merge together.

The cut on the fraction of energy in the hadronic SpaCal
behind the lepton candidate cluster, $E_h/\ee<0.15$, 
rejects purely hadronic background. The cut does not
reject background for $R_{sp}> 60$~cm because of the limited acceptance of the 
hadronic SpaCal. As a compromise between signal efficiency and background rejection,
an extra cut $R_{\log}<4.5$~cm is introduced for $R_{sp}>60$~cm.
% for the  \lermer\ analyses.

The photoproduction background is suppressed further by requiring  cluster validation
by a track (``track link''). The \hqcjc\ analysis uses tracks reconstructed solely in the CJC tracker.
The other analyses use a dedicated reconstruction algorithm which combines information
obtained from the CJC, BST, event vertex and the SpaCal (``BC'' algorithm~\cite{sebastian}). 
The tracks are extrapolated to 
the SpaCal position and required to match the SpaCal cluster within $D_{\rm CJC}<6$~cm for 
the CJC reconstruction and within $D_{\rm BC}<3$~cm for the BC 
algorithm.
The tighter cut on the 
track-cluster matching 
for the BC compared to the CJC algorithm is possible for low $R_{sp}$ because of accurate 
BST $\thetae$ reconstruction
and for higher $R_{sp}$ because the SpaCal cluster is used in the BC algorithm
and  pulls the track towards the cluster.
As discussed below,  in \Sec~\ref{sec:background}, the measured scattered lepton 
charge is required to match
the beam charge for $y>\ytrans$. The sample for which the charges are 
different is 
used to estimate the remaining background. For $\ee <7$~GeV, the momentum reconstruction 
is accurate enough and  $|\ee/P_e|>0.5$ is required, where
$P_e$ is the track momentum of the electron candidate.

The total energy reconstructed in the hadronic section of the SpaCal,
$E_{h,{\rm tot}}$, is required to be below $16$~GeV for the \lermer\ data.
This  avoids a trigger inefficiency arising from a veto on the total
energy deposited in the 
hadronic SpaCal.

Events with high energy  initial state photon radiation  are rejected by requiring 
$\empz >35$~GeV. This cut is also efficient against the photoproduction 
background. 
The QED Compton process, $ep \to ep\gamma$,
is suppressed using a topological cut against events with two 
back-to-back electromagnetic clusters reconstructed in the SpaCal.

\begin{figure}
\epsfig{file=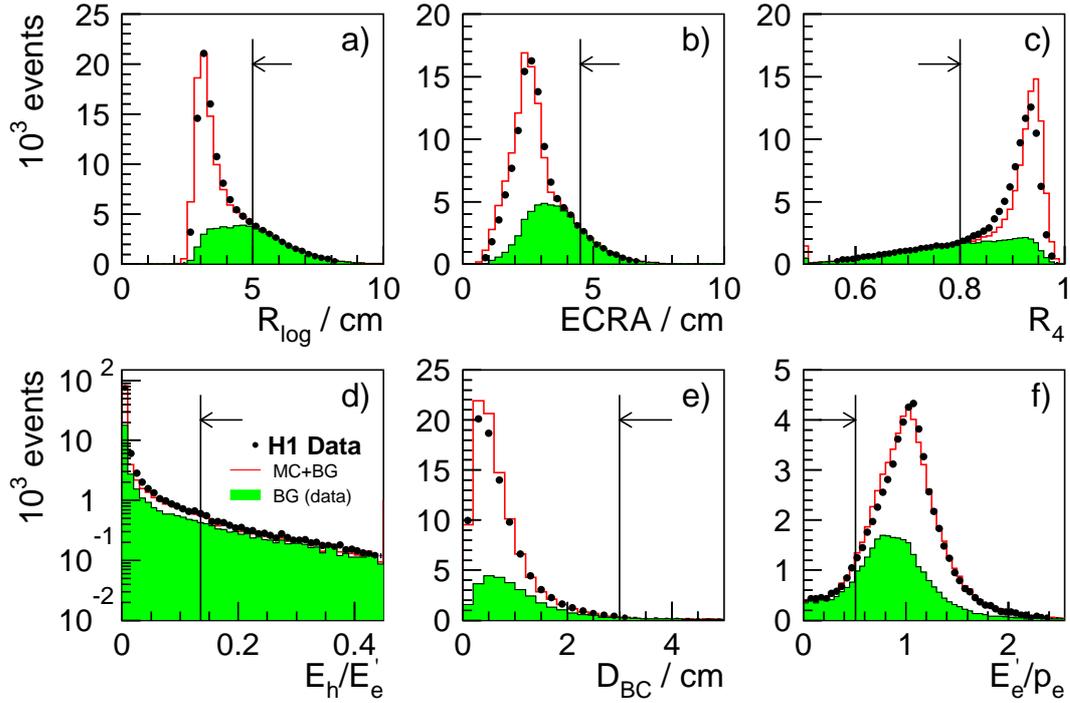,width=\linewidth}
\caption{\label{fig:eid}Distribution of variables used 
for the scattered lepton identification, for the \ler\ sample: 
a) logarithmic cluster radius
$R_{\log}$; b) square root weighted cluster radius {\tt ECRA}; c) 
fraction of energy in the four hottest cells of the cluster $R_4$;
d) ratio of the energy in the hadronic section of the SpaCal in the cone 
behind the electron candidate to the electron candidate energy $E_h/E'_e$;
e) distance between BC track extrapolated to the SpaCal cluster $Z$ and the 
cluster $D_{BC}$; f) ratio of the lepton candidate energy to the track momentum
$\ee/p_e$. The  data are shown as dots, 
the background as shaded histograms and
the sum of the signal MC simulation and the background as open histograms. 
The vertical lines indicate the value of the
electron identification cuts. The arrow points in the
 direction of the part of the distribution kept by the cut.
The data are presented for $\ee<10$~GeV except for the $\ee/p_e$ plot which is shown for $\ee<7$~GeV.
%Data are shown as dots with error bars, 
%the shaded histograms show the data driven estimation of the background
%and the open histograms represent
%the simulation of  DIS signal.
}
\end{figure}

Distributions of the variables which are used for the scattered lepton 
identification are shown for the \ler\ sample in \Fig~\ref{fig:eid}. 
While the shapes observed in the data are sometimes not perfectly reproduced by the simulation,
those differences occur far from the cut values.
The electron identification selection 
criteria are designed to have high efficiency for the signal while rejecting 
a significant amount of the background.

\subsection{Background Subtraction}
\label{sec:background}
\begin{figure}
\end{figure}
At low $E'_e$, corresponding to high $y$, 
the background contribution after the event
selection is of a size comparable to the  DIS signal.
To reduce the systematic uncertainty,  
the background determination in this analysis
         relies on data.
Two distinct methods are applied depending on the event inelasticity $y$.
For high  $y\ge \ytrans$, 
the background estimation is based on
a sample of events, in which the charge of the lepton candidate
is opposite to the beam charge (``wrong charge  method''). 
For lower $y < \ytrans$  the background contamination is small, however, the uncertainty 
due to  the charge determination becomes large, 
and an
alternative
method is employed. In this method the background is estimated using 
a sub-sample of events, in which the scattered lepton
is detected in the electron tagger (``tagger method'').

The wrong charge subtraction method relies on the approximate charge
symmetry of the background and a good charge reconstruction
with the tracker at low  momenta. 
The residual charge asymmetry of the background is defined as
\begin{equation}
\kappa_+ = \frac{\textstyle N^{\rm bg}_+}{\textstyle N^{\rm bg}_-},~~~~~~~\kappa_- = 1/\kappa_+ = \frac{\textstyle N^{\rm bg}_-}{\textstyle N^{\rm bg}_+}\,,
\end{equation}
where $N^{\rm bg}_{\pm}$ is the number of background events in which the
lepton candidate is associated with a positively (for $+$) and a negatively 
(for $-$) charged track. The charge asymmetry of the background
arises from the different response of the SpaCal to particles compared
to antiparticles (in particular $p$ and $\bar{p}$) and detector misalignments. 
The asymmetry depends  on the electron identification cuts
since they alter the ratio of the electromagnetic to hadronic 
components
of the background.

The charge asymmetry is measured directly from the data
by comparing  background  estimates from the $e^+p$ and $e^-p$ scattering periods
as well as using clean background samples with the scattered
lepton measured in the electron tagger. The charge asymmetry of the 
background is found
to deviate from unity by $5\%$ and $1\%$ for scattering angles between
$155^{\circ}$ and $174^{\circ}$. 

The scattered lepton charge may be misidentified which leads to an
overestimation of the background at $y>\ytrans$.
The charge reconstruction is studied in the background free sample at
$\ee > 15$~GeV by comparing events with correctly and wrongly reconstructed
lepton charge. The fraction of wrongly reconstructed events depends on
both the energy and the angle of the scattered electron. This dependence is
well reproduced by the simulation. The fraction is smaller at small $\ee$ due
to a larger track curvature and at small $\thetae$ 
since the CJC has a better momentum resolution than the BST.

\begin{figure}
\epsfig{file=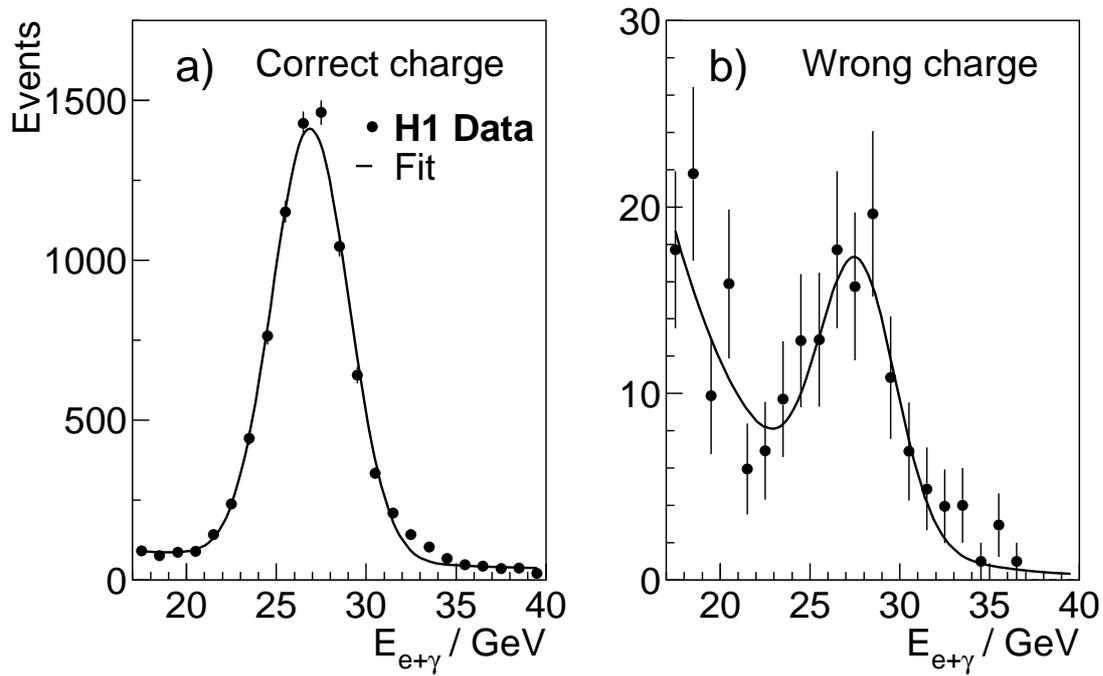,width=0.9\linewidth}
\caption{\label{fig:charge} 
Sum of energies of the scattered
electron and the photon for correct (a) and wrong (b) charge of 
the lepton candidate,
for the dedicated selection of events
with hard initial state radiation. 
The distributions are fitted by the sum of a 
Gaussian and exponential distributions, the fit result is shown by the line.
The data are taken from the \lermer\ samples.}
\end{figure}

Charge reconstruction at low energy ($E'_e < 15$~GeV) is 
studied using 
events with initial state radiation in which 
the radiative photon is detected in the photon tagger. For these events,
the sum of the scattered electron and photon energies, $E_{e+\gamma}$, 
peaks at the beam energy which allows the estimation of the residual background
using a side-band method. This procedure is illustrated for the combined
\lermer\ dataset   in \Fig~\ref{fig:charge}.
The DIS signal is approximated by a Gaussian while the background is assumed
to follow an exponential distribution. 
The Gaussian width of the signal distribution is fixed to be the same 
for both lepton candidate charges. The data are fitted by the sum of signal
and background hypotheses.
From these fits, the fraction of events with wrongly
reconstructed charge in the data is  
determined to be $(1.1 \pm 0.2_{\rm stat}) \% $, compared
to $0.6\%$ in the simulation. The simulation is corrected for the $0.5\%$
difference and a systematic uncertainty of $0.5\%$ is used for the charge
determination of the lepton candidate.

The tagger method of background estimation used at low $y<\ytrans$ relies on an accurate determination
of the tagger acceptance, $A_{\rm tag}$, which is defined in this analysis
as the fraction
of background events in which the scattered electron is tagged. The 
acceptance is measured by comparing all wrong charge events with $\ee<8$~GeV passing nominal selection cuts 
to those in which a scattered electron candidate is detected
in the electron tagger.
For this selection, the wrong charge sample
is almost entirely comprised of photoproduction events with a small admixture 
of DIS events with charge misidentification, which is subtracted using the MC estimate.  
The tagger acceptance is seen to vary between
$(16.9\pm0.2_{\rm stat})\%$ for the \hqcjc\ $e^-$ sample, taken in 
the year $2005$,
and $(20.5\pm 0.4_{\rm stat})\%$ for the \mer\ sample. This difference 
in acceptance may be  explained by differences in the beam optics.
Stability of the tagger acceptance for different kinematic ranges is studied 
by varying the $\thetae$ and $\ee$ cuts.
A systematic uncertainty of $20\%$ is assigned to the tagger acceptance.
This uncertainty  also accounts 
for a potential variation of the acceptance as a function
of $E'_e$ and $\thetae$.
 Finally, to avoid a subtraction of overlapping DIS and Bethe-Heitler
events containing energy deposits in the electron tagger, in the background
estimation the tagged events
are also required to have a charge opposite to  the lepton beam charge.
Neglecting the background charge asymmetry, this 
reduces the number of tagged background events by a factor of two.

To summarise, the number of signal events for the $e^+p$ and $e^-p$ 
running periods is estimated as  
\begin{equation}
 N^{e^{\pm}p}_{\rm sig} = \left\{ 
\begin{array}{ll}
\textstyle N^{e^{\pm}p}_{\pm} - \kappa_{\pm} N^{e^{\pm}p}_{\mp}~&\mbox{for }y \ge \ytrans\,,\\
\textstyle N^{e^{\pm}p}    - \frac{\textstyle 2}{\textstyle A_{\rm tag}} N^{e^{\pm}p}_{\mp \rm tag}~&\mbox{for }y < \ytrans\,.
\end{array}
\right.
\end{equation}
Here $N^{e^{\pm} p}_{\pm}$ ($N^{e^{\pm}p}_{\mp}$) is the number of events
with the charge of the lepton candidate  the same as (opposite to) 
the lepton beam charge 
and $N^{e^{\pm}p}_{\mp \rm tag}$ is the number of tagged events with the
charge of the lepton candidate  opposite to the
lepton beam charge.

\subsection{Efficiencies}
The efficiency of the electron identification cuts is studied in the data and MC, and the
simulation is adjusted accordingly.  
\subsubsection{Online Selection and Vertex Efficiency}
The efficiencies of the triggers used in the analysis 
are determined using events collected with independent triggers. The 
efficiency
of the L1 energy condition  is checked with  tracker-based triggers.
It is found to be fully efficient for  $\ee>3$~GeV. 

The efficiency of the L1  tracking condition (CIP or BST for \lermer\ data) is checked
using events triggered by  SpaCal-based triggers without tracking conditions.
The efficiency of the L1 tracking condition is correlated with the vertex
efficiency. To avoid biases, 
a combined  efficiency is calculated as 
\begin{equation}
 \epsilon_{\rm (CIP||BST) \& VTX} = \epsilon_{\rm CIP||BST} \cdot 
\epsilon_{\rm VTX|_{(CIP||BST)}}
\end{equation}  
where $\epsilon_{\rm CIP||BST}$ is the CIP or BST L1
condition efficiency based on events
without vertex cut and $\epsilon_{\rm VTX|_{(CIP||BST)}}$ is the vertex efficiency
for the events passing the CIP or BST tracking condition. A similar decomposition
is used for the $\epsilon_{\rm CIP  \& VTX}$ condition, used for $E_p=920$~GeV data.

The {\rm CIP} L1 condition is found to have uniform efficiency for 
$R_{\rm sp}>30$~cm, i.e. within the CIP acceptance. The efficiency
varies for different periods between $95\%$ and $98.5\%$.
For lower radii, the efficiency decreases by about $2\%$ 
since the scattered lepton leaves the CIP acceptance. This region
is covered by the BST pad detector and for the 
combined ${\rm CIP||BST}$ condition, there is no drop of the efficiency. 

The efficiency $\epsilon_{\rm VTX|_{(CIP||BST)}}$  decreases at low $R_{\rm sp}<40$~cm
and high $y>0.5$ since the scattered lepton as well as part of the 
hadronic final state leave the CJC acceptance.
The decrease in the efficiency occurs mostly for  diffractive events
which have a rapidity gap between the proton remnant and the struck quark.
For $R_{\rm sp}<20$~cm at 
$y=0.85$, the inefficiency in the data reaches $8\%$ compared to $5\%$ in 
the Monte Carlo simulation.
The difference in the efficiency is parameterised as a function of 
$R_{\rm sp}$ and $\ee$ and is applied to the simulation.

\subsubsection{Offline Selection Efficiency}

\label{sec:eid}
The determination of the offline electron selection efficiencies at high 
$y$ is complicated due to the large background contamination.
%This contamination increases if the electron identification cuts are not 
% applied. 
Thus an accurate estimation of the background is a matter of paramount
importance for this analysis. As discussed in \Sec~\ref{sec:background},
this estimation is provided by the wrong charge
data sample, for which  a 
track link is required for the lepton candidate.

The track-link  efficiency  is measured using a background free sample with 
$17 < \ee < 22$~GeV and no track condition, as the fraction of events satisfying
the track link requirement. 

For the \hqcjc\ sample, the SpaCal cluster is 
linked with a 
track from the CJC. It is observed that the track-link efficiency
has a radial dependence which is well reproduced 
by the MC simulation. 
For the  overall level of inefficiency, however, the MC prediction
has to be
downgraded by  $3.6\%$, $3.3\%$ and $5\%$ for the $2003$-$2004$, 
$2004$-$2005$ and the $2006$-$2007$ running  periods, respectively. 
The track-link efficiency for the {\lqbst}, \lermer\ analyses
has radial and azimuthal dependencies in the BST and BST-CJC overlap
regions. It has 
a typical value of $90\%$ but drops in some regions to $75\%$. 
The simulation is corrected in radial steps of $2$~cm in
the $18\le R_{sp}\le 45$~cm range
for the $12$~BST azimuthal sectors individually. This correction does not 
exceed $10\%$.
The correction factors to the MC
%, which are determined at intermediate $\ee$,
are applied for all lepton energies. 
A cross check for low $\ee$ events is performed using a sample of ISR events
for which a good agreement between the data and corrected MC is observed. 

\begin{sloppypar}
The transverse and longitudinal distributions of the electromagnetic
shower energies (``show\-er shapes'')
are affected by the amount of material passed through by the
scattered lepton before entering the calorimeter. The total amount of material
before the SpaCal depends on the scattering angle and
varies between $1.7$ and $1.2$ radiation lengths. A detailed map
of the detector material is included in the simulation.
The distribution of the material
in the CT is checked using reconstructed photon conversions and
nuclear interactions. A
significant contribution  to the material budget is due to the BST sensors,
cooling circuit and readout electronics. The contribution due to the readout
electronics 
%can not be checked using secondary verticies, this contribution
is determined using the transverse shower shapes. The method exploits the fact
that the BST was removed from the 
H1 detector for repair during the $2005$ data taking.
A comparison of the shower shapes in the data 
and MC simulation for the 2005 and
2006 data taking periods thus allows a check of the BST material
contribution with high accuracy.
\end{sloppypar}

The signal efficiencies of the other offline selection cuts 
listed in \Tab~\ref{tab:selection} are studied after the background 
subtraction described in \Sec~\ref{sec:background}. 
%These efficiencies were studied for
%different $y$ and $Q^2$ intervals and it was observed that MC describes 
%inefficiency in data within 2~$\%$.
Since the background is large and its level  
typically varies significantly  upon applying these cuts, 
a variation of the background asymmetry   
is also considered for each cut of the electron identification. 
The background subtraction is the 
dominant systematic uncertainty for the efficiency determination.
It is measured with $3\%$ accuracy for $y>0.8$, with $1.5\%$ for $0.7<y<0.8$ and
 with $1\%$ for $y<0.7$.
%An efficiency is 
%determined as a ratio of signal events after and before the cut as,
%\begin{equation}
%\epsilon = {{N^{cut} - N^{BGcut}} \over {N^{all} - N^{BGall}}} 
%\end{equation}
%where $N^{cut}$ and $N^{all}$ are number of events with a correct-sign 
%track linked to a cluster after and before a cut. $N^{BGcut}$ and 
%$N^{BGall}$ are number of 
%remaining events with a correct-sign track after and before a cut
%defined as,
%\begin{eqnarray}
% N^{BGcut} &=& N^{WCcut} \cdot k^{cut}  ~ ,\\
% N^{BGall} &=& N^{WC} \cdot k  ~ ,
%\end{eqnarray}  
%with $k^{cut}$ and $k$ being background asymmetry factors after 
%and before a cut. $N^{WCcut}$ and $N^{WC}$ are number of events with a 
%wrong-charge track linked to a SpaCal cluster after and before a cut.
%Since background contribution, and possibly the charge-asymmetry factor, 
%vary with inelasticity, $y$, $Q^2$ dependence of selection cut 
%efficiencies was obtained for different $y$ intervals. From this study, no
%$Q^2$ dependence of efficiency is observed and MC efficiency describes 
%efficiency from data within 2$\%$.  
\subsection{Calibration and Alignment}
\label{sec:calib}
The alignment and calibration of the H1 detector follows a procedure
similar to that  described in~\cite{H1:2009bp}. The alignment starts with the internal
alignment of the CT and proceeds to the backward detectors,
SpaCal and BST. The alignment of the BST sensors is performed with the
minimisation package Millepede~\cite{Blobel:2002ax} by using
position information from the central tracker and the SpaCal. 
The global alignment  of the BST is refined by requiring a matching between
the momentum and energy measurements in the BST, CJC and the SpaCal~\cite{sebastian}. 

The calibration of the SpaCal electromagnetic energy scale
uses the double angle method\,\cite{H1:2009bp}. 
The linearity of the SpaCal energy response is checked using $\pi^0\to \gamma\gamma$,
$J/\psi \to e^+e^-$ and QED Compton events.

The hadronic final state is reconstructed using 
information from the central tracker, the
 LAr calorimeter and the SpaCal. 
The tracker momentum scale is checked by reconstructing narrow resonances such as $K^0_S \to \pi^+\pi^-$ and 
$\Lambda \to p^{\pm}\pi^{\mp}$ decays. The hadronic calibration of
the LAr calorimeter employs the
 transverse energy balance between the scattered electron and 
the hadronic  final state as described in~\cite{H1:2009bp}. 
The hadronic calibration  of the SpaCal employs the longitudinal momentum balance. 
The relative contribution of the SpaCal to $\empz$ becomes large at high $y$
and  the absolute calibration is obtained for $ \ee < 15$~GeV by requiring 
$\empz$ to peak at $2 E_e$ for both the data and the
Monte Carlo simulation. 
The calibration constants are determined separately for 
the electromagnetic and hadronic sections of the SpaCal.
% by comparing data and Monte Carlo distributions.
%%\subsection{Background Subtraction}
%%\input{background}
\subsection{Radiative Corrections}
\label{sec:radcor}
For large inelasticity $y>0.5$ and low $x$ the kinematics reconstruction using
the electron method is prone to large radiative corrections which can reach
a level of more than $50\%$ of the Born cross section. Studies 
based on the DJANGO and HECTOR programs show that the largest radiative
contribution arises because of hard initial state radiation 
from the incoming lepton.

The hard ISR process is strongly suppressed by the cut $\empz>35$~GeV.
After this cut, the radiative corrections amount to about $10\%$ of the Born
cross section with no strong dependence on $y$. A slight increase in
 the corrections occurs at the highest $y>0.7$ due to QED Compton events. 
These events
are efficiently rejected using the topological cut against two back-to-back
clusters in the SpaCal.

\begin{figure}
\epsfig{file=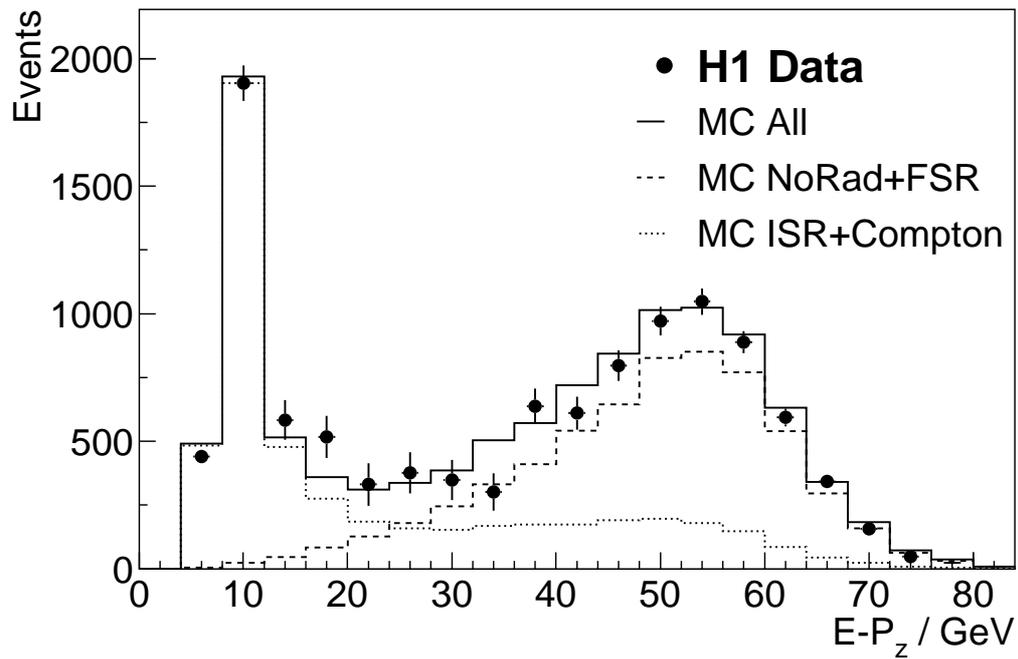,width=0.9\linewidth}
\caption{\label{fig:empzrad}
$\empz$ distribution for the \mer\ sample passing all cuts excluding cut on $\empz$
for $\ee<5$~GeV. The dots show the background subtracted data with statistical errors,
 the solid line is the total MC prediction, the dashed line, labeled MC NoRad+FSR, 
is the sum non-radiative 
and final state radiation components and the dotted line, labeled MC ISR+Compton,
 shows
the sum of initial state radiation  and QED-Compton contributions. 
}
\end{figure}

Events rejected by the cut $\empz>35$~GeV can be used to study the description 
by the simulation of the hard ISR. This is illustrated in \Fig~\ref{fig:empzrad} which shows the background subtracted
$\empz$ distribution for  events passing all cuts excluding the $\empz$ cut for the \mer\ sample. The sample is restricted to $\ee<5$~GeV which corresponds to $y>0.8$.
A prominent peak for $\empz \approx 10$~GeV corresponds to the hard ISR process.
The data in this kinematic region are well described by the simulation.
%\subsection{Electron Identification Efficiency} -- move to efficiencies
%\input{eid}
\subsection{Control Distributions}
\begin{figure}
\vspace{1cm}
\centerline{\epsfig{file=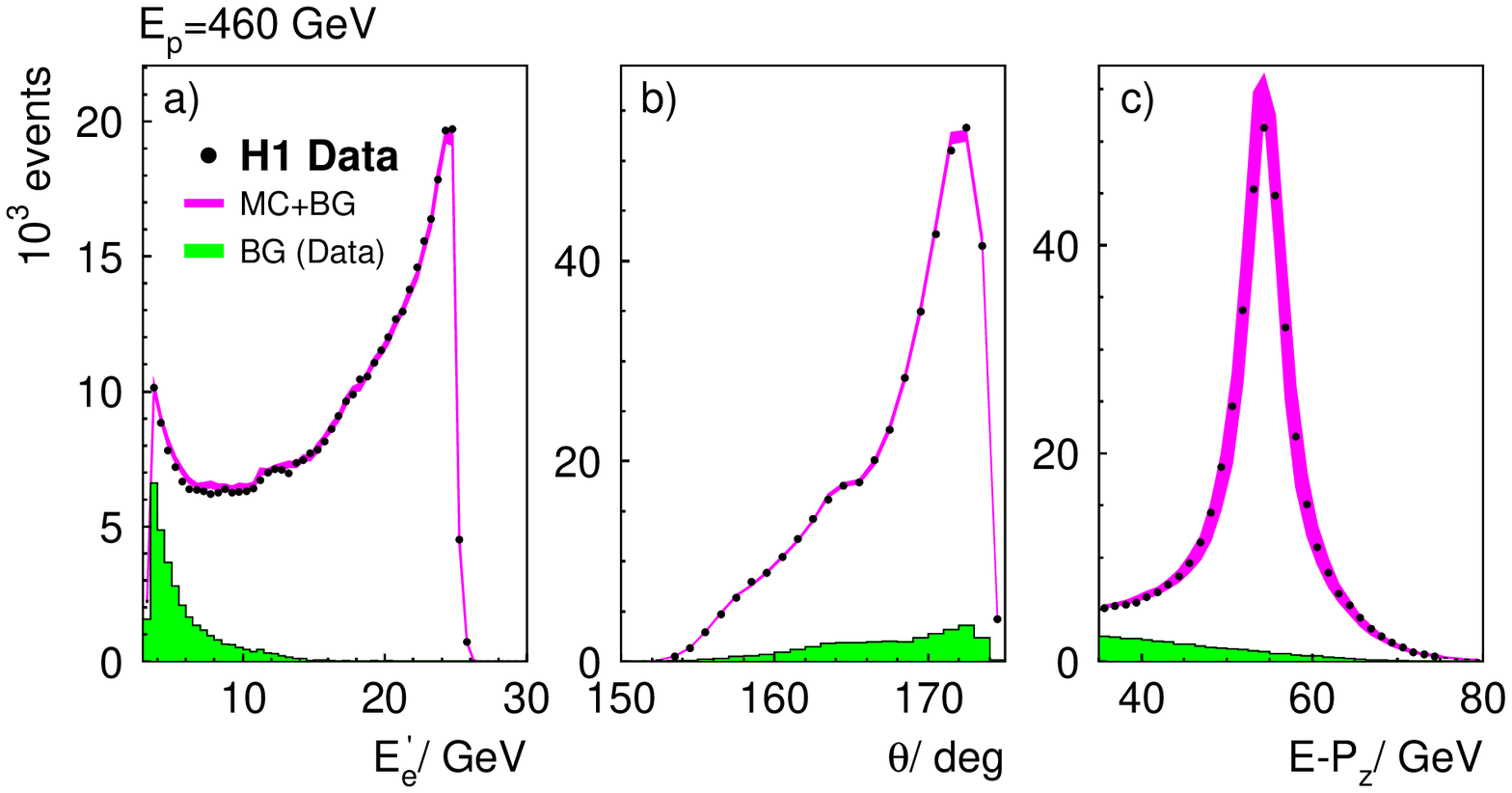,width=0.9\linewidth}}
\vspace{-7.cm}
\centerline{\epsfig{file=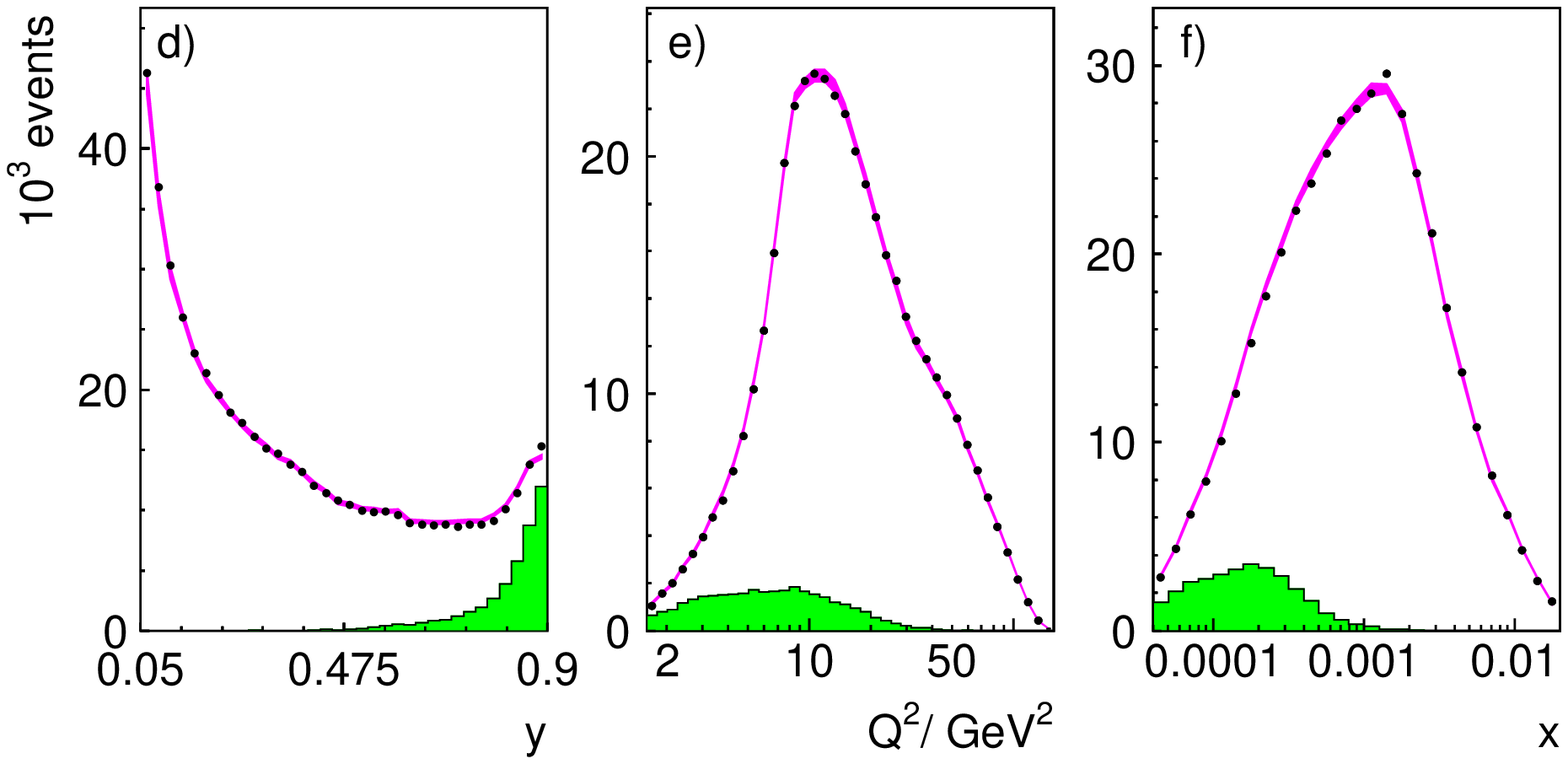,width=0.9\linewidth}}
\vspace{-7.cm}
\caption{
Distributions of the scattered electron energy $E'_e$ (a), 
polar angle $\thetae$ (b), $\empz$ (c) and of the kinematic variables $y$ (d), 
$Q^2$ (e), $x$ (f) for events
passing all analysis cuts from \ler\ sample.
Data are shown as dots with statistical errors, 
the shaded histograms show the data driven estimation of the background
and the shaded bands represent
the simulation of DIS signal 
with statistical and systematic uncertainties added
in quadrature.
%Data are shown as dots with error bars, 
%open histogram represents the simulation of DIS
%signal and shaded histogram shows the data driven estimation of the background.
 \label{fig:cont460}}
\end{figure}

\begin{figure}
\vspace{1cm}
\centerline{\epsfig{file=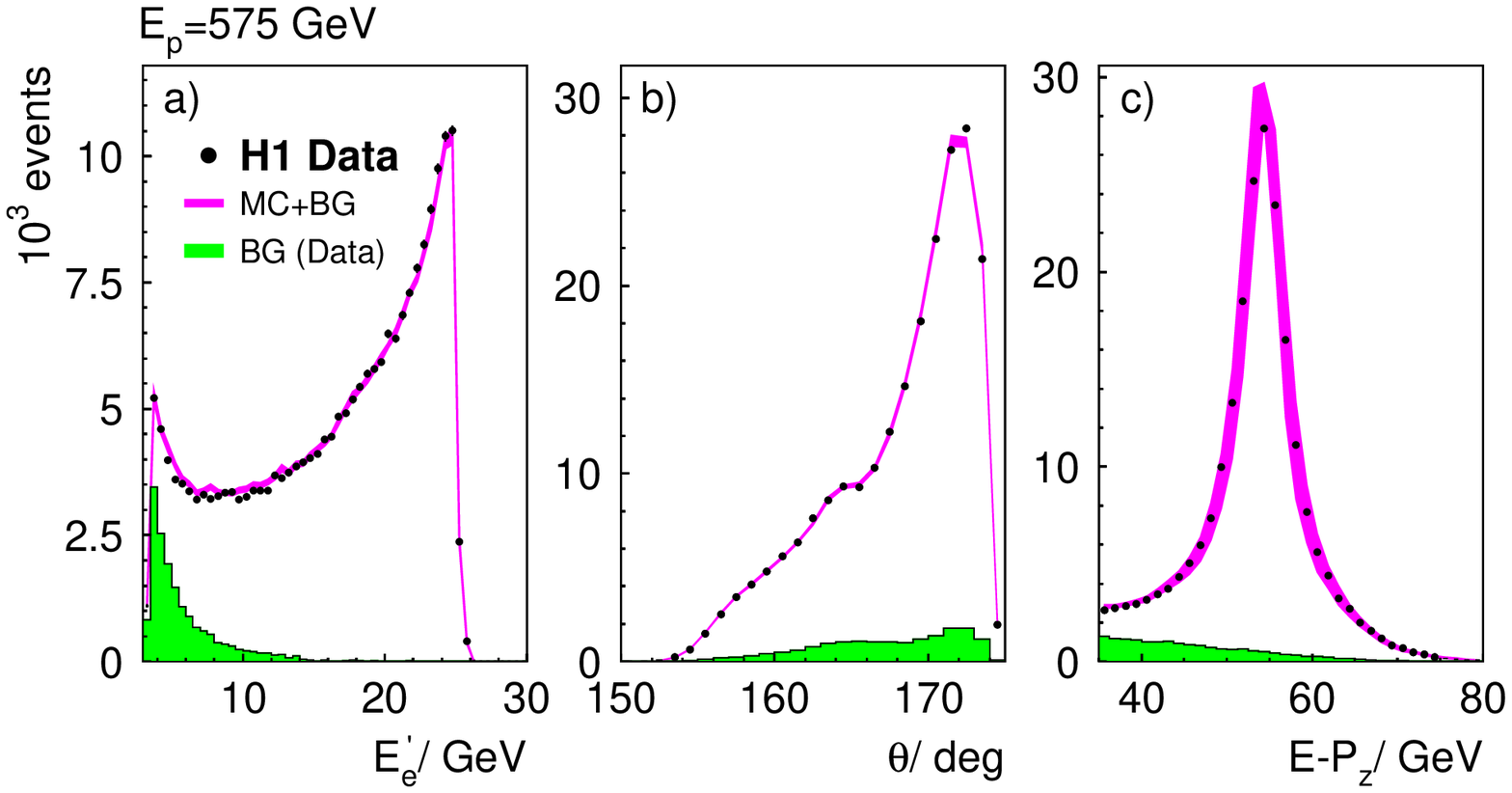,width=0.9\linewidth}}
\vspace{-7.cm}
\centerline{\epsfig{file=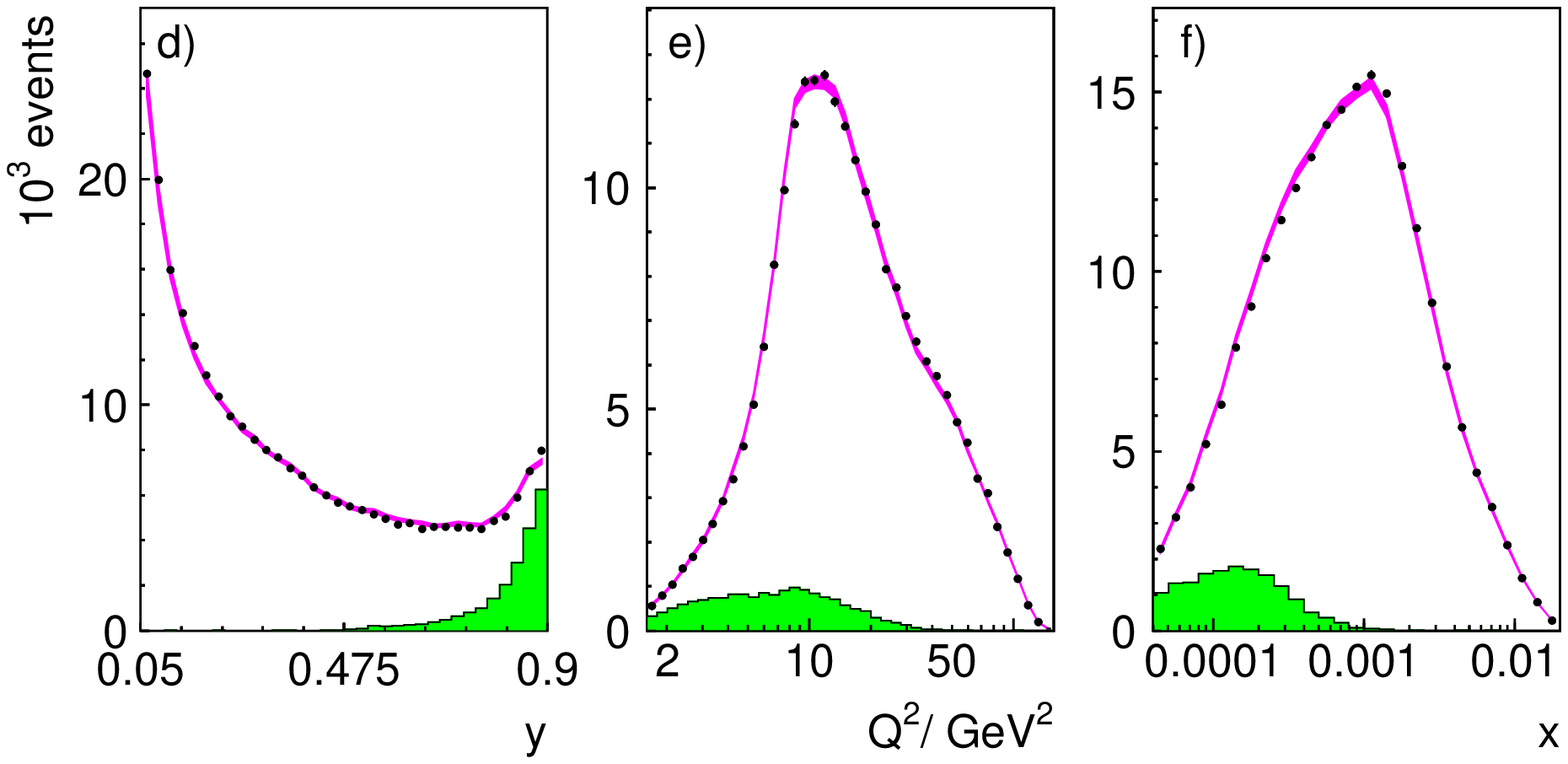,width=0.9\linewidth}}
\vspace{-7.cm}
\caption{
Distributions of the scattered electron energy $E'_e$ (a), 
polar angle $\thetae$ (b), $\empz$ (c) and of the kinematic variables $y$ (d), 
$Q^2$ (e), $x$ (f) for events
passing all analysis cuts from \mer\ sample.
Data are shown as dots with statistical errors, 
the shaded histograms show the data driven estimation of the background
and the shaded bands represent
the simulation of DIS signal 
with statistical and systematic uncertainties added
in quadrature.
%Data are shown as dots with error bars, 
%open histogram represents the simulation of DIS
%signal and shaded histogram shows the data driven estimation of the background.
 \label{fig:cont575}}
\end{figure}

\begin{figure}
\vspace{1cm}
\centerline{\epsfig{file=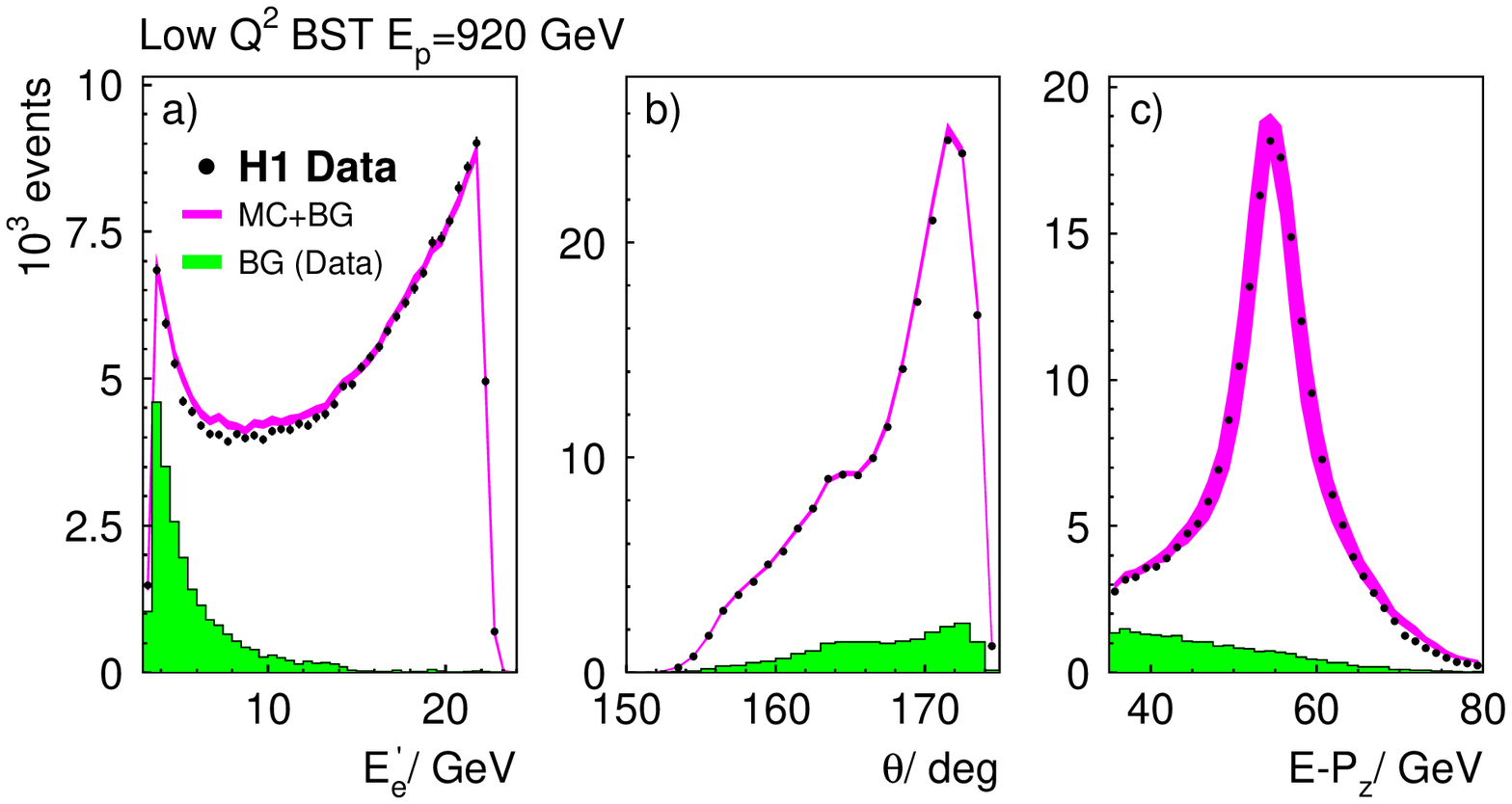,width=0.9\linewidth}}
\vspace{-7.cm}
\centerline{\epsfig{file=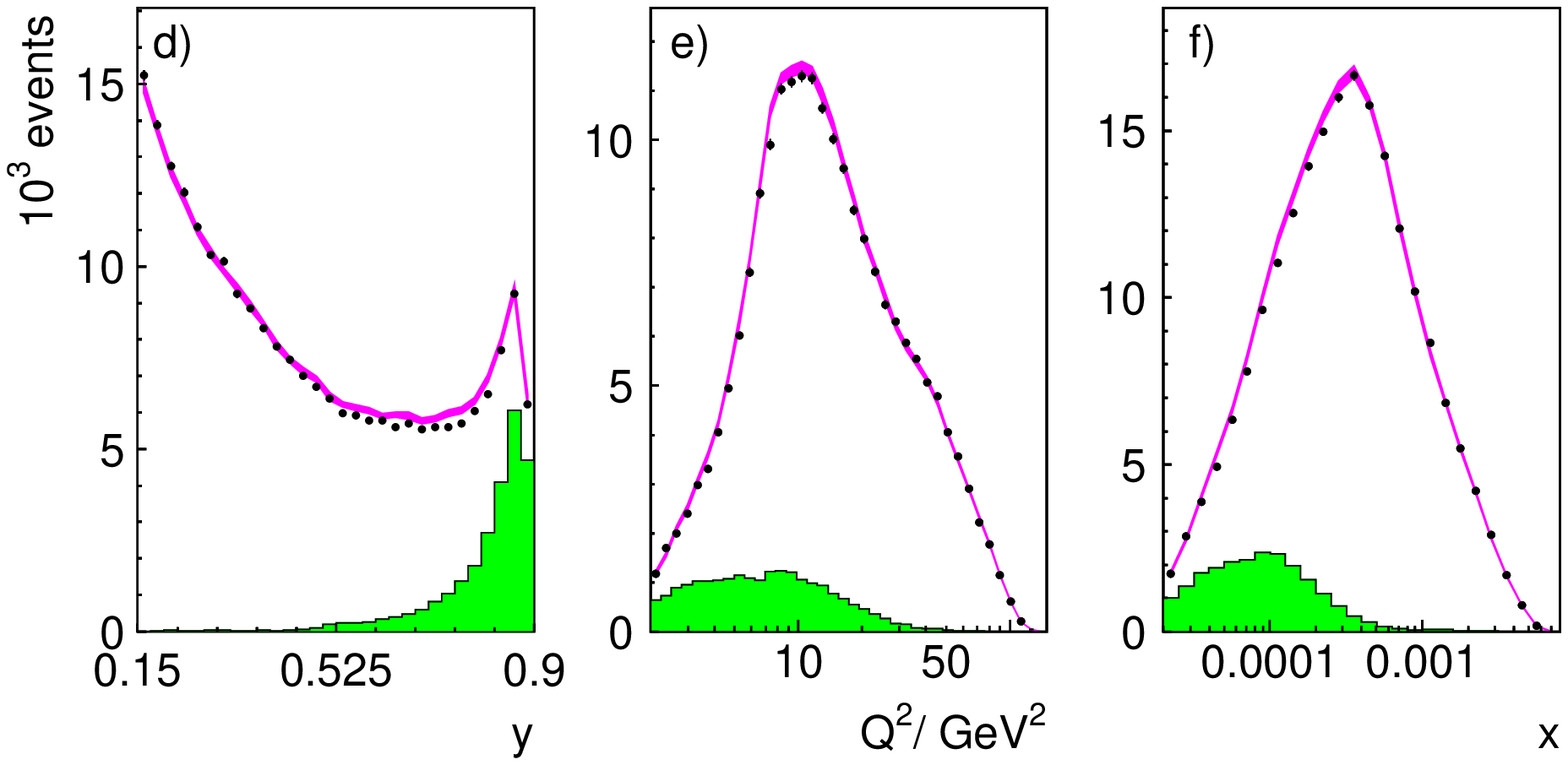,width=0.9\linewidth}}
\vspace{-7.cm}
\caption{
Distributions of the scattered electron energy $E'_e$ (a), 
polar angle $\thetae$ (b), $\empz$ (c) and of the kinematic variables $y$ (d), 
$Q^2$ (e), $x$ (f) for events
passing all analysis cuts from \lqbst\ sample.
Data are shown as dots with statistical errors, 
the shaded histograms show the data driven estimation of the background
and the shaded bands represent
the simulation of DIS signal 
with statistical and systematic uncertainties added
in quadrature.
%Data are shown as dots with error bars, 
%open histogram represents the simulation of DIS
%signal and shaded histogram shows the data driven estimation of the background.
 \label{fig:cont920a}}
\end{figure}

\begin{figure}
\vspace{1cm}
\centerline{\epsfig{file=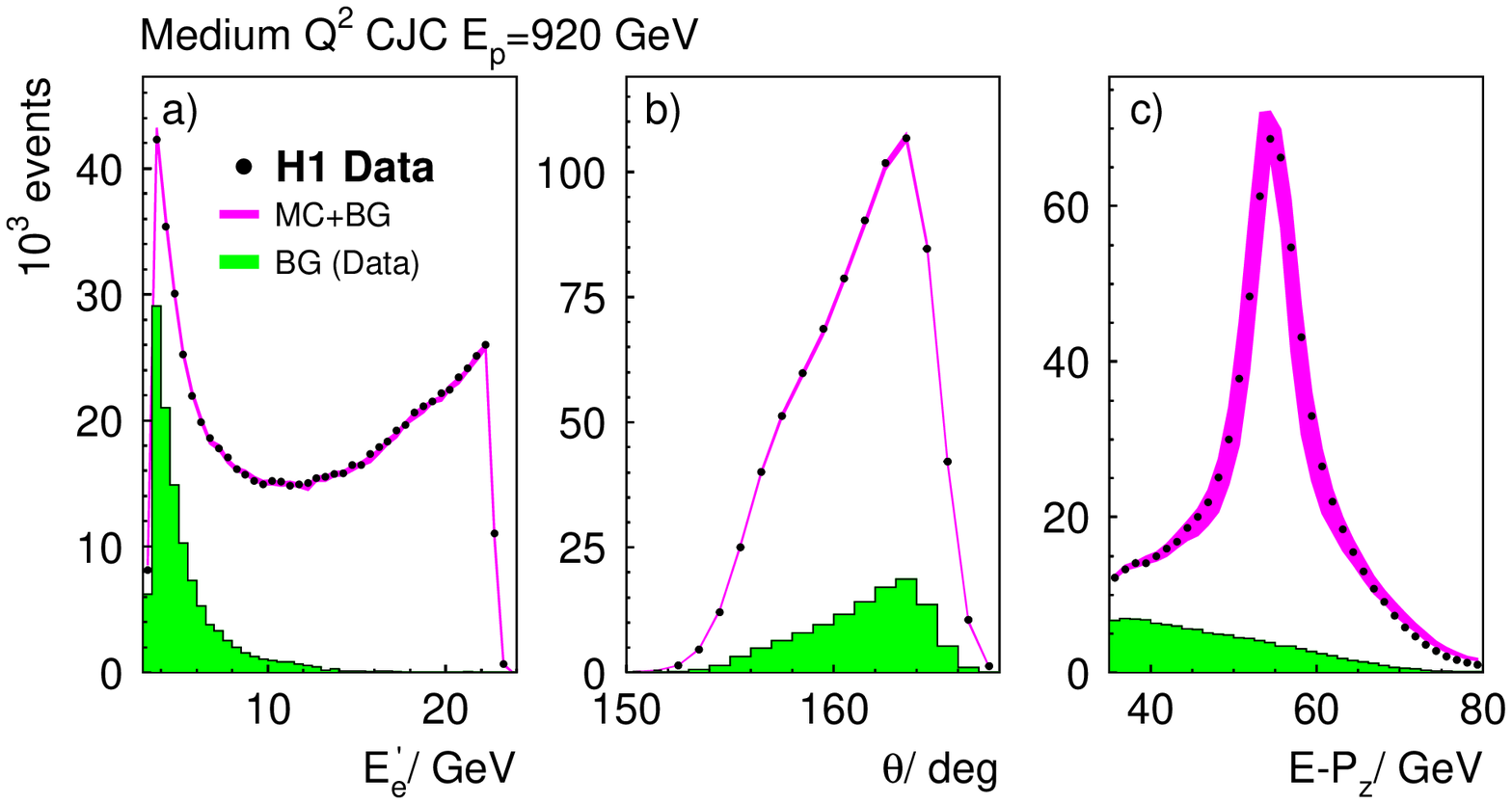,width=0.9\linewidth}}
\vspace{-7.cm}
\centerline{\epsfig{file=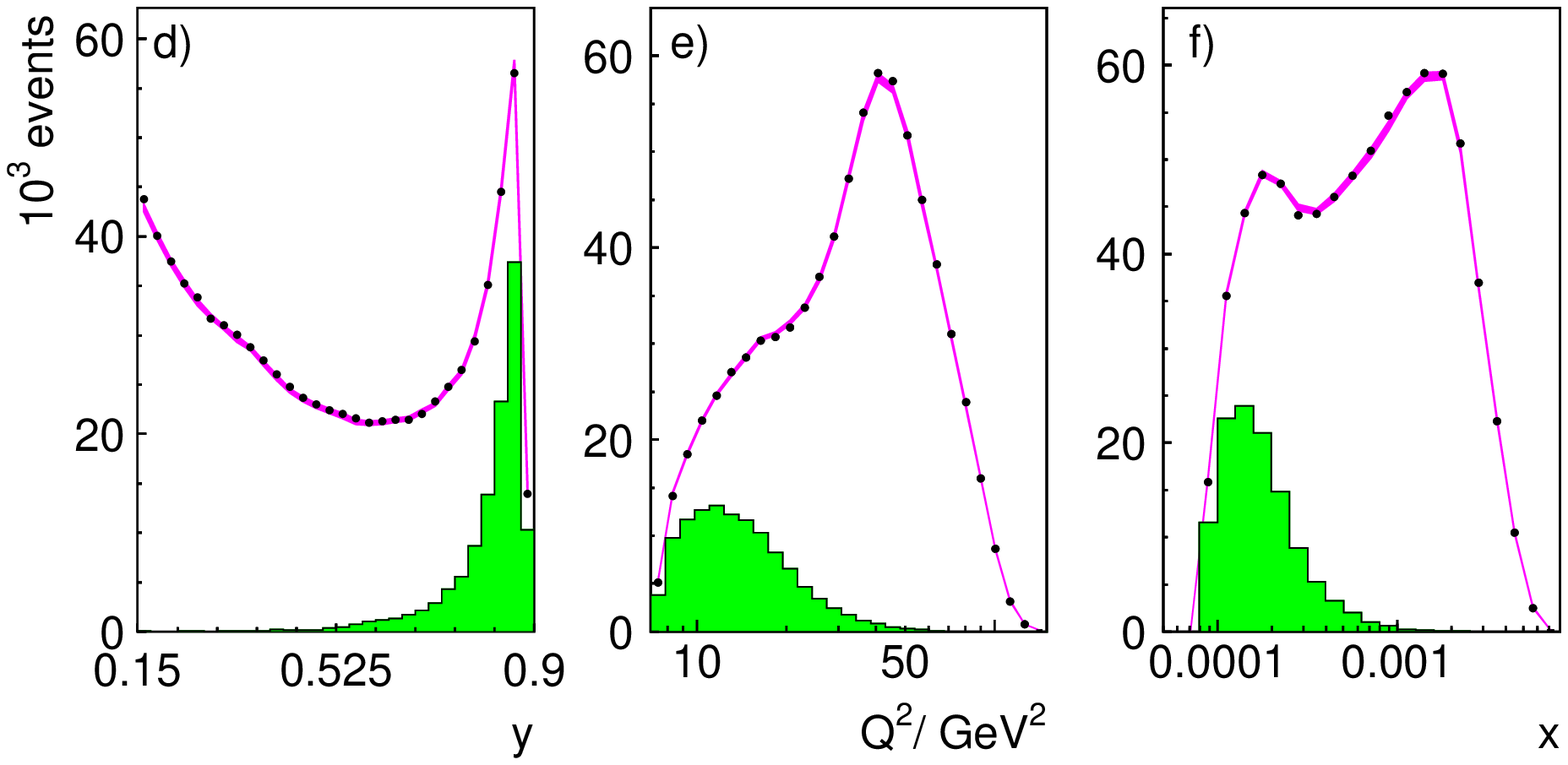,width=0.9\linewidth}}
\vspace{-7.cm}
\caption{
Distributions of the scattered electron energy $E'_e$ (a), 
polar angle $\thetae$ (b), $\empz$ (c) and of the kinematic variables $y$ (d), 
$Q^2$ (e), $x$ (f) for events
passing all analysis cuts from \hqcjc\ sample.
Data are shown as dots with statistical errors, 
the shaded histograms show the data driven estimation of the background
and the shaded bands represent
the simulation of DIS signal 
with statistical and systematic uncertainties added
in quadrature.
%Data are shown as dots with error bars, 
%open histogram represents the simulation of DIS
%signal and shaded histogram shows the data driven estimation of the background.
 \label{fig:cont920b}}
\end{figure}

Data and MC distributions of the main quantities used
to reconstruct the event kinematics for the events passing all selection 
criteria are compared in \Figs~\ref{fig:cont460} to \ref{fig:cont920b}
for all data sets included in the analysis.
The MC distributions are normalised to the integrated luminosity and corrected for selection
efficiency differences, as explained above. 
The control distributions illustrate the considerable level
 of background for low $\ee$,
that is estimated from the data. 
The DIS simulation uses the H1PDF2009  set of parton distributions~\cite{Aaron:2009kv}.
There is a good overall agreement observed between the measurements and predictions.
The local residual 
differences, visible for the energy distribution in the lowest
$Q^2$, BST sample near to $\ee \simeq 10$\,GeV 
and corresponding to $y_e\simeq 0.7$ (\Fig~\ref{fig:cont920a}a and \Fig~\ref{fig:cont920a}d ), 
do not affect the cross-section measurement.
%\subsection{Cross Checks}
%\input{checks}
\subsection{Systematic Uncertainties}

\label{sec:systematics}

The systematic uncertainty on the cross-section measurements arises
from several contributions.  Besides the global normalisation uncertainty,
these contributions are classified as  correlated
uncertainties, which affect measurements at different $Q^2,x$ in a correlated
manner, and as uncorrelated ones, for which each of the measurements is affected
individually. The summary of all systematic uncertainties is given in
table~\ref{tab:systematics}.

The global normalisation uncertainty  
is $3\%$ for the $E_p=920$~GeV period and $4\%$ for the \lermer\ analyses.
The uncertainty includes the uncertainty of the luminosity measurement as
well as global trigger and reconstruction efficiency uncertainties.

The uncertainty on the SpaCal electromagnetic
energy scale is determined to be $0.2\%$ at 
$\ee=27.5$~GeV increasing to $1\%$ at $\ee=2$~GeV for all but 
the \hqcjc\ analysis. The latter covers a large 
period of runs, from the year $2003$ to $2005$, and therefore is
prone to variations of the SpaCal performance. For this 
analysis the scale uncertainty is $0.3\%$ at $\ee=27.5$ increasing to 
$1\%$ at $\ee=2$~GeV. The uncertainty at around
$27.5$~GeV is estimated from 
the difference between the result of the double-angle
calibration and the position of the kinematic peak. 
The uncertainty at $\ee=2$~GeV is obtained using $J/\psi\rightarrow e^+e^-$ and 
$\pi^0\rightarrow\gamma\gamma$ decays~\cite{H1:2009bp}.

The uncertainty on the lepton polar angle is $0.5$\,mrad, which covers
uncertainties of the
alignment of the SpaCal as well as of the cluster position
determination.

The hadronic energy scale has an uncertainty of $4\%$. Apart from reconstruction 
in the LAr calorimeter and in the tracker, this value covers
the uncertainty of the 
hadronic energy scale of the SpaCal, which is important at high $y$.
The uncertainty of the LAr electronic noise and beam related background is $20\%$.
These uncertainties have %a small 
little impact on the cross-section measurement 
which 
is based on the electron method since they enter only
via the $\empz$ cut.

The background charge asymmetry is determined with a precision of
 $2\%$. It affects
 only the data for $y\ge \ytrans$ where the wrong charge subtraction
method is used. Its uncertainty has negligible impact on the \hqcjc\
and \lqbst\ measurements since these are based on both $e^+p$ and $e^-p$ HERA
running periods and have a charge symmetric background sample. 
For the \lermer\ runs, the impact on the cross section
reaches $3.5\%$ at $y=0.85$.

\renewcommand{\arraystretch}{1.2}

\begin{table}
\begin{center}
\begin{tabular}{l|r}
%\multicolumn{2}{c}{\bf Correlated systematic errors} \\
\hline
{\bf Correlated uncertainty source}   & {\bf Uncertainty} \\
\hline
Global normalisation  & $3\%$ for \ner\ run \\
                      & $4\%$ for \lermer\ run\\
\hline
$\ee$ energy scale    & $0.2\%$ at $27.5$ to $1\%$ at $2$~GeV \\
                      &                  (all, but \hqcjc)\\
                      & $0.3\%$ at $27.5$ to $1\%$ at $2$~GeV \\
                      &                        (\hqcjc) \\
Polar angle $\thetae$ & $0.5$~mrad   \\
Hadronic energy scale & $4\%$  \\
LAr noise  & $20\%$  \\
Background charge asymmetry & $2\%$ \\
Electron tagger acceptance  & $20\%$ \\
\hline
%\multicolumn{2}{c}{\bf Uncorrelated  systematic errors} \\
{\bf Uncorrelated uncertainty source}   & {\bf Uncertainty} \\
\hline
 Trigger efficiency                 & $1\%$\\
 Track-cluster link efficiency    & $1.5\%$\\
 Lepton charge determination      & $1\%$ \\
 Electron identification efficiency &  $1-3\%$\\
 Radiative corrections              &  $1\%$\\
\hline
\end{tabular}
\end{center}
\caption{\label{tab:systematics}
Summary of systematic uncertainties. For the correlated error sources,
the uncertainties are given in terms of the uncertainty in the corresponding
source. For the uncorrelated error sources, the uncertainties are given
in terms of the effect on the measured cross section.}
\end{table}
\renewcommand{\arraystretch}{1.}

 The electron tagger acceptance is known to $20\%$. This uncertainty
is applied for $y< \ytrans$ only and, since the background at low $y$ is small,
this source does not have a significant impact on the measurement.

The uncorrelated systematic uncertainties include the Monte Carlo statistical
errors and the following sources:
the uncorrelated part of the trigger efficiency, known to $1\%$;
the track-cluster link efficiency,  known to $1.5\%$;
the uncertainty of the lepton charge determination of $0.5\%$ leading
to $1\%$ uncertainty of the cross section, for $y\ge \ytrans$ only;
the electron identification uncertainty varies from $3\%$ for $y>0.8$
to $1\%$ for $y<0.6$;
the uncertainty due to the radiative corrections is determined to be $1\%$.
\section{Cross Section Determination}
\label{sec:xs}
%
%\subsection{This Measurement}
%\input{xsec}
\subsection{Method}
\label{xsec}
At low $Q^2$ the contributions to the NC scattering process are completely
dominated by photon exchange with negligible differences between
the $e^+p$ and $e^-p$ scattering cross sections.
The background
determination at high $y$ is based on the measured lepton-candidate charge. 
In order to reduce the sensitivity to the background charge asymmetry,
the cross section is determined for a charge symmetric data sample for the
\hqcjc\ and the \lqbst\ samples. The reduced cross section is calculated in this case
for each $x,Q^2$ bin as
\begin{equation}
\sigma_r(x,Q^2) = \frac{\textstyle N^{e^-p}_{\rm sig} + N^{e^+p}_{\rm sig}
\frac{\textstyle {\cal L}^{e^-p}} {\textstyle {\cal L}^{e^+p}}}
{\textstyle N^{e^-p}_{\rm sig,MC} \frac{\textstyle {\cal L}^{e^-p}} 
{\textstyle {\cal L}^{e^-p}_{\rm MC}}
+ N^{e^+p}_{\rm sig,MC} \frac{\textstyle {\cal L}^{e^-p}} {\textstyle {\cal L}^{e^+p}_{\rm MC}}}\, \sigma_r^{\rm MC}(x,Q^2)\,.  \label{eq:chsym}
\end{equation}
Here 
 ${\cal L}^{e^{\pm} p}$ (${\cal L}^{e^{\pm}p}_{\rm MC}$) is the 
integrated luminosity for the data (MC), $N^{e^{\pm} p}_{\rm sig,MC}$ is the number of signal events in the 
MC and
$\sigma_r^{\rm MC}(x,Q^2)$ is the value of the reduced cross section in the MC.

\EEq~\ref{eq:chsym} is rather insensitive to the uncertainty of the background
charge asymmetry $\kappa_{\pm}$ since for $y\ge \ytrans$, the total background is  
estimated as $N_{bg} = \kappa_{-}N^{e^{-}p}_+ + \kappa_+N^{e^+p}_- ({\cal L}^{e^-p}
/{\cal L}^{e+p})$. The statistical accuracy of \Eq~\ref{eq:chsym} is limited
by the sample with the smaller luminosity, therefore the data taking strategy
was tuned to obtain $e^+p$ and $e^-p$ samples of about equal size.

For the \lermer\ samples, the 
absence of $e^-p$ data does not allow for
usage of \Eq~\ref{eq:chsym}, and a more standard cross section 
determination formula is used as
\begin{equation}
\sigma_r(x,Q^2) = \frac{\textstyle N^{e+p}_{\rm sig}}{\textstyle N^{e+p}_{\rm sig,MC} \frac{\textstyle {\cal L}^{e^+p}} 
{\textstyle {\cal L}^{e^+p}_{\rm MC}}
}\, \sigma^{\rm MC}_r(x,Q^2)\,.
\end{equation}
These cross sections are therefore more sensitive to the uncertainty 
% Steve wants ``in'' not ``on'':
in
$\kappa_{\pm}$. 
\begin{table}
%\begin{small}
\begin{center}
\begin{tabular}{l|llllllllll}
\hline
 {\bf Sample} & \multicolumn{10}{c}{\bf Bin boundaries in $\boldsymbol{y}$} \\
\hline
 $E_p=460$~GeV &         &        &  $0.9$ & $0.8$  & $0.7$  & $0.6$  & $0.38$  & $0.24$  & $0.15$ & $0.095$ \\
 $E_p=575$~GeV & $0.896$ & $0.80$ & $0.72$ & $0.64$ & $0.56$ & $0.48$ & $0.304$ & $0.192$ & $0.12$ & $0.076$\\
 $E_p=920$~GeV & $0.56$  & $0.50$ & $0.45$ & $0.40$ & $0.35$ & $0.30$ & $0.19$ & $0.12$ & $0.075$ & $0.0475$\\
\hline
\end{tabular}
\end{center}
%\end{small}
\caption{\label{tab:bins}Bin boundaries in $y=Q^2/sx$
for the cross-section analyses at different $E_p$ 
used to measure the structure function $F_L$.}
\end{table}

The \hqcjc\ and \lqbst\ samples extend the published H1 measurements to high $y$ and for them the same mixed $(Q^2,x) - (Q^2,y)$ binning
is adapted as used in~\cite{H1:2009bp}. 
The \lermer\ samples are used to measure the structure function $F_L$. For this measurement,
an optimal binning is in $(Q^2,y)$ with the $y$ boundaries 
of the bins adjusted so that the corresponding
 $x=Q^2/(4 E_e \cdot E_p \cdot y)$ values agree  for 
different $E_p$. This binning is given in \Tab~\ref{tab:bins}. Bin centres are calculated
as an arithmetic average of the bin boundaries.
 Apart from the \lermer\ samples, the binning is also
employed in the reanalysis of  the published H1 data at
 $E_p=920$~GeV for the $F_L$ measurement, as is discussed 
below, in \Sec~\ref{sec:fl}.
The purity and stability~\cite{H1:2009bp} of the cross-section measurements are typically above $70\%$ at highest $y$ reducing
to about $50\%$ at lowest $y$.
\subsection{Results}
\begin{figure}
\centerline{\epsfig{file=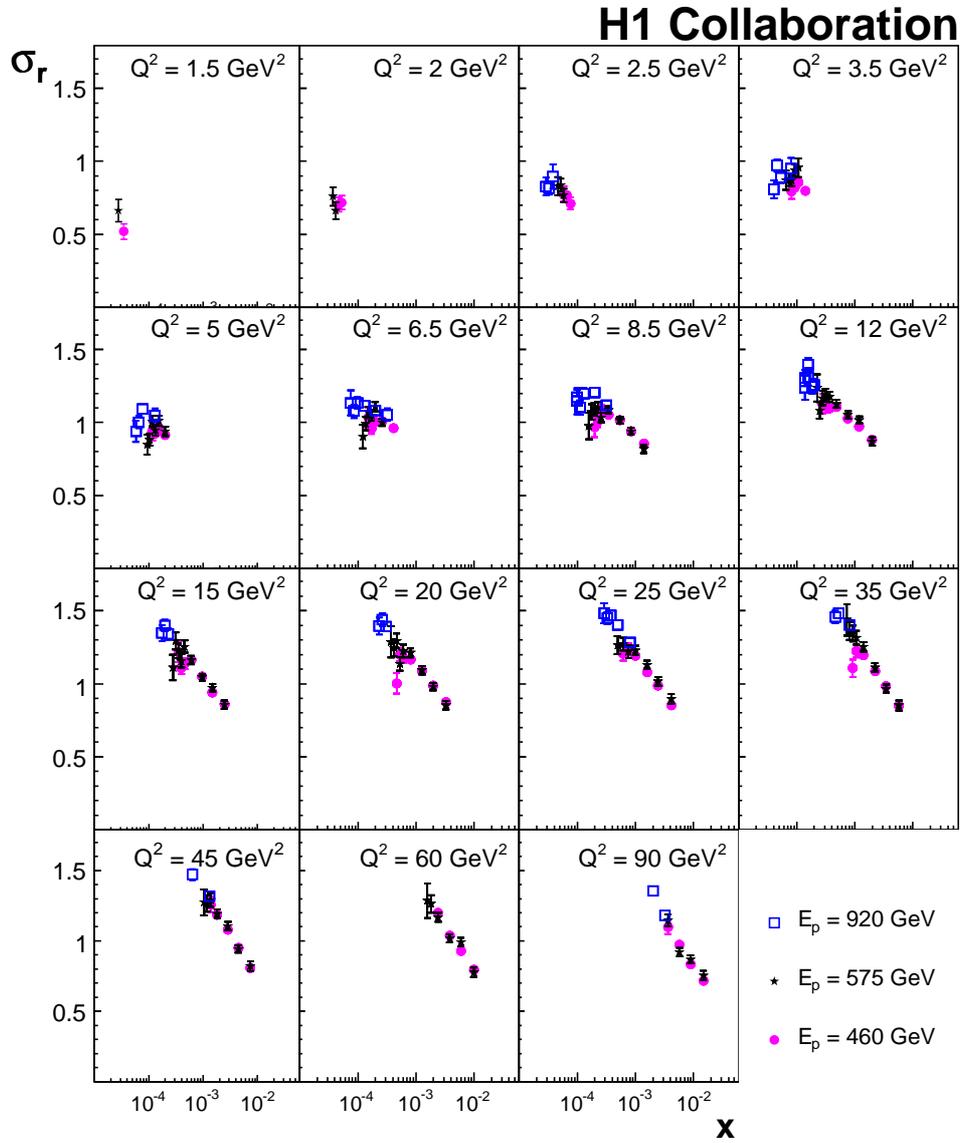,width=\linewidth}}
\caption{Results on the reduced cross section
$\sigma_r$ as determined from the
$E_p=920$~GeV, $E_p=575$~GeV and $E_p=460$~GeV samples. 
The error bars represent statistical and 
systematic uncertainties added in quadrature. \label{fig:newdata}}
\end{figure}

The cross-section measurements are given in 
\Tabs~\ref{tab:bst920}-\ref{tab:h1575last} and shown in \Fig~\ref{fig:newdata}.
The new data 
cover the range between $1.5$~GeV$^2$ and $90$~GeV$^2$ in $Q^2$ 
reaching values of inelasticity $y$ as high as $0.85$. 

For the $E_p=920$~GeV sample, the new data can be compared to the previous
H1 results~\cite{H1:2009bp,Aaron:2009kv}.
For the high $y$ region, the precision of the new data 
is significantly better than that of  the previous H1 result, apart from the
global normalisation uncertainty that is  larger for the new result. 
This uncertainty is
significantly reduced by combining the H1 measurements.

\subsection{Combination of Data}
\label{sec:combination}
\def\h19500s1v{-0.38}
\def\h19500s1e{0.91}
\def\h19500s2v{-0.01}
\def\h19500s2e{1.00}
\def\h19500s3v{0.03}
\def\h19500s3e{0.99}
\def\h19500s4v{-0.16}
\def\h19500s4e{0.94}
\def\h19500s5v{-0.04}
\def\h19500s5e{1.00}
\def\h19500s6v{0.04}
\def\h19500s6e{0.99}
\def\h19500s7v{-0.09}
\def\h19500s7e{0.99}
\def\h19500s8v{-0.10}
\def\h19500s8e{1.00}
\def\h19500s9v{-0.04}
\def\h19500s9e{0.98}
\def\h19500s10v{0.14}
\def\h19500s10e{0.97}
\def\h19500s11v{0.05}
\def\h19500s11e{0.98}
\def\h19500s12v{0.06}
\def\h19500s12e{0.99}
\def\h19500s13v{-0.05}
\def\h19500s13e{0.98}
\def\h19500s14v{-0.06}
\def\h19500s14e{0.99}
\def\h19500s15v{0.15}
\def\h19500s15e{0.97}
\def\h19500s16v{-0.10}
\def\h19500s16e{0.99}
\def\h19500s17v{0.15}
\def\h19500s17e{0.97}
\def\h19500s18v{0.10}
\def\h19500s18e{0.91}
\def\h19500s19v{-0.15}
\def\h19500s19e{0.96}
\def\h19500s20v{0.04}
\def\h19500s20e{0.99}
\def\h19500s21v{-0.03}
\def\h19500s21e{1.00}
\def\h19500s22v{0.04}
\def\h19500s22e{1.00}
\def\h19500s23v{-0.14}
\def\h19500s23e{0.99}
\def\h19500s24v{-0.03}
\def\h19500s24e{1.00}
\def\h19500s25v{-0.10}
\def\h19500s25e{0.99}
\def\h19500s26v{-0.03}
\def\h19500s26e{1.00}
\def\h19600s1v{-0.42}
\def\h19600s1e{0.87}
\def\h19600s2v{-0.46}
\def\h19600s2e{0.92}
\def\h19600s3v{0.58}
\def\h19600s3e{0.87}
\def\h19600s4v{0.08}
\def\h19600s4e{0.99}
\def\h19600s5v{0.16}
\def\h19600s5e{0.98}
\def\h19600s6v{-0.01}
\def\h19600s6e{0.99}
\def\h19600s7v{-0.01}
\def\h19600s7e{1.00}
\def\h19600s8v{-0.05}
\def\h19600s8e{1.00}
\def\h19600s9v{-0.05}
\def\h19600s9e{0.99}
\def\h19600s10v{0.15}
\def\h19600s10e{0.98}
\def\h19600s11v{0.00}
\def\h19600s11e{1.00}
\def\h19600s12v{-0.03}
\def\h19600s12e{1.00}
\def\h19600s13v{-0.02}
\def\h19600s13e{1.00}
\def\h19600s14v{-0.01}
\def\h19600s14e{1.00}
\def\heratwoev{-0.25}
\def\heratwoee{0.85}
\def\heratwothv{0.14}
\def\heratwothe{0.84}
\def\heratwolarv{-0.03}
\def\heratwolare{0.99}
\def\heratwonoisev{0.00}
\def\heratwonoisee{0.99}
\def\heratwolumiv{0.03}
\def\heratwolumie{0.96}
\def\heratwolumiBSv{-0.30}
\def\heratwolumiBSe{0.37}
\def\chiave{15.4}
\def\dofave{ 36}
\def\ElowPv{-0.00}
\def\ElowPe{0.91}
\def\thlowPv{-0.09}
\def\thlowPe{0.92}
\def\ehlowPv{-0.00}
\def\ehlowPe{1.00}
\def\noilowPv{-0.00}
\def\noilowPe{1.00}
\def\asylowPv{0.00}
\def\asylowPe{1.00}
\def\gplowPv{0.03}
\def\gplowPe{1.00}
\def\lum460v{0.08}
\def\lum460e{0.71}
\def\lum575v{-0.08}
\def\lum575e{0.71}
\def\chiavelowP{17.2}
\def\dofavelowP{ 27}

For the  proton beam energy $E_p=920$, the new data cover a 
phase space similar to previous H1 results~\cite{H1:2009bp,Aaron:2009kv} which 
are based on HERA-I data, collected in the years $1994$ to $2000$.
Therefore the data are combined,
following the procedure described in~\cite{H1:2009bp,glazov}.
\begin{sloppypar}
Four data sets are considered in this combination: the combined H1
results from HERA-I \cite{H1:2009bp,Aaron:2009kv}, reported for $E_p=820$~GeV
and $E_p=920$~GeV, and the two new data sets, \hqcjc\ and \lqbst. The systematic
uncertainties are assumed to be uncorrelated between the HERA-I and 
HERA-II
measurements, apart from a $0.5\%$ overall normalisation uncertainty due to the 
theoretical uncertainty on the Bethe-Heitler process cross section 
used for the luminosity measurement. 
In total, there are $46$ independent sources of systematic uncertainty.
%The treatment of correlated, uncorrelated and statistical
%uncertainties follows \Eq~31 of~\cite{H1:2009bp}. 
For $Q^2\ge12$~GeV$^2$,
the new data extend the kinematic coverage towards high $y>0.6$. 
At low $y<0.6$ and for all values of $y$ at low $Q^2$,  
there is a sizable region of overlap.  
\end{sloppypar}

The combined cross-section measurements are given in 
\Tabs~\ref{tab615a1} to \ref{tab615a4}.
The full information about correlation
between cross-section measurements can be found elsewhere~\cite{fullcorr}.
The data show very good compatibility, with $\chi^2/\dof = \chiave/\dofave$.
At low $y$, the previous H1 data from~\cite{H1:2009bp,Aaron:2009kv} have a higher
precision than the new result. In particular, the global normalisation
uncertainty was significantly smaller: about $1\%$ at HERA-I 
compared to $3\%$ at HERA-II. Therefore, in the combination,
the new HERA-II data are effectively
normalised to the HERA-I result and their global normalisation uncertainties are reduced
significantly. \TTab~\ref{tab:syst920}
lists those few systematic sources of the HERA-II analyses,
which are noticeably altered by the averaging procedure.
All alterations stay within one standard deviation
of the estimated error.

\begin{table}
\begin{center}
\begin{tabular}{l|rr}
\hline
{\bf Systematic Source} & {\bf Shift in $\boldsymbol{\sigma}$} & {\bf Uncertainty in $\boldsymbol{\sigma}$} \\
\hline
$\ee$ scale & $\heratwoev$ & $\heratwoee$\\
$\thetae$   & $\heratwothv$ & $\heratwothe$\\
${\cal L}_{CJC}$ & $\heratwolumiv$& $\heratwolumie$\\
${\cal L}_{BST}$ & $\heratwolumiBSv$& $\heratwolumiBSe$\\
\hline
\end{tabular}
\end{center}
\tablecaption{\label{tab:syst920} Shifts of the central
values and reduction of the uncertainty of the systematic error sources in the
combination of the \hqcjc\ and \lqbst\ data sets
with HERA-I data,
expressed as fractions of the original uncertainty.}
\end{table}

The systematic errors of the HERA-I data are not significantly affected
by the combination. At low $y$,
the gain in the combined data precision  compared
to the HERA-I result is small. The uncertainties are reduced by at most $5\%$ 
of their size and the shift of central values does not exceed $0.2\%$.  
At high $y$, however,  there is a significant gain in the precision
achieved by the data combination.
For the region $2.5\le Q^2<12$~GeV$^2$ and $y=0.8$, for example,
the accuracy of the $E_p=920$~GeV data
is improved by about a factor of two. For medium $Q^2$,
$12\le Q^2\le 35$~GeV$^2$, 
the new high $y$ measurements, corresponding to $E_p=920$~GeV, 
exceed the accuracy of the HERA-I 
%points
data, corresponding to $E_p=820$~GeV, by a factor $1.5$ to $2$.
The $E_p=920$\,GeV measurement at HERA-I was limited to $y \leq 0.6$.

The \lermer\ data sets are measured using an identical grid of $(Q^2,x)$ bin centres.
At low $y$, the influence of the structure function $F_L$ is small. Therefore
the two data sets are combined for all $(Q^2,x)$ points satisfying 
$y_{460}=Q^2/(4 E_e\cdot 460~{\rm GeV}\cdot x )<0.35$ after a small correction of 
the cross-section values to $E_p=575$~GeV. 
At higher $y$ the measurements
are kept separately but they are affected by the combination procedure. 
The data show   good compatibility, with 
$\chi^2/\dof = \chiavelowP/\dofavelowP$, and the combined reduced
cross-section values are given in 
\Tabs~\ref{tablowea1} and \ref{tablowea2}. This combined reduced $E_p$ 
set, together with the combined nominal $E_p$ set, is used for the 
phenomenological analysis presented in \Sec~\ref{sec:theory}.
\section{Determination of the Structure Function $\mathbold{F_L}$}
\label{sec:fl}
\subsection{Procedure}
\def\rchisq{113.8}
\def\rval{0.260}
\def\rndf{  150}
\def\rerr{0.050}
The structure function $F_L$ is determined using 
the separate \lermer\ samples and the published
$920$~GeV data  from~\cite{H1:2009bp,Aaron:2009kv}. 
To determine $F_L$, common values of the $(x,Q^2)$ grid centres are required 
for all centre-of-mass energies. 
The published $920$~GeV data have 
therefore been reanalysed using the binning adopted
for the $F_L$ analysis, see \Tab~\ref{tab:bins}. 
To determine $F_L$, the data measured at high $y$ for 
$E_p=460$~GeV are combined with the data at intermediate $y$ for $E_p=575$~GeV and low $y$ for
$E_p=920$~GeV.
The usage of the published $920$~GeV data compared
to a new analysis of the HERA-II data is motivated by a wider $Q^2$
acceptance at low $y$, extending to $Q^2=1.5$~GeV$^2$. In addition, as 
%it is 
discussed
in \Sec~\ref{sec:combination}, adding the HERA-II data does not improve 
the precision at low $y$.
%The cross section measurements at three different 
%centre-of-mass energies are shown in \Fig~\ref{fig:flextract}.

\begin{figure}
\centerline{\epsfig{file=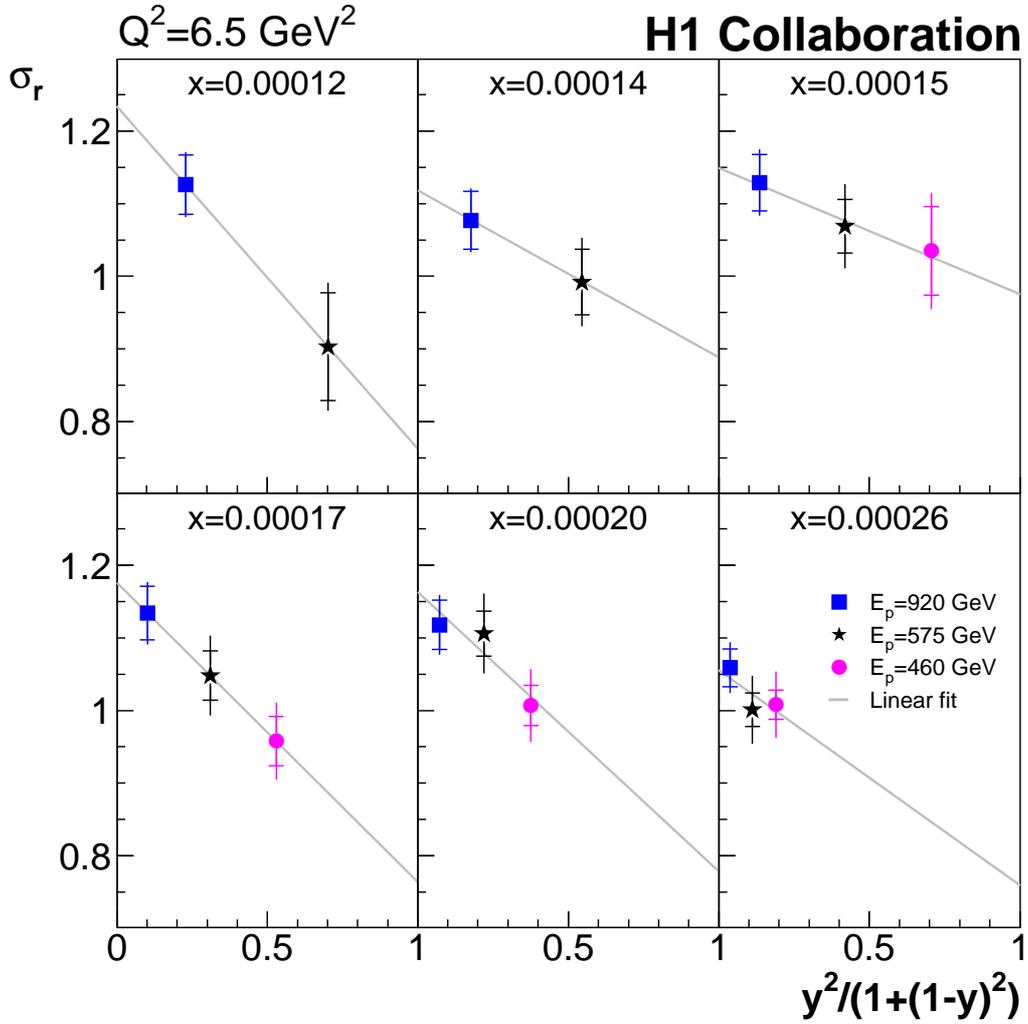,width=0.9\linewidth}}
\caption{\label{fig:rosen}The reduced DIS cross section $\sigma_r$
as a function of $y^2/(1+(1-y)^2)$ for six values of $x$ at $Q^2=6.5$~GeV$^2$,
measured for proton beam energies of $E_p=920$, $575$ and $460$~GeV.
The inner error bars denote the statistical error, the outer error bars
show statistical and 
systematic uncertainties added in quadrature. The luminosity uncertainty is not included in the error bars. The slope of the straight-line fits 
is determined by the structure function $F_L(x,Q^2)$.}
\end{figure}

The determination of the structure function $F_L$ depends on the treatment
of the relative normalisations and systematic uncertainties of the data sets.  
A straightforward but simplified procedure was adopted in~\cite{h1fl} where the data sets
were normalised to each other at low $y$. The values of the 
structure function $F_L$ were
determined in  straight-line fits to the reduced cross section as
a function of $y^2/(1+(1-y)^2)$ in each $(x,Q^2)$ bin using the statistical and uncorrelated 
systematic uncertainties. The correlated systematic errors were 
determined using an offset method. An illustration
of this procedure, applied to the cross-section data from the current analysis,
is shown in \Fig~\ref{fig:rosen}.
The procedure adopted in~\cite{h1fl} does not fully take into account 
correlations between the low and high $y$ regions, used for the cross-section 
normalisation and the $F_L$ computation. The offset method does not allow
for shifts of the central values of the correlated systematic error sources.
Thus the information on the  goodness of the straight-line fits to the 
cross-section measurements at the three centre-of-mass energies is not 
fully employed. 

The procedure for the $F_L$ determination is improved  in the current analysis. The new method extends the
averaging procedure of~\cite{H1:2009bp}. For additive 
uncertainties it
%\footnote{Extension to more complex
%dependence of the uncertainties can be performed in analogous way as 
%in~\cite{H1:2009bp}.}
is based on the minimisation of the  function
\begin{equation} \label{eq:flex}
\chi^2_0 
\left({\bf F_2,F_L,b}\right) = \sum_i 
\frac{\textstyle \left[ \left(F_2^i - f(y^i) F_L^i\right) - 
\sum_j \Gamma^i_j b_j - \mu^i\right]^2}{\textstyle \Delta^2_i} + \sum_j b^2_j\,.
\end{equation} 
Here $\mu^i$ is the measured central value of the reduced 
cross section at a $(Q^2,x;s)$ point $i$ with a combined
statistical and uncorrelated systematic uncertainty 
$\Delta_i = (\Delta_{i,{\rm stat}}^2+\Delta^2_{i,{\rm uncor}})^{1/2}$. The effect of 
correlated error sources $b_j$ on
the cross-section measurements is approximated by the systematic 
error matrix $\Gamma^i_j$. The function $\chi^2_0$ depends quadratically 
on the
structure functions $F^i_2$ and $F^i_L$ ( denoted as vectors $\bf F_2,F_L$)
as well as on $b_j$. Minimisation of $\chi^2_0$ with respect to these
variables leads to a system of linear equations.
% which can be solved.

For low $y\le 0.35$, the coefficient $f(y)$ is small 
compared to unity 
and thus $F_L$ can not be accurately measured. In this kinematic
domain, the constraint $0\le F_L\le F_2$ provides an even
better bound on the value
of $F_L$ than the experimental data. Furthermore, the ratio
$R$ is not expected to vary strongly as a function of $x$
in the limited $x$ range of sensitivity to $F_L$. For the 
kinematic range studied in this paper it is measured to be consistent
with $R\sim 0.25$. %see \cite{h1fl} and below. 
To avoid unphysical values for $F_L$, an extra prior
is introduced for the $\chi^2$ minimisation:
\begin{equation} \label{eq:flfull}
\chi^2\left({\bf F_2,F_L,b}\right) 
= \chi^2_0\left({\bf F_2,F_L,b}\right) + \sum_i \left(\frac{\textstyle F^i_L - \frac{\textstyle R}{\textstyle R+1} F_2^i }
{\textstyle \Delta_{F_L}}\right)^2\,,
\end{equation}
where  $R=0.25$ and the width $\Delta_{F_L} = 3$  is chosen
such that it has a negligible influence for $y > 0.35$. The additional
prior preserves the quadratic dependence of the $\chi^2$ function
on $F^i_2$ and $F^i_L$. The prior has a significant contribution at low $y$
only and is very similar to 
imposing a common cross-section normalisation at low $y$ used in~\cite{h1fl}.
Since $\Delta_{F_L}$ is chosen to be large, the prior affects only points with large uncertainty on
$F_L$.  
The bias introduced by the prior is investigated by varying the value of
$R$ between $0$ and $0.5$ and $\Delta_{F_L}$ between $1$ and $5$, and found to be negligible, for the points chosen for the $F_L$ determination.
\subsection{Results}
\label{sec:flresults}
\begin{figure}
\centerline{\epsfig{file=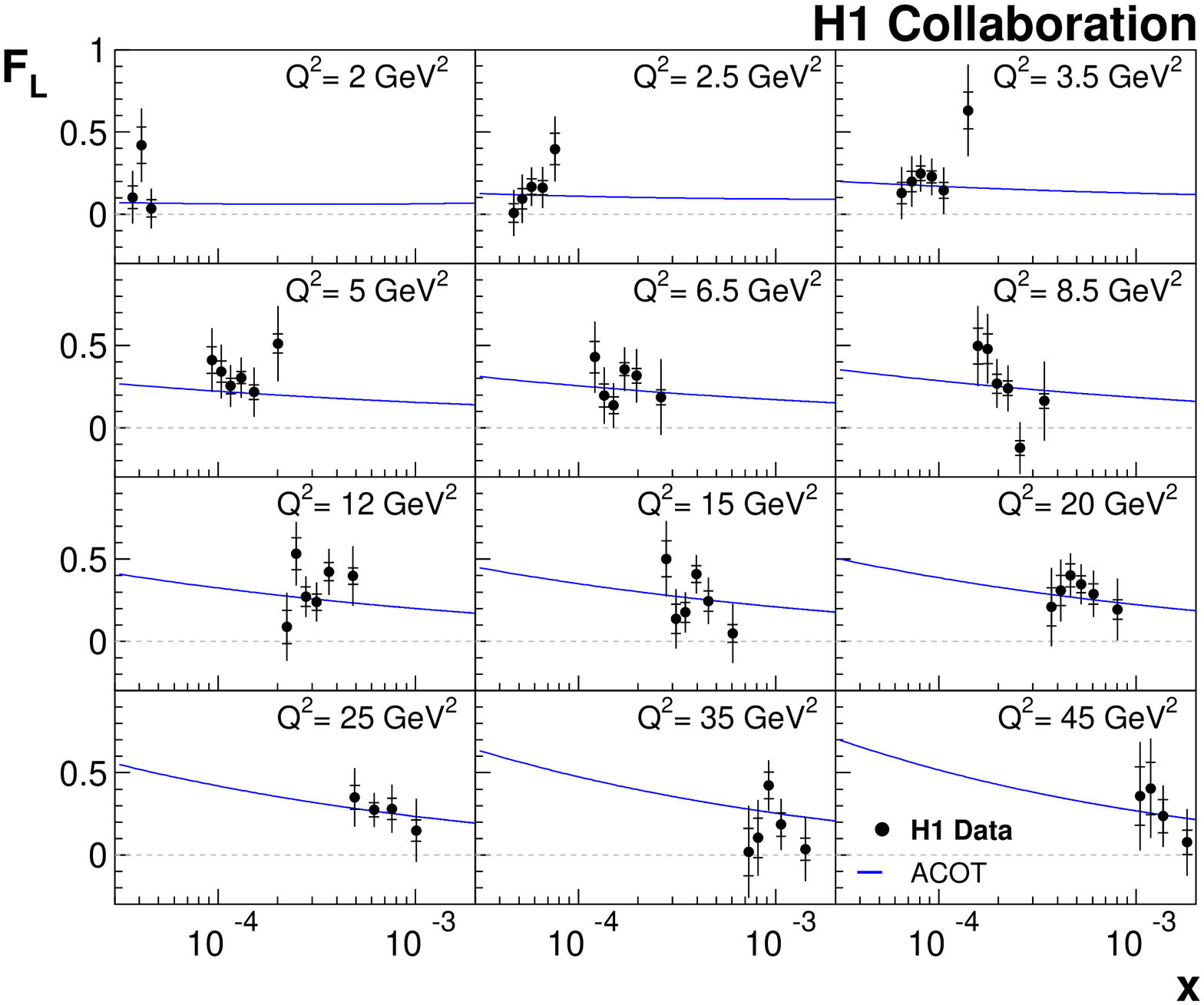,width=\linewidth}}
\caption{The proton structure function $F_L(x,Q^2)$. The inner error bars
represent statistical error, the full error bars include the statistical and 
systematic uncertainties added in quadrature,
excluding $0.5\%$ global normalisation uncertainty. 
The curves represent predictions of the DGLAP fit in the ACOT scheme.
\label{fig:flxq}}
\end{figure}

\begin{figure}
\centerline{\epsfig{file=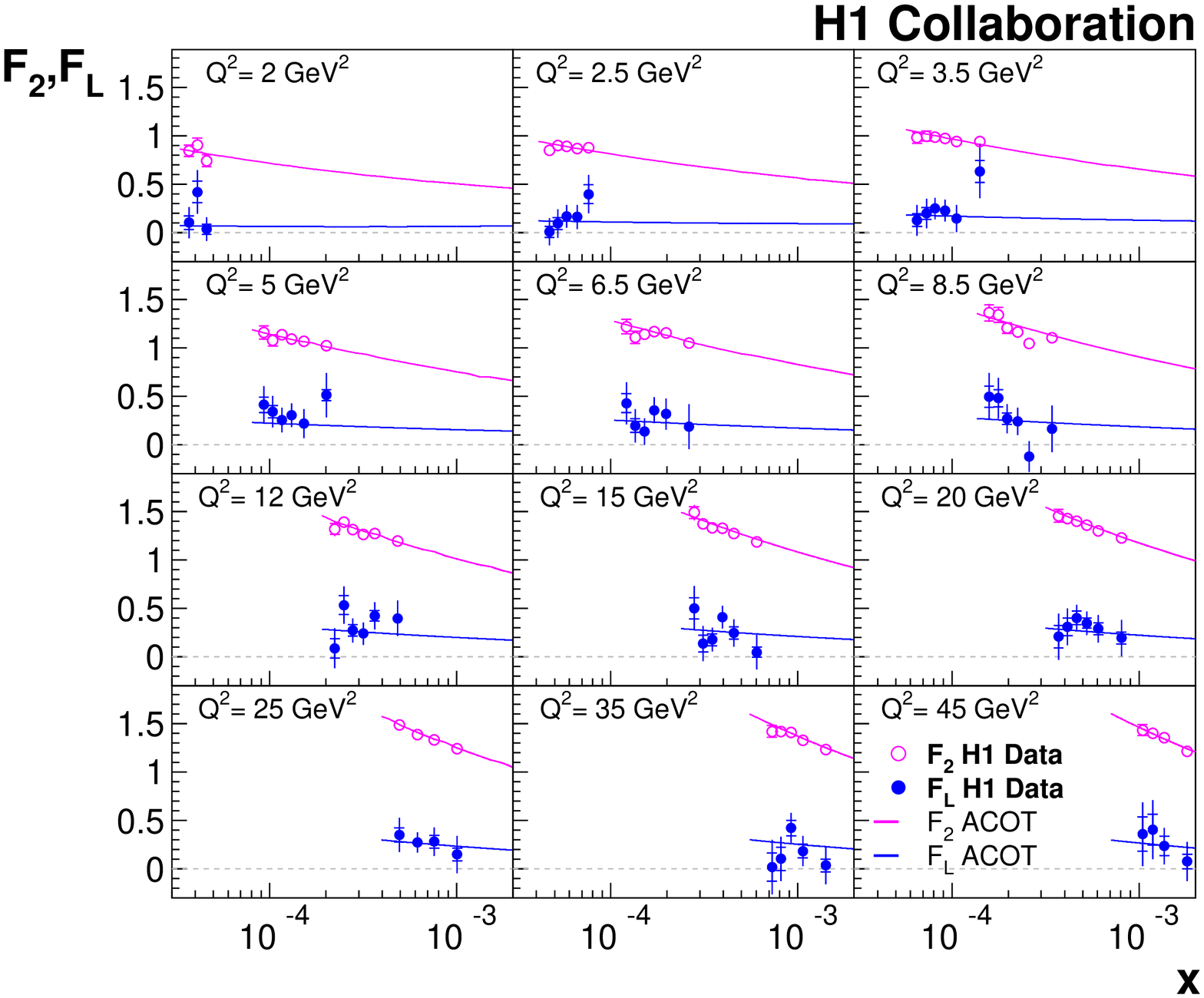,width=\linewidth}}
\caption{The proton structure functions  $F_2(x,Q^2)$ and $F_L(x,Q^2)$. 
The inner error bars
represent statistical error, the full error bars include the statistical and 
systematic uncertainties added in quadrature,
excluding $0.5\%$ global normalisation uncertainty. 
The curves represent predictions of the DGLAP fit in the ACOT scheme
for the structure functions  $F_2$ and $F_L$.
\label{fig:flf2xq}}
\end{figure}

The measured structure function $F_L(x,Q^2)$ is given in \Tab~\ref{tab:flf2xq} 
and shown in \Fig~\ref{fig:flxq}.
By convention, only measurements with total uncertainties
below $0.3$ for $Q^2\le 35$~GeV$^2$  and below $0.4$ for $Q^2=45$~GeV$^2$
are presented. The selection on the total uncertainty removes the bias due to the prior in \Eq~\ref{eq:flfull}.
The measurement spans over two decades in $x$ at low $x$,
from $x=0.00002$ to $x=0.002$. The data are compared to the result
of the DGLAP ACOT fit, which is described in \Sec~\ref{sec:dglap}.
The structure function $F_2$ measured for the corresponding bins is 
given in \Tab~\ref{tab:flf2xq} and shown together with $F_L$ in \Fig~\ref{fig:flf2xq}.
Note that compared
to the previous determinations of $F_2$ by the H1 collaboration, 
this measurement represents a model independent determination without extra
assumptions on $F_L$.

\begin{figure}
\centerline{\epsfig{file=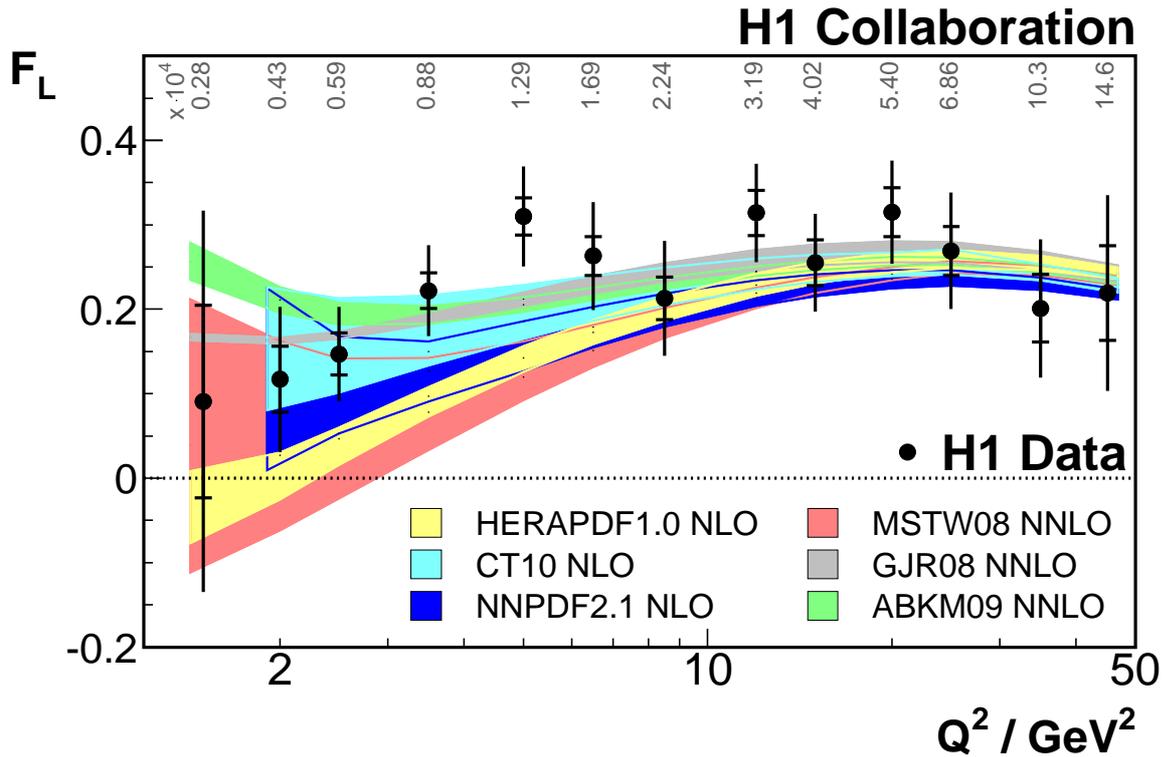,width=\linewidth}}
\caption{\label{fig:flaveth}The proton structure function $F_L$ 
shown as a function of $Q^2$. The average $x$ values for each $Q^2$  are indicated. 
The inner error bars
represent statistical error, the full error bars include the statistical and 
systematic uncertainties added in quadrature. The bands represent 
predictions based  on HERAPDF1.0, CTEQ6.6 and NNPDF2.1 NLO 
as well as 
DGLAP ACOT and RT as well as MSTW08, GJR08 and ABKM09 NNLO 
calculations.
}
\end{figure}

The values of $F_L(x,Q^2)$ resulting from averages over $x$ at fixed $Q^2$
are  given in \Tab~\ref{tab:flave} and presented in \Fig~\ref{fig:flaveth}.
The average is performed taking into account correlations. 
The measured structure function $F_L$ is compared with theoretical 
predictions from HERAPDF1.0~\cite{h1zeus:2009wt}, CT10~\cite{Lai:2010vv},
NNPDF2.1~\cite{Ball:2010de,Forte:2010ta}, MSTW08~\cite{Martin:2009iq}, GJR08~\cite{Gluck:2007ck,Gluck:2008gs}
and ABKM09~\cite{Alekhin:2009ni} sets. Depending on the PDF set, the calculations are performed at
 NLO or NNLO in perturbative QCD. 
Within the uncertainties all predictions describe the data reasonably well.
 
\begin{figure}
\centerline{\epsfig{file=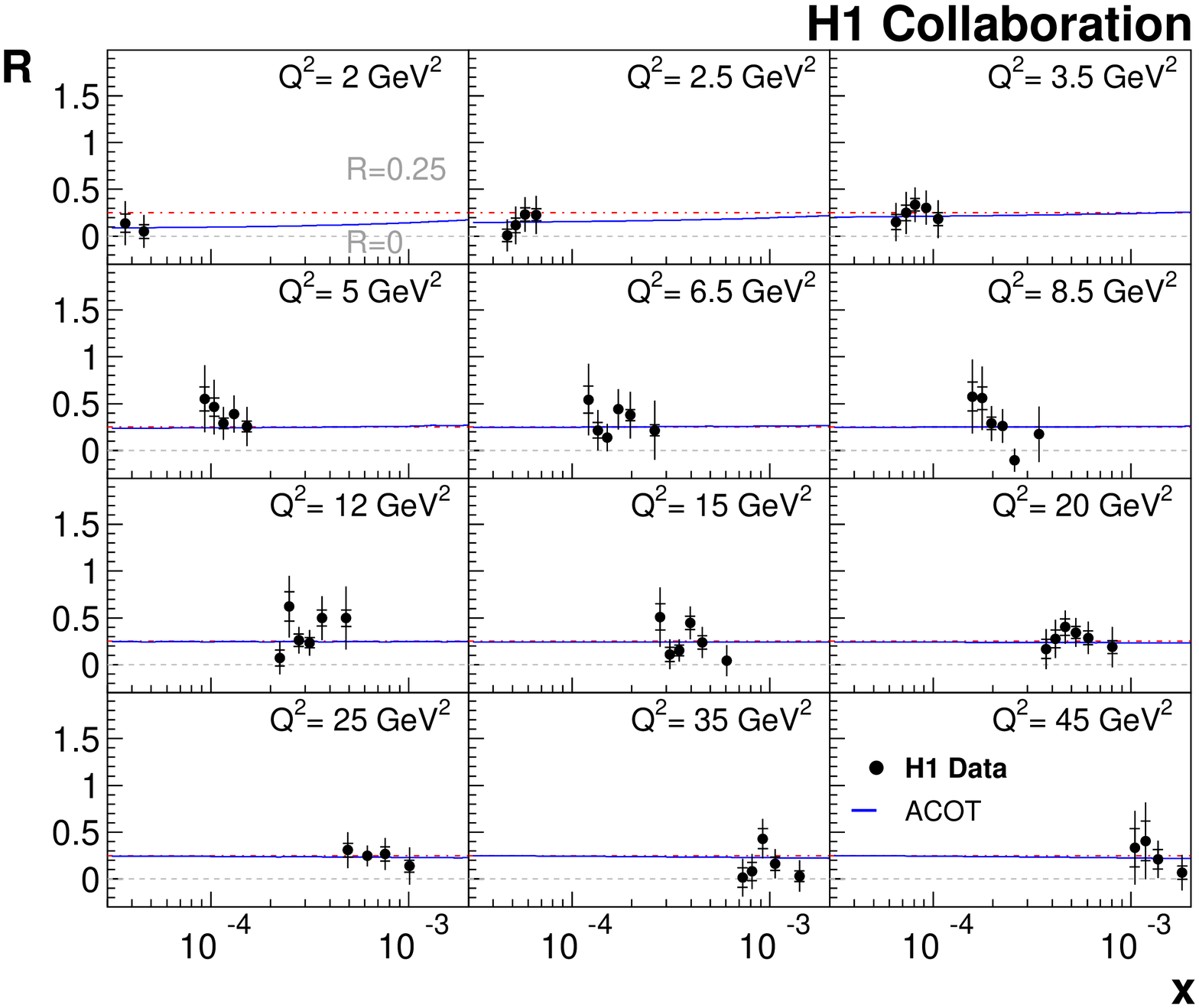,width=\linewidth}}
\caption{The ratio $R=F_L(x,Q^2)/(F_2(x,Q^2)-F_L(x,Q^2))$. The inner error bars
represent statistical error, the full error bars include the statistical and 
systematic uncertainties added in quadrature.
 The solid curves represent predictions of the DGLAP fit in ACOT
scheme. 
\label{fig:flr}}
\end{figure}

The measurement of the structure functions $F_2$ and $F_L$ can be used to determine
the ratio $R$ (see \Eq~\ref{ratio}). This ratio is shown
 in \Fig~\ref{fig:flr}. Apart from conditions applied to select $F_L$ 
results, only measurements with total uncertainties 
below $0.6$ 
are included.
 
\begin{figure}
\centerline{\epsfig{file=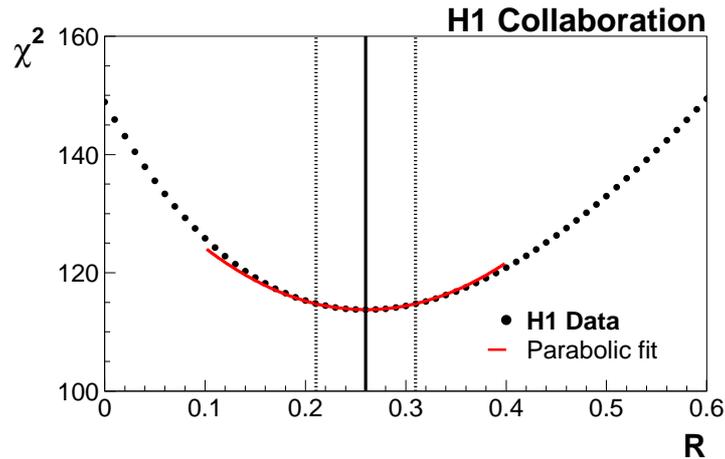,width=0.6\linewidth}}
\caption{\label{fig:rscan} $\chi^2$ for combination of the data
taken at $E_p=920$, $575$ and $460$~GeV as a function
of $R$ assuming $R$ being constant. The solid line shows 
parabolic fit around the $\chi^2$ minimum.
The solid vertical  line shows the value of $R_{\rm min}$ at which $\chi^2$ has the minimum  and the dotted vertical lines
correspond to the total uncertainty of $R_{\rm min}$.}
\end{figure}

For $Q^2\ge3.5$~GeV$^2$, the ratio $R$ is consistent with a constant 
behaviour. This hypothesis is tested by a simultaneous determination of the
values of the structure function $F_2(x,Q^2)$ at all $(x,Q^2)$  data points 
under the assumption that $R$ is constant.
In this procedure, values of $R$ are scanned between $R=0$ and $R=0.6$ in
$\Delta R=0.01$ steps, and  each of the cross-section measurements is 
used
to calculate the structure function $F_2(x,Q^2)$ using \Eq~\ref{eq:rrr}.
The measurements of $F_2(x,Q^2)$ from different $E_p$ are then combined using the standard averaging programme~\cite{H1:2009bp} taking into account correlations
of the systematic uncertainties. 
\FFig~\ref{fig:rscan} shows the results of this scan represented as 
$\chi^2$ for each average as a function of $R$. The minimum is found at
$R_{\rm min} = \rval \pm \rerr$ where
$\chi^2_{\rm min}/\dof=\rchisq/\rndf$ 
suggesting a conservative error estimation.
It is remarkable that all the low $7\cdot 10^{-5} < x < 2\cdot 10^{-3}$, low $3.5\le Q^2\le 45$~GeV$^2$ data are consistent
with the hypothesis that $R$ is constant.
\section{Phenomenological Analysis}
\label{sec:theory}
The  combined cross-section data for $E_p=460,575$ and $E_p=820,920$~GeV
are used for several phenomenological analyses. The fits are applied to the 
combined reduced cross-section measurements
accounting for correlations between the data points.

In the following, the quality of different fits is compared in terms of $\chi^2/\dof$.  
Since the systematic uncertainties dominate over statistics, and they are estimated
conservatively, in several cases 
$\chi^2/\dof$ is observed to be less than unity. 
This, however, does not prevent
the comparison of quality among different fits with the same number of degrees
of freedom in terms of $\Delta \chi^2$ since the average error overestimation, approximated 
as $\sqrt{\chi^2/\dof}$, does not exceed $5-10\%$.
%
%\subsection{Rise of $\mathbold{F_2}$ Towards Low $\mathbold{x}$}
\subsection{$\boldsymbol{\lambda}$ Fit}
The increase of the structure function  $F_2$ for $x\to 0$ can be approximated
by a power law in $x$, $F_2 = c(Q^2)  x^{-\lambda(Q^2)}$. 
This simple parameterisation was shown to describe previous H1 data rather well for 
$x<0.01$~\cite{adloff-2001-520}.
In the recent H1 analysis~\cite{H1:2009bp}, a fit was performed to
the measured reduced cross section, $\sigma_r$, 
represented as
\begin{equation}
  \sigma_{r}(Q^2,x) 
= c(Q^2) x^{\textstyle -\lambda (Q^2)}\left[1 - \frac{\textstyle y^2}{\textstyle 1+(1-y)^2} 
\frac{\textstyle R}{1+\textstyle R} \right]\,  \label{eq:lamfit}
\end{equation}
by allowing $R$ to float for each $Q^2$ bin independently.
At low $Q^2\le 10$~GeV$^2$, this lead to surprisingly large values of 
$R \approx 0.5$, which are incompatible with the  result $R \approx 0.26$ 
obtained in 
\Sec~\ref{sec:flresults}.
A similar behaviour is observed when equation\,\ref{eq:lamfit}
is applied to the present data. This points to some inconsistency in
the simple power law for the rise of $F_2$ towards low $x$ 
and the measured value of $R$.
A different approach is therefore adopted in this analysis. 
It is generally assumed that $R=0.26$ for all $Q^2$ bins.
A fit 
termed the $\lambda$ fit
is made with only  $c(Q^2)$ and $\lambda(Q^2)$ as free parameters.
This is extended in a subsequent step to allow for possible deviations in the
behaviour of $F_2$ from the simple $\lambda$ fit formula.

\begin{figure}
%\epsfig{file=figs/c_lambda_vs_q2.eps,width=\linewidth}
%\put(-40,-3){\large\bf $Q^{2}/GeV^{2}$}
%\put(-110,-3){\large\bf $Q^{2}/GeV^{2}$}
%\put(-88,50){\begin{sideways}\large\bf $\lambda$\end{sideways}}
%\put(-157,50){\begin{sideways}\large\bf C\end{sideways}}
\begin{tabular}{cc}
\begin{minipage}{0.5\linewidth}
\hspace*{-1cm}\epsfig{file=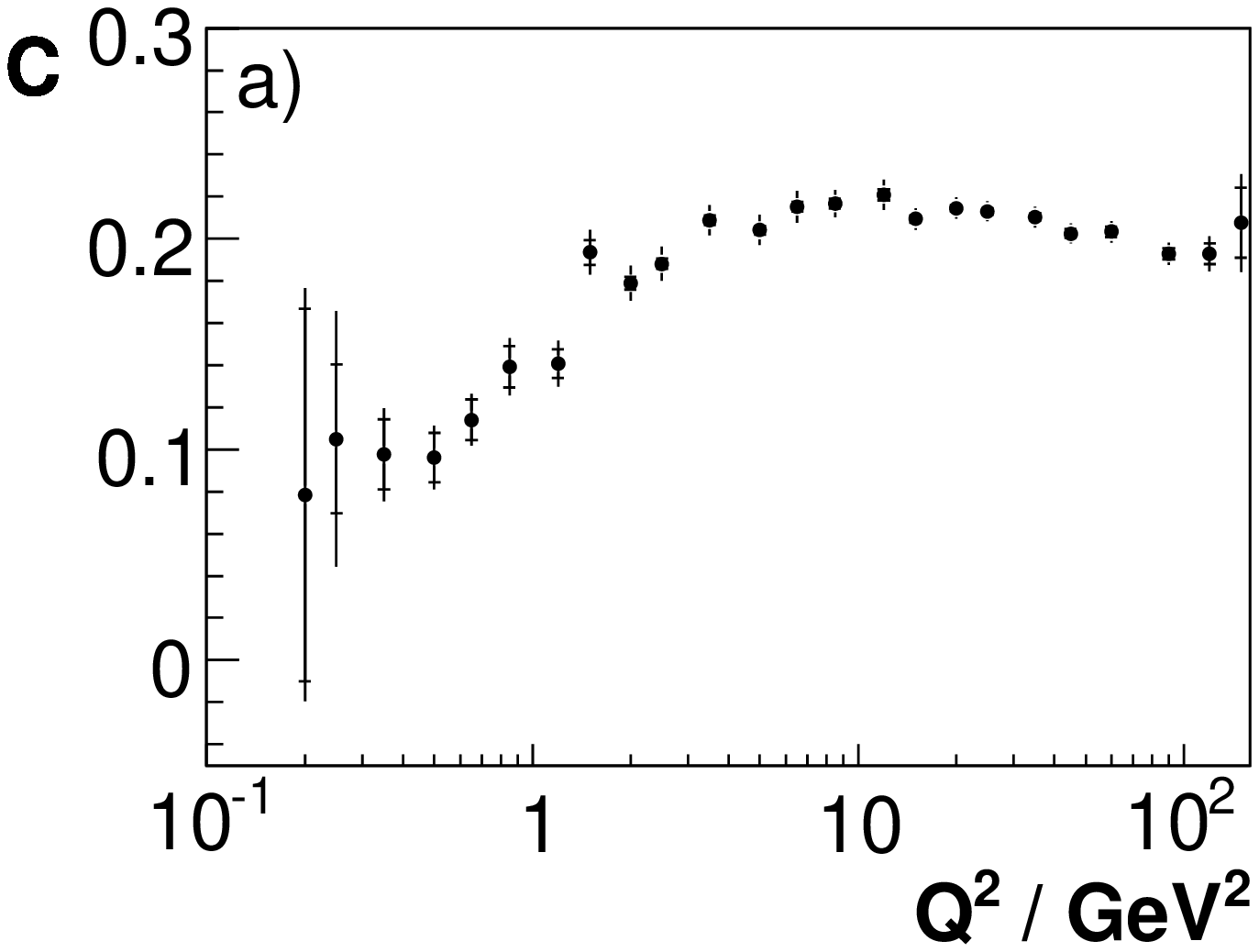,width=1.2\linewidth} 
\end{minipage}
&
\begin{minipage}{0.5\linewidth}
\hspace*{-1cm}\epsfig{file=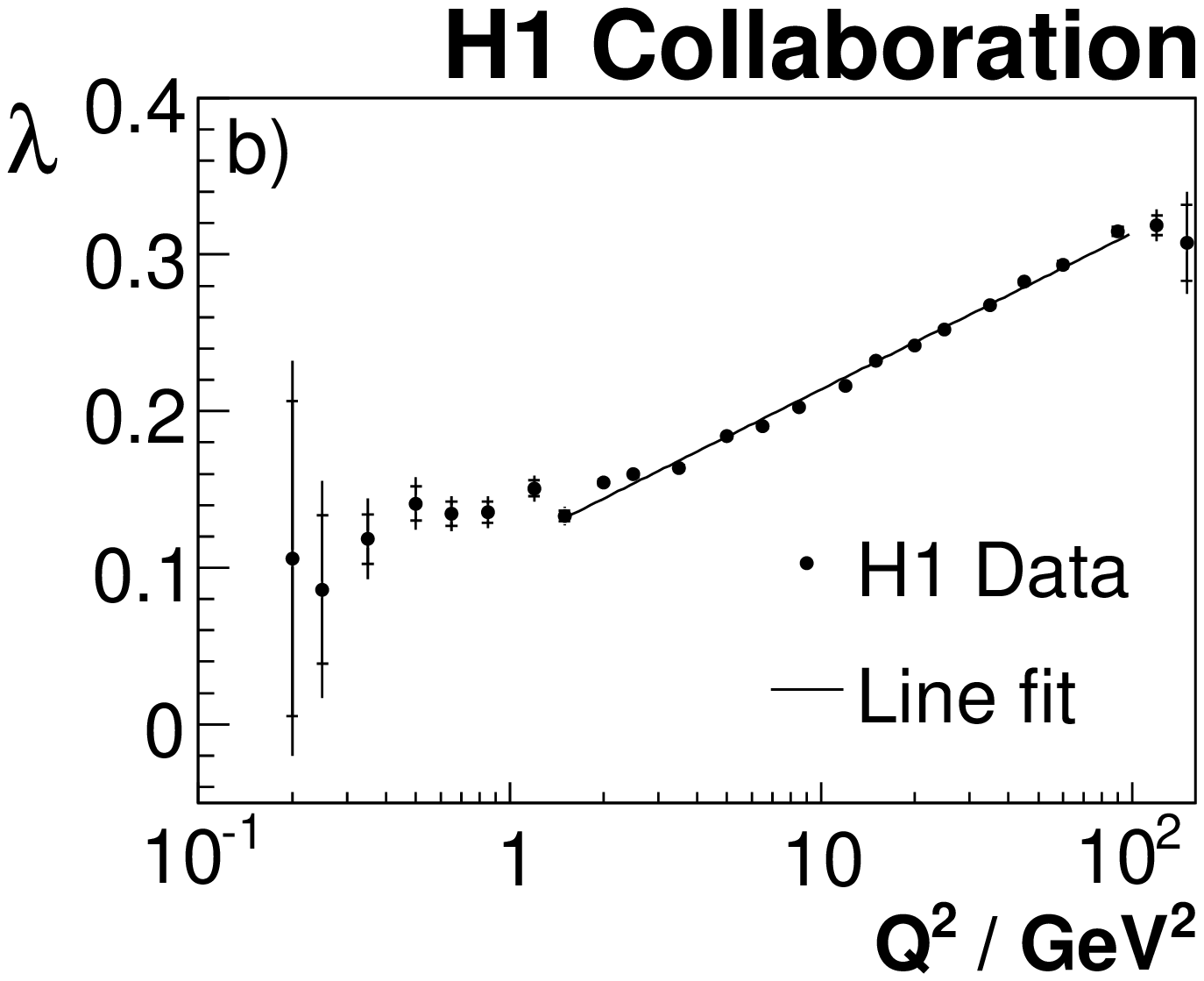,width=1.2\linewidth} 
\end{minipage}
\end{tabular}
\caption{Coefficients $c$ and $\lambda$, as defined in \Eq~\ref{eq:lamfit}, determined from a fit to the  data as a function
of $Q^2$. The inner error bars represent statistical uncertainties. The outer error
bars contain the statistical and 
systematic uncertainties added in quadrature. The line in b) is from a straight-line fit for $Q^2\ge 2$~GeV$^2$.
 \label{fig:xlamlc}}
\end{figure}

\begin{figure}
\begin{tabular}{cc}
\begin{minipage}{0.5\linewidth}
\hspace*{-1cm}\epsfig{file=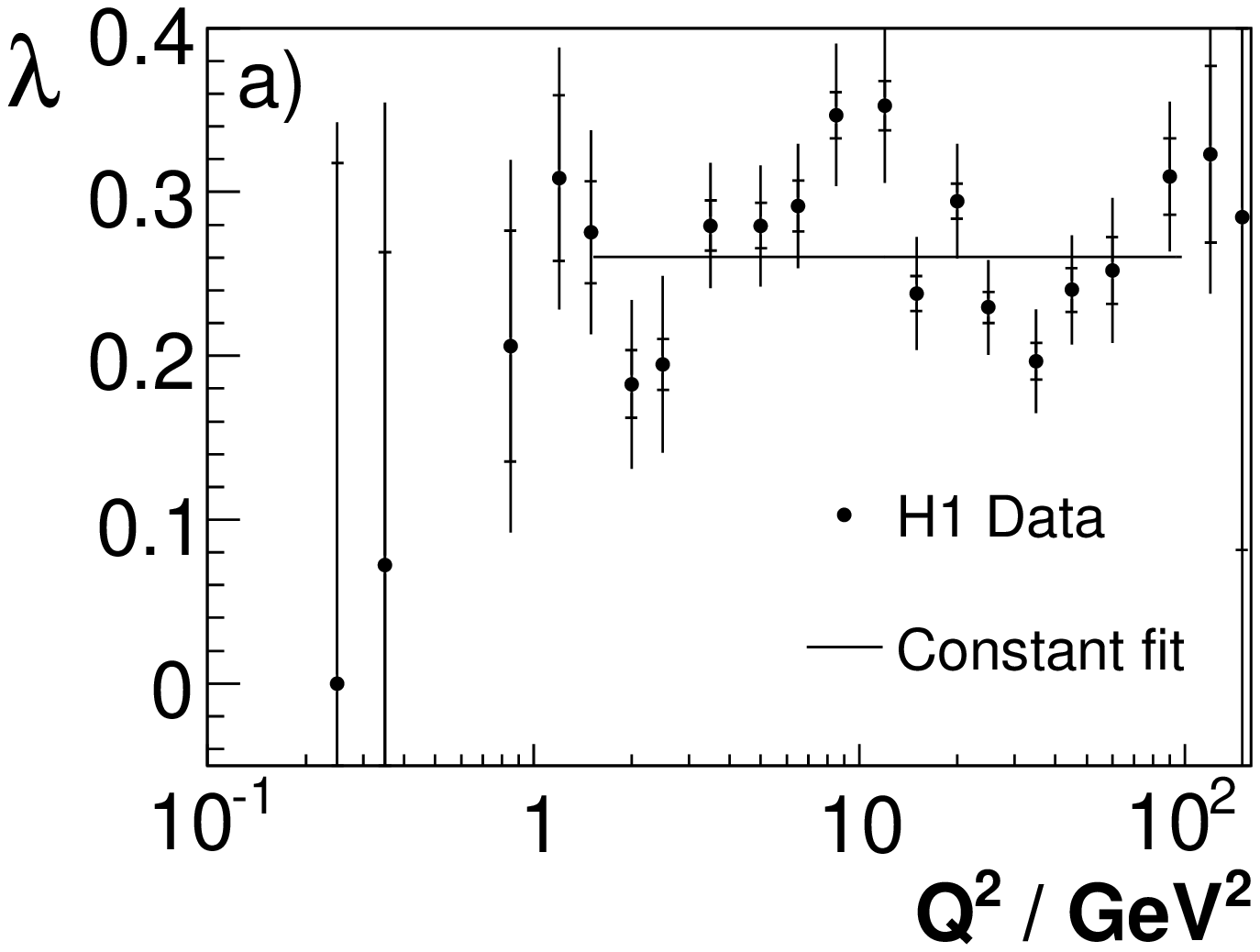,width=1.2\linewidth} 
\end{minipage}
&
\begin{minipage}{0.5\linewidth}
\hspace*{-1cm}\epsfig{file=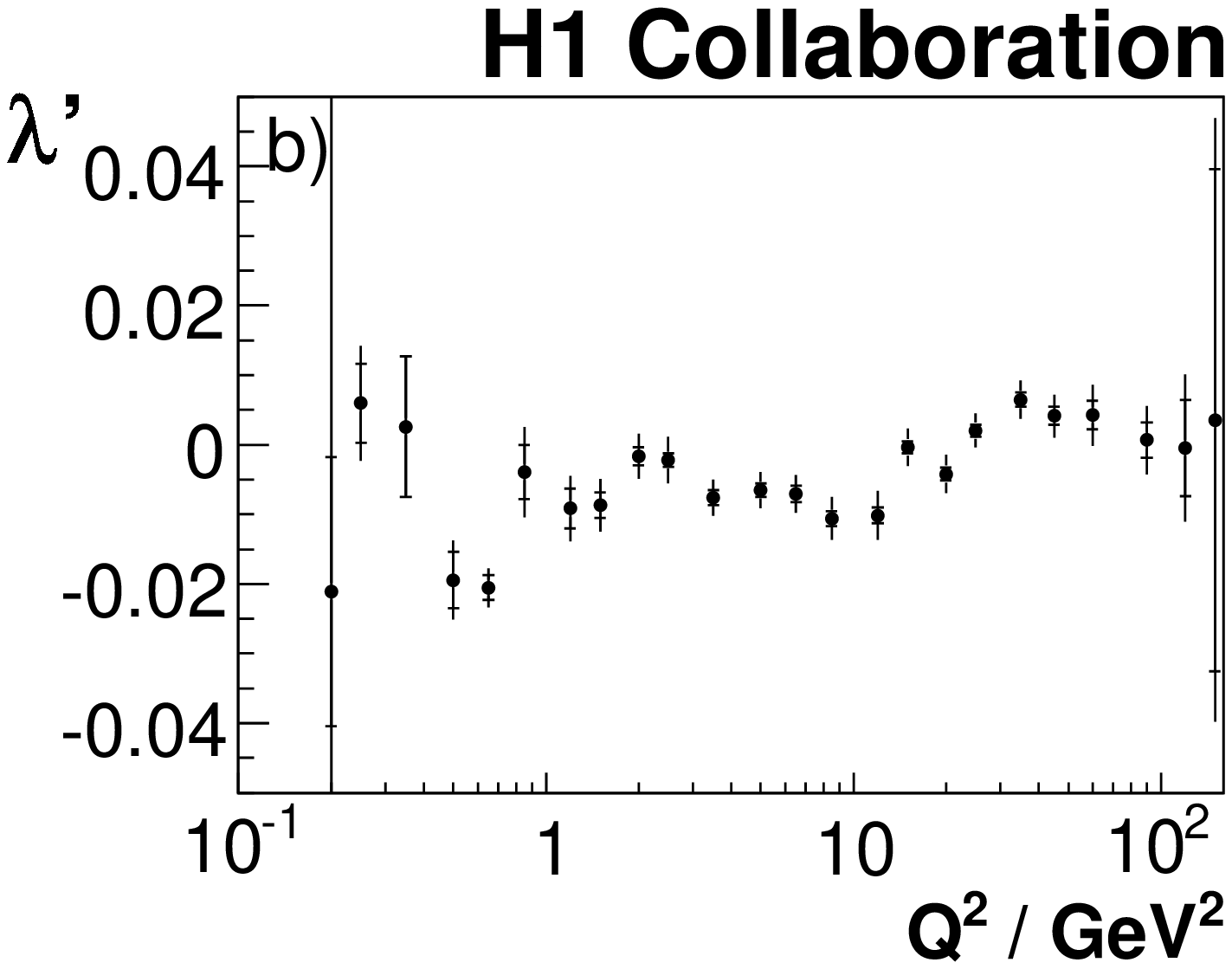,width=1.2\linewidth} 
\end{minipage}
\end{tabular}
\caption{
Coefficients $\lambda$ and $\lambda'$, as defined in \Eq~\ref{eq:lamprim}, determined from a fit to the H1 data as a function
of $Q^2$. The inner error bars represent statistical uncertainties. The outer error
bars contain the statistical and 
systematic uncertainties added in quadrature. The line in a) is from a constant fit for $Q^2\ge 2$~GeV$^2$.
 \label{fig:lamlamprim}
}
\end{figure}

\begin{figure}
\begin{tabular}{cc}
\begin{minipage}{0.5\linewidth}
\hspace*{-1cm}\epsfig{file=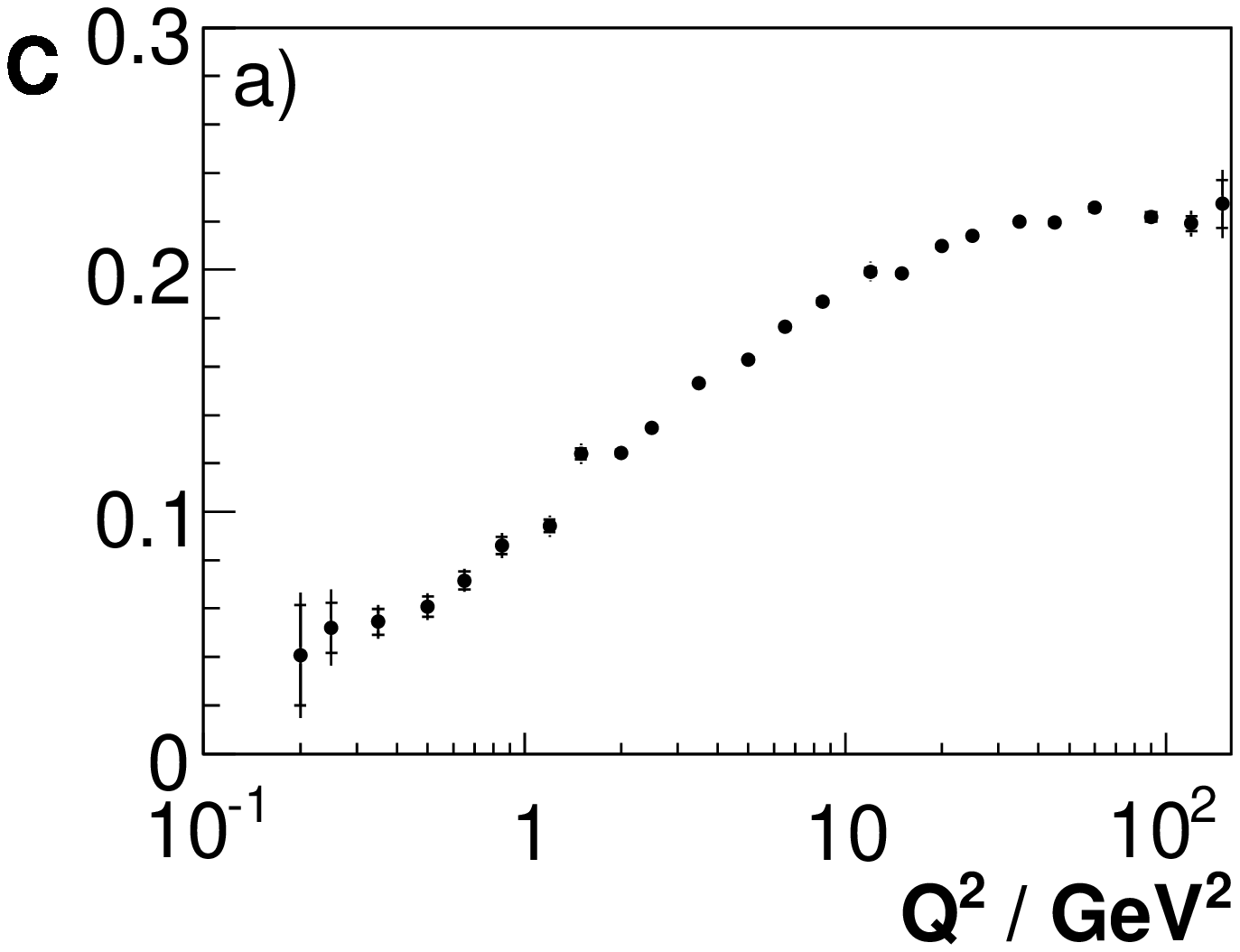,width=1.2\linewidth} 
\end{minipage}
&
\begin{minipage}{0.5\linewidth}
\hspace*{-1cm}\epsfig{file=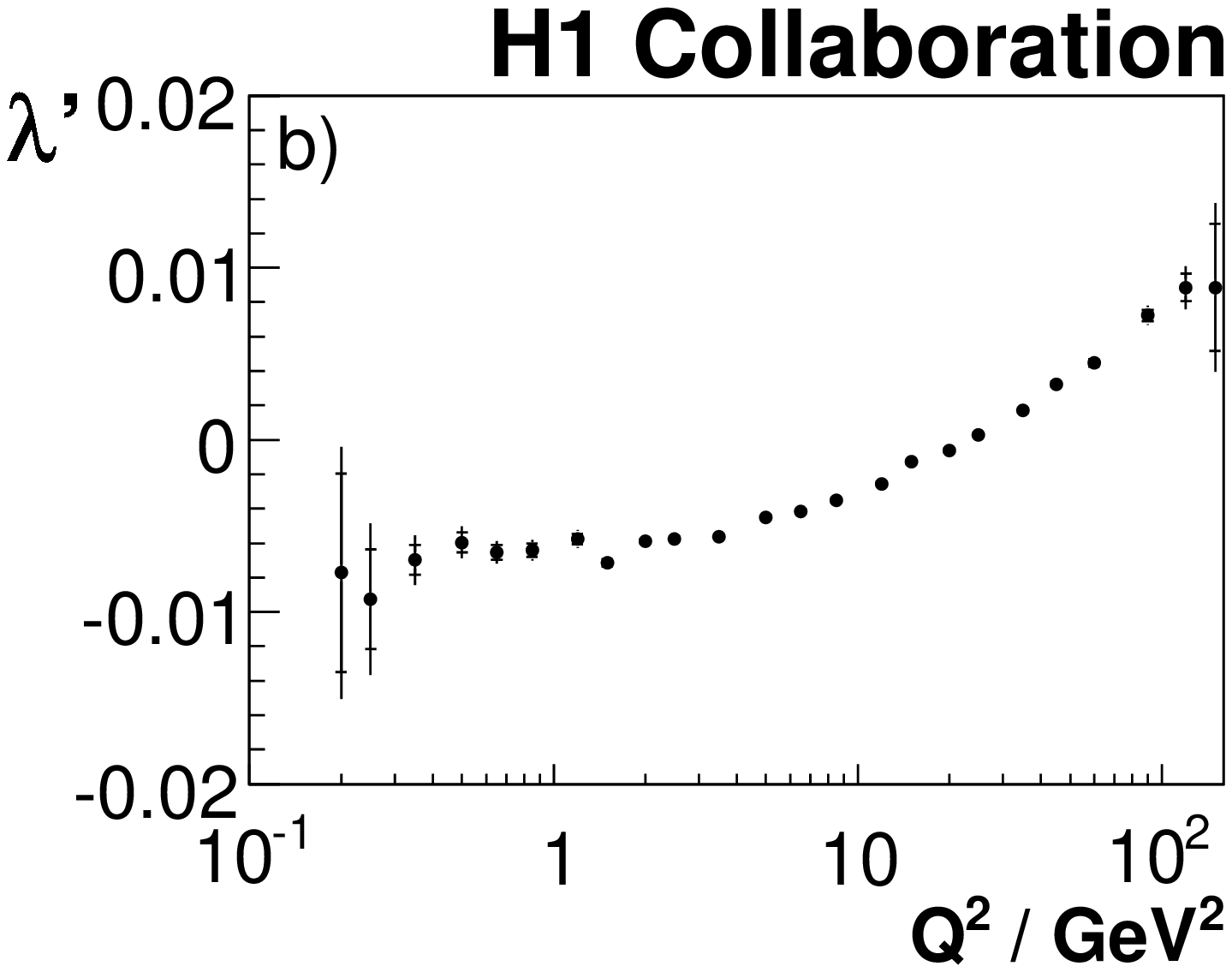,width=1.2\linewidth} 
\end{minipage}
\end{tabular}
\caption{Coefficients $c$ and $\lambda'$, as defined in \Eq~\ref{eq:lamprim}, determined from a fit to the H1 data as a function
of $Q^2$ with fixed $\lambda=0.25$. The inner error bars represent statistical uncertainties. The outer error
bars contain the statistical and 
systematic uncertainties added in quadrature. 
 \label{fig:clamprim}}
\end{figure}

\begin{figure}[t]
\begin{center}
\epsfig{file=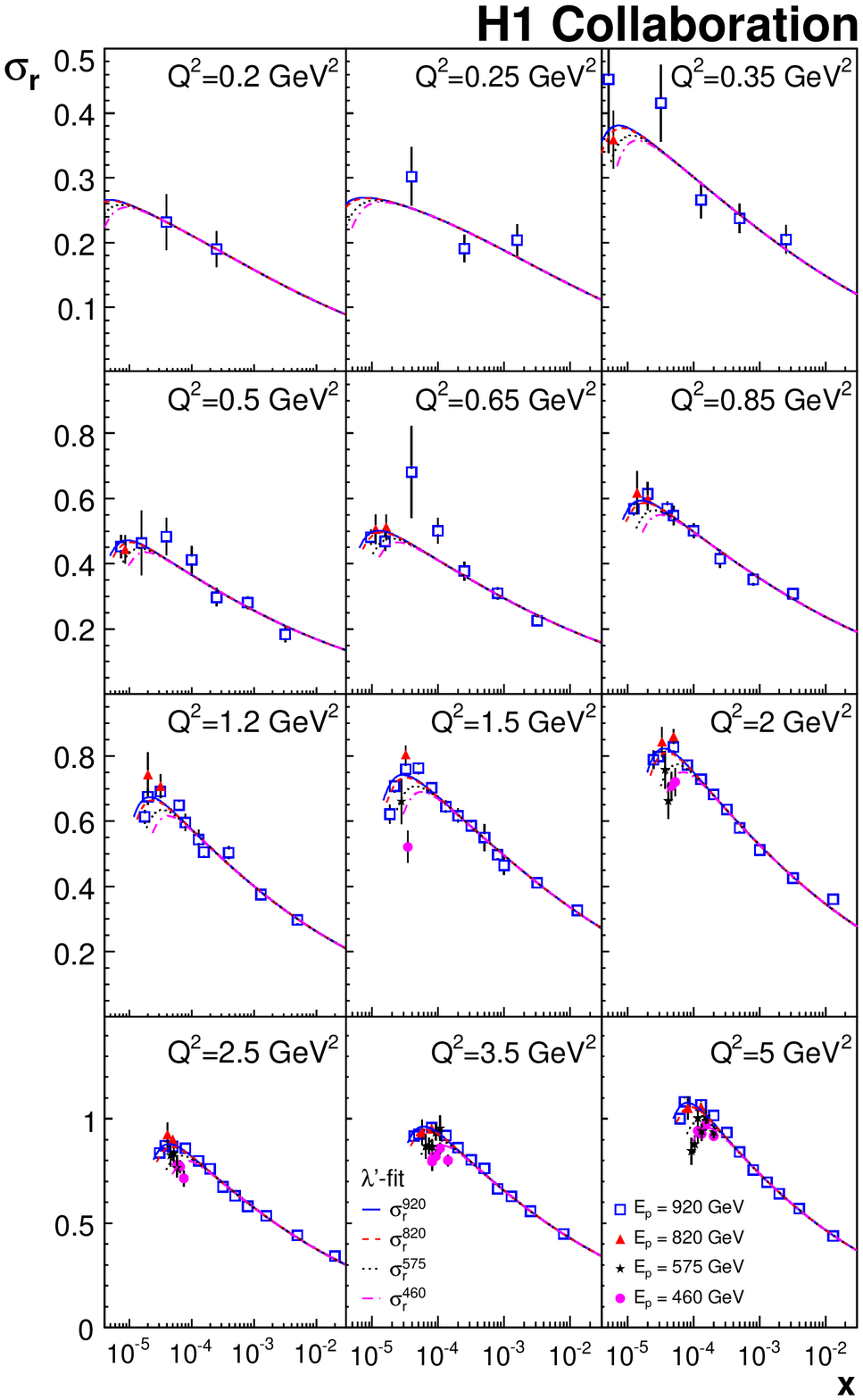,width=0.9\linewidth}
\end{center}
\caption{
Reduced cross section $\sigma_{r}$  as a function 
of $x$    for different $Q^2$ bins for $Q^2\le 5$~GeV$^2$. 
The H1 data are compared to the
$\lambda'$ fit result (shown by curves)
for different proton beam energies $E_p$.
\label{fig:xlam1}}
\end{figure}

\begin{figure}[t]
\begin{center}
\epsfig{file=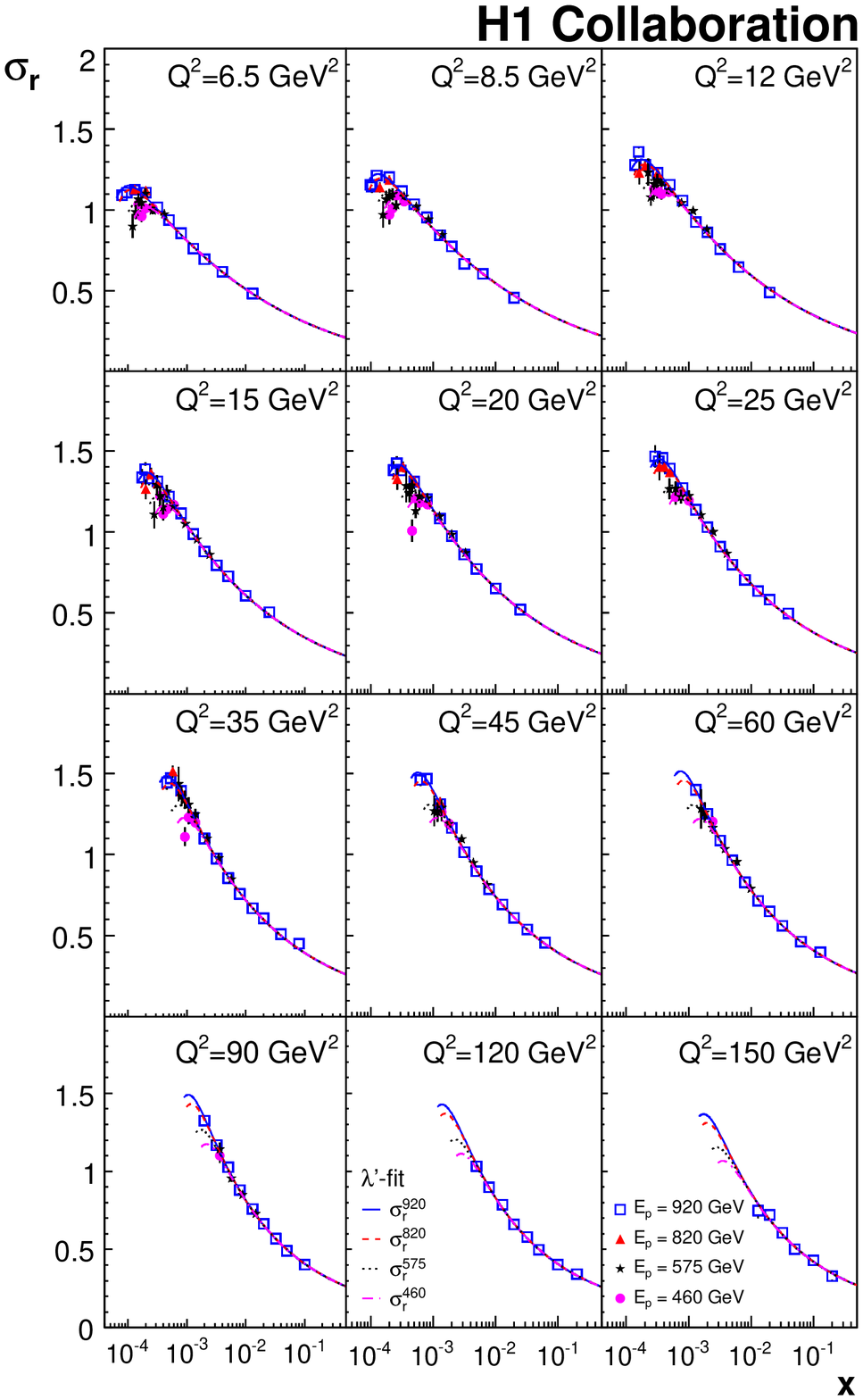,width=0.9\linewidth}
\end{center}
\caption{
Reduced cross section $\sigma_{r}$  as a function 
of $x$    for different $Q^2$ bins for $Q^2> 5$~GeV$^2$. 
The H1 data are compared to the
$\lambda'$ fit result (shown by curves)
for different proton beam energies $E_p$.
\label{fig:xlam2}}
\end{figure}

\begin{figure}
\centerline{\epsfig{file=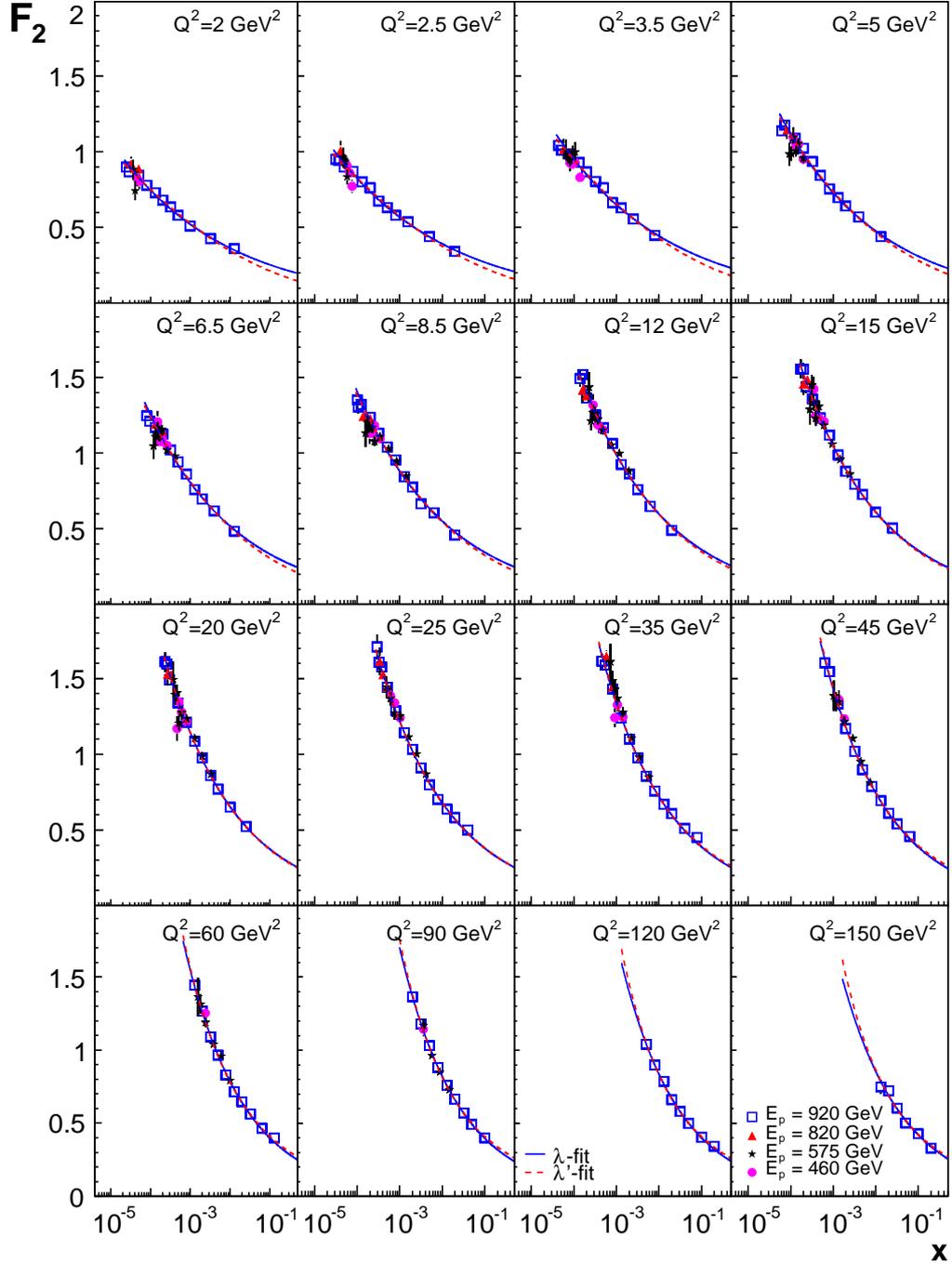,width=\linewidth}}
\caption{
Structure function $F_2(x,Q^2)$  as a function of $x$ calculated from 
the reduced cross section using $R=0.26$ for different $Q^2$ bins. The H1 data
for different proton beam energies $E_p$  are compared to the
$\lambda$ and $\lambda'$ fit results.
\label{fig:f2}}
\end{figure}

The combined H1 data are fitted using the offset method to evaluate systematic uncertainties. 
The parameters obtained in the fits as a function of $Q^2$ 
 are shown in \Fig~\ref{fig:xlamlc}.
The parameter $\lambda$ exhibits an approximately linear increase as a function
of $\ln Q^2$ for $Q^2 \ge 2$~GeV$^2$. For lower $Q^2$, the variation of 
$\lambda$ deviates from that linear dependence. 
The normalisation coefficient $c(Q^2)$ 
rises with increasing $Q^2$ for $Q^2 < 2$\,GeV$^2$ and is consistent with
a constant behaviour for higher $Q^2$, as in\,\cite{adloff-2001-520}.
The total $\chi^2$ of the fit is $\chi^2/\dof=538/350$
when the uncertainties are taken as the statistical and uncorrelated
systematic uncertainties added in quadrature. 
Values of $\chi^2/n_{dof}$ significantly larger than unity may arise in the 
offset method because it does not take into account the 
correlated systematic uncertainties.
Studies show that 
the largest contribution to the  $\chi^2$ arises 
from the $1<Q^2 < 10$~GeV$^2$ domain. In order to further investigate 
this behaviour, the parameterisation of the structure function
$F_2$ is extended by one additional parameter 
\begin{equation} \label{eq:lamprim}
 F_2(x,Q^2)  = c(Q^2) x ^{\textstyle -\lambda (Q^2) + \lambda'(Q^2) \ln x} = c(Q^2) \exp\left[ -\lambda(Q^2)\ln x + \lambda'(Q^2) \ln^2 x\right] 
\end{equation} 
to allow for deviations from a single power law. This fit returns a significantly improved
$\chi^2/\dof = 405/326$. The  parameters $\lambda$ and $\lambda'$ are shown in 
\Fig~\ref{fig:lamlamprim}. 
From this \Fig, it is interesting to observe that
the parameter $\lambda$ becomes consistent with having  a
constant value of $\lambda=0.25$.
The two parameters  $\lambda$ and $\lambda'$ are strongly correlated
since for each $Q^2$ bin the data span over a limited range in $x$.
Therefore, fits are performed,
termed $\lambda'$ fits, for which $\lambda=0.25$ is fixed. 
The quality of these fits with a total $\chi^2/\dof = 464/350$  is 
better than of the original $\lambda$ fits. 
The fitted parameters $c$ and $\lambda'$ are shown in \Fig~\ref{fig:clamprim}.
A comparison of the $\lambda'$ fit result with  the H1 reduced cross-section data 
is given in \Figs~\ref{fig:xlam1} and \ref{fig:xlam2}. 
\FFig~\ref{fig:f2}
shows comparison of the $\lambda$ and $\lambda'$ fits for $Q^2\ge 2$~GeV$^2$
with the structure function $F_2$ which is calculated from the reduced
cross sections  assuming $R=0.26$. 
The parameter $\lambda'$ is negative and  shows a constant behaviour 
for $Q^2 < 5$~GeV$^2$, with a smooth transition  
and a linear rise with $\ln Q^2$ for $Q^2 > 20$~GeV$^2$. 
Therefore, for the low $Q^2$  domain, 
the $\lambda'$ fit shows somewhat softer increase towards low $x$
compared to the $\lambda$ fit opposite to that  observed
for higher $Q^2$ values.
%The $\lambda$ and $\lambda'$  fits are compared to the H1 data in \Fig~\ref{fig:f2}
%for $Q^2\ge 2$~GeV$^2$.
%For this comparison, the measured cross 
%sections are corrected to the structure function 
%$F_2$ assuming $R=0.25$.
% \FFig~\ref{fig:2f2} shows results of the 
%fits and their ratio for $Q^2=5$~GeV$^2$ and $Q^2=90$~GeV$^2$ bins.
%For the low $Q^2$  domain, 
%the $\lambda'$ fit shows somewhat softer increase towards low $x$
%compared to the $\lambda$ fit opposite to that  observed
% for higher $Q^2$ values.
The present measurement of $R$ therefore leads to a refined understanding
of the rise of $F_2$ towards low $x$, which appears to be tamed at
low $Q^2$, and correspondingly lower $x$, compared to a pure power-law 
behaviour, as it was predicted in~\cite{DeRujula:1974rf}. 

In order to consolidate the observations obtained with the offset method, an analysis is 
performed in which the errors are evaluated
using the Hessian method following the $\chi^2$  definition given
in~\cite{Aaron:2009kv}.
The fitted parameters are $2\times 24$ coefficients
$c(Q^2),\lambda(Q^2)$ or $c(Q^2),\lambda'(Q^2)$ for the $24$ $Q^2$ bins
and 
the parameters for the sources of the  systematic uncertainty.
The $\lambda'$ fit returns $\chi^2/\dof = 345.7/350$ compared to 
a worse $\chi^2/\dof = 370.5/350$ of the $\lambda$ fit.
\subsection{DGLAP Fit}
\label{sec:qcdfit}
%\input{qcd}

%\subsubsection{Settings}
The new combined H1 data are used together with the previously published high $Q^2\ge 90$~GeV$^2$ 
H1 data~\cite{Adloff:1999ah,Adloff:2000qj,Adloff:2003uh}
as input to  a DGLAP pQCD fit analysis to NLO, 
with the main objective of studying  $F_L$
predictions.
The HERA measurement regions  are  limited by $W^2_{min} = 300$~GeV$^2$  and  $x_{max} = 0.65$, 
such  that target mass corrections and higher twist 
contributions can be assumed to be small. 
In addition, in order to restrict to a region where perturbative QCD 
is valid, only data with
$Q^2 \ge Q^2_{min} = 3.5$~GeV$^2$ are used in the central fit. 
The influence of this value is discussed further in this section.
The internal consistency of the input data set 
%and its small systematic uncertainties 
enables a 
calculation of the experimental uncertainties on the PDFs using the 
$\chi^2$ tolerance criterion of $\Delta\chi^2=1$.  
The data are fitted using  the program  QCDNUM~\cite{qcdnum} and
the complete error correlation information 
in a $\chi^2$ fit as in~\cite{Aaron:2009kv} using MINUIT~\cite{minuit} as the minimisation program.

The fit procedure begins with  parameterising the input parton 
distribution functions (PDFs) at a starting 
scale  $Q^2_0=1.9~ \rm GeV^2$, chosen to be below the charm mass threshold.
The PDFs are then evolved using the DGLAP evolution
equations~\cite{Gribov:1972ri,Gribov:1972rt,Lipatov:1974qm,Dokshitzer:1977sg,Altarelli:1977zs} 
at NLO~\cite{Curci:1980uw,Furmanski:1980cm} in the $\overline{MS}$ scheme with the
renormalisation and factorisation scales set to $Q^2$, and the strong coupling to 
$\alpha_s(M_Z) =  0.1176$~\cite{pdg}.
The QCD predictions for the structure functions are 
obtained by convoluting  the PDFs with the calculable NLO coefficient functions.
Those are calculated 
using the general mass variable-flavour scheme of ACOT~\cite{Kramer:2000hn}
and cross checked against the  RT scheme~\cite{Thorne:1997ga,Thorne:2006qt}. 
The ACOT and RT schemes differ  in the inclusion of various terms at higher
orders in $\alpha_s$ for the 
computation of the heavy quark structure functions, and for the
structure function $F_L$.

For the QCD fit, the following independent input PDFs are chosen:
the valence quark distributions $xu_v$ and  $xd_v$, the gluon distribution $xg$ and 
anti-quark distributions 
$x\bar{U}$ and $x\bar{D}$. The conditions $x\bar{U} = x\bar{u}$, and
$x\bar{D} = x\bar{d} +x\bar{s}$ are imposed at the starting scale $Q^2_0$. 
A standard generic functional form is used to parameterise these PDFs:
\begin{equation}
 xf(x) = A x^{B} (1-x)^{C} (1 + D x + E x^2).
\label{eqn:pdf}
\end{equation}
The normalisation parameters, $A_{u_v}, A_{d_v}$ and  $A_g$,  
are constrained by  
the fermion number and momentum sum rules. 
The up and down quark type $B$ parameters  
%$B_{\bar{U}}$ and $B_{\bar{D}}$ 
are set equal, $B_{\bar{U}}=B_{\bar{D}}$, such that 
there is only a single $B$ parameter for the sea distributions, which governs
the PDFs at low $x$. 

The strange quark distribution 
is already present at the starting scale, and 
it  is  assumed  that
$x\bar{s}= f_s  x\bar{D}$ at $Q^2_0$. 
The  strange fraction is chosen to be $f_s=0.31$, which is
consistent with determinations 
of this fraction using neutrino induced di-muon production
data~\cite{Martin:2009iq,Nadolsky:2008zw}. 
In addition, to ensure that $x\bar{u} \to x\bar{d}$ 
as $x \to 0$,  the constraint
$A_{\bar{U}}=A_{\bar{D}} (1-f_s)$ is applied.

The initial fits are performed using the same
parameterisation type as for the HERAPDF1.0 fit~\cite{h1zeus:2009wt}, which has
only one  free polynomial parameter, $E_{u_v}$. 
For this parameterisation, the ACOT and RT heavy flavour schemes are compared.
Both fits give good descriptions of the data, but the ACOT fit, which has
$\chi^2/\dof = \chacotone/\ndfdefone$, is superior to the RT fit, with  
$\chi^2/\dof = \chrtone/\ndfdefone$, by about $50$~units. 
Therefore, the ACOT fit is chosen for further more detailed investigations.

\begin{figure}
\centerline{\epsfig{file=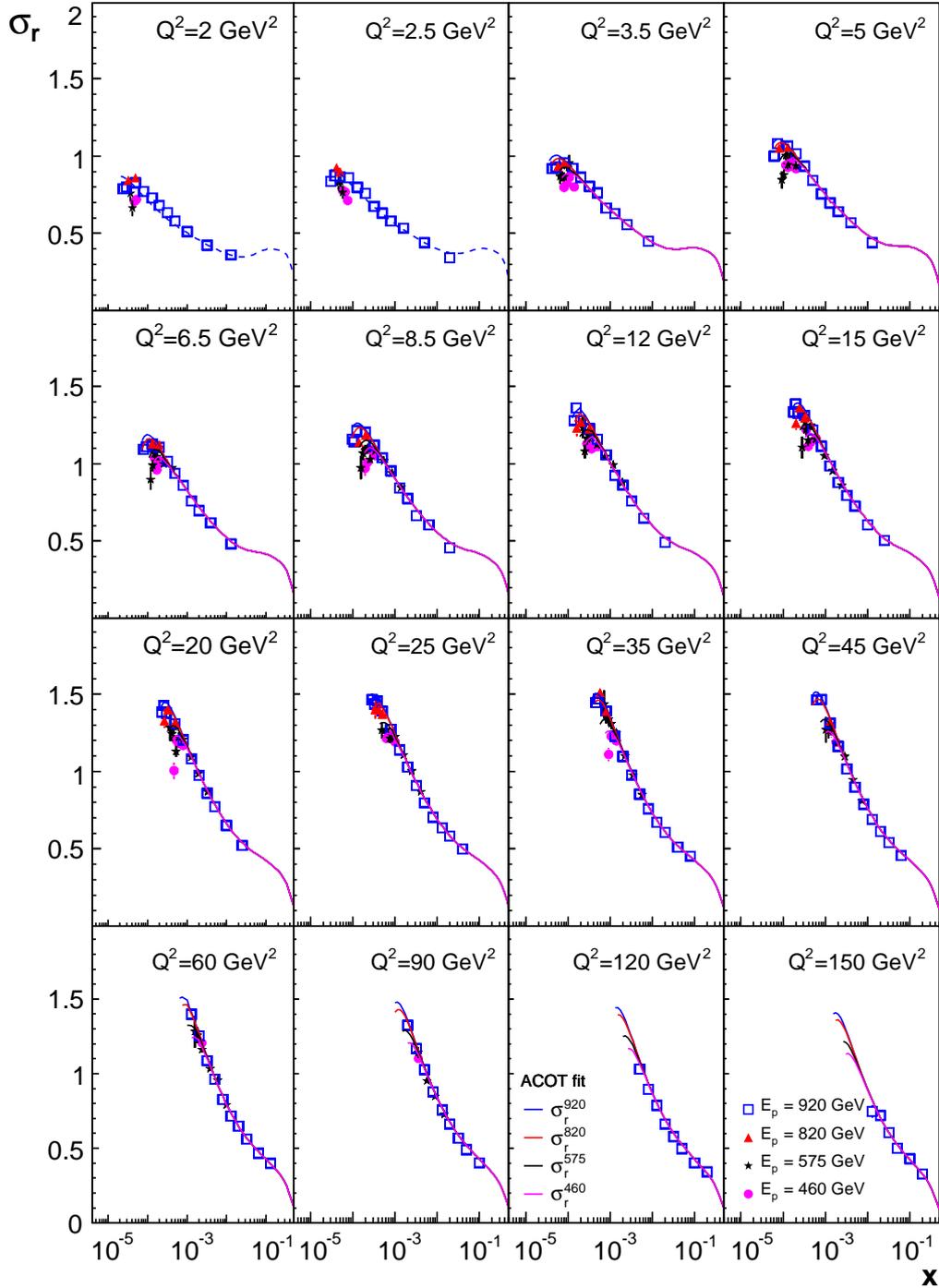,width=\linewidth}}
\caption{ \label{fig:acotcent}
The reduced cross-section measurements taken at different proton beam energies $E_p$ 
compared to the DGLAP fit in the ACOT scheme (shown by curves) for $2.0 \le Q^2\le 150$~GeV$^2$. The dashed line for $Q^2\le 2.5$~GeV$^2$ corresponds to the
fit extrapolation.  
}
\end{figure}

The central fit 
 is chosen
in a $\chi^2$ optimisation procedure, as previously used by H1~\cite{Aaron:2009kv},
in which all extra parameters $D,E$ are
first set to zero, leading to a nine parameter fit. They are then added,
one at a time until no significant improvement in $\chi^2$ is observed.
In addition, 
the assumption that $B_{u_v} = B_{d_v}$ is 
removed and a flexible parameterisation for 
the gluon density with two extra 
parameters~\cite{h1zeus:2009wt} is also tried but both variations
do not lead to significant fit improvements.
The parameterisation procedure also requires
for the central fit that all 
PDFs are positive definite.
The best fit is obtained with the extra free parameters 
$D_{u_v}$ and $E_{u_v}$ 
% and the valence quark 
%approximation $d_{v} > \bar{d}$ and $ u_{v} > \bar{u}$  
%is satisfied at large $x$. 
%The values
%of the parameters are given in \Tab~\ref{tab:acottab}.
%For this fit,
resulting in a $\chi^2/\dof = \chacot/\ndfdef$. 
\FFig~\ref{fig:acotcent} compares the fit result to the low $Q^2$ H1 data.
As a consistency check, a fit using the RT 
heavy flavour scheme is repeated. A similar increase  
in $\chi^2$ of about $50$ units is observed in this case.
\label{sec:dglap}

\begin{table}
\begin{center}
\begin{tabular}{l|cccccc}
\hline
$\boldsymbol{Q^2_{min}}$ {\bf / GeV}$\boldsymbol{^2}$  & $1.5$ & 
$2$ & $2.5$   & $3.5$    &$5$        & $7.5$\\
$\boldsymbol{\chi^2/\dof}$    & $824.8/834$ & $777.9/818$ & $748.7/801$ & $\chacot/\ndfdef$ & $677.6/759$ & $626.9/712$ \\ 
\hline
\end{tabular}
\end{center}
\tablecaption{\label{tab:q2scan}Values of $\chi^2/\dof$ for QCD fits 
with different $Q^2_{min}$ values.}
\end{table}

\begin{table}
\begin{center}
\begin{tabular}{l|cccccc}
\hline
$\boldsymbol{A_S}$                & $0.2$ & $0.3$ & $0.5$   & $0.7$   &$1.0$    &$1.5$\\
$\boldsymbol{\chi^2/\dof}$    & $709.5/777$ & $696.1/762$ & $643.1/734$ & $617.3/709$ & $594.4/690$ & $554.1/654$ \\ 
\hline
\end{tabular}
\end{center}
\tablecaption{\label{tab:satscan}Values of $\chi^2/\dof$ for QCD fits 
with different values of the parameter $A_S$ for saturation-inspired cut on data, see \Eq~\ref{eq:asasas}.}
\end{table}

\begin{figure}
\centerline{\epsfig{file=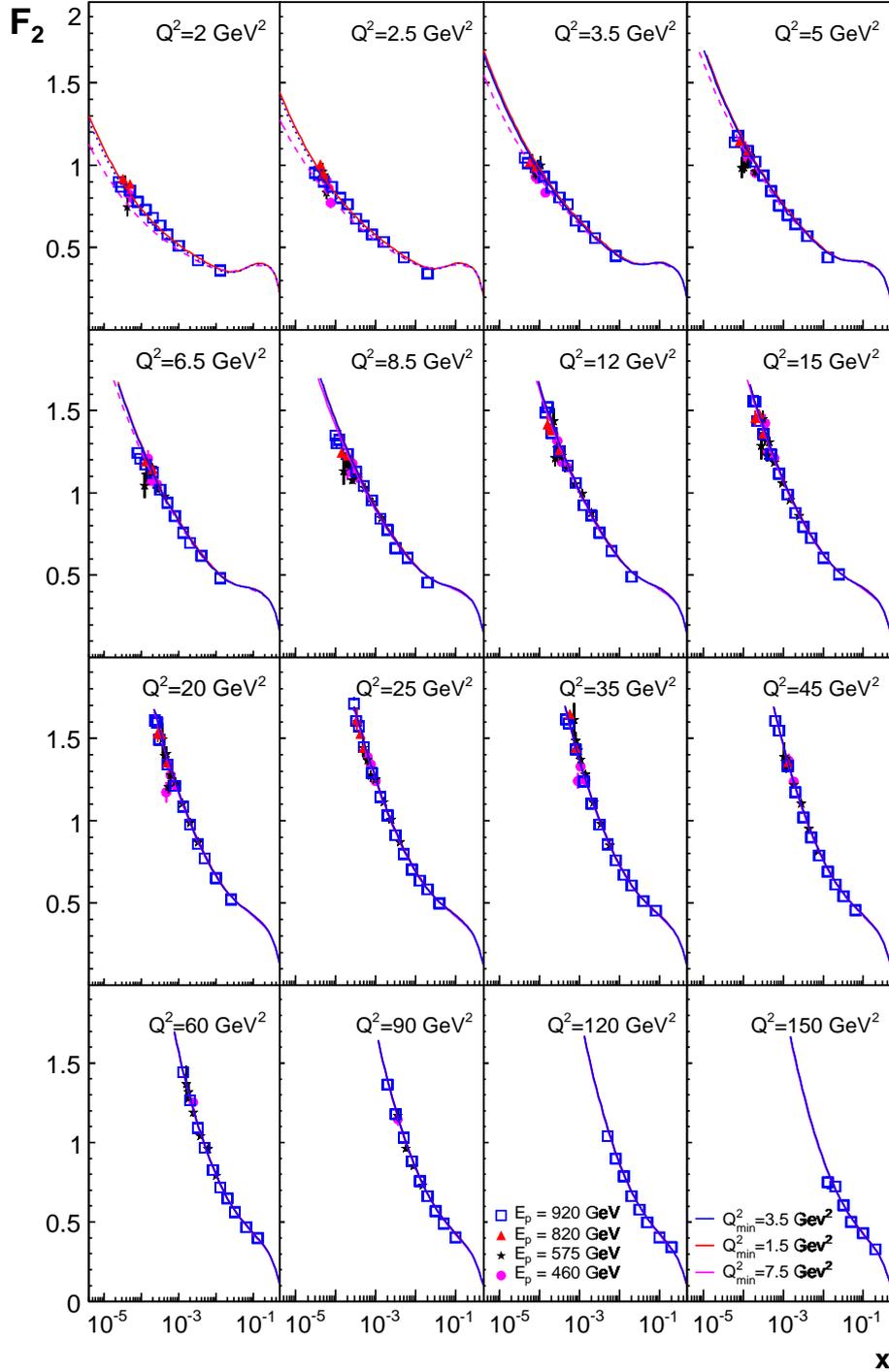,width=0.9\linewidth}}
\caption{ \label{fig:acotf2}
The structure function $F_2$ as a function of $x$ calculated from 
the reduced cross section using $R=0.26$ for different $Q^2$ bins. The H1 data
for different proton beam energies $E_p$  are compared to the
DGLAP fit in the ACOT scheme with different values of the $Q^2_{min}$ cut.
The dashed and dotted lines correspond to the  extrapolation of the fits
to lower $Q^2$ values.
}
\end{figure}

\begin{figure}
\centerline{\epsfig{file=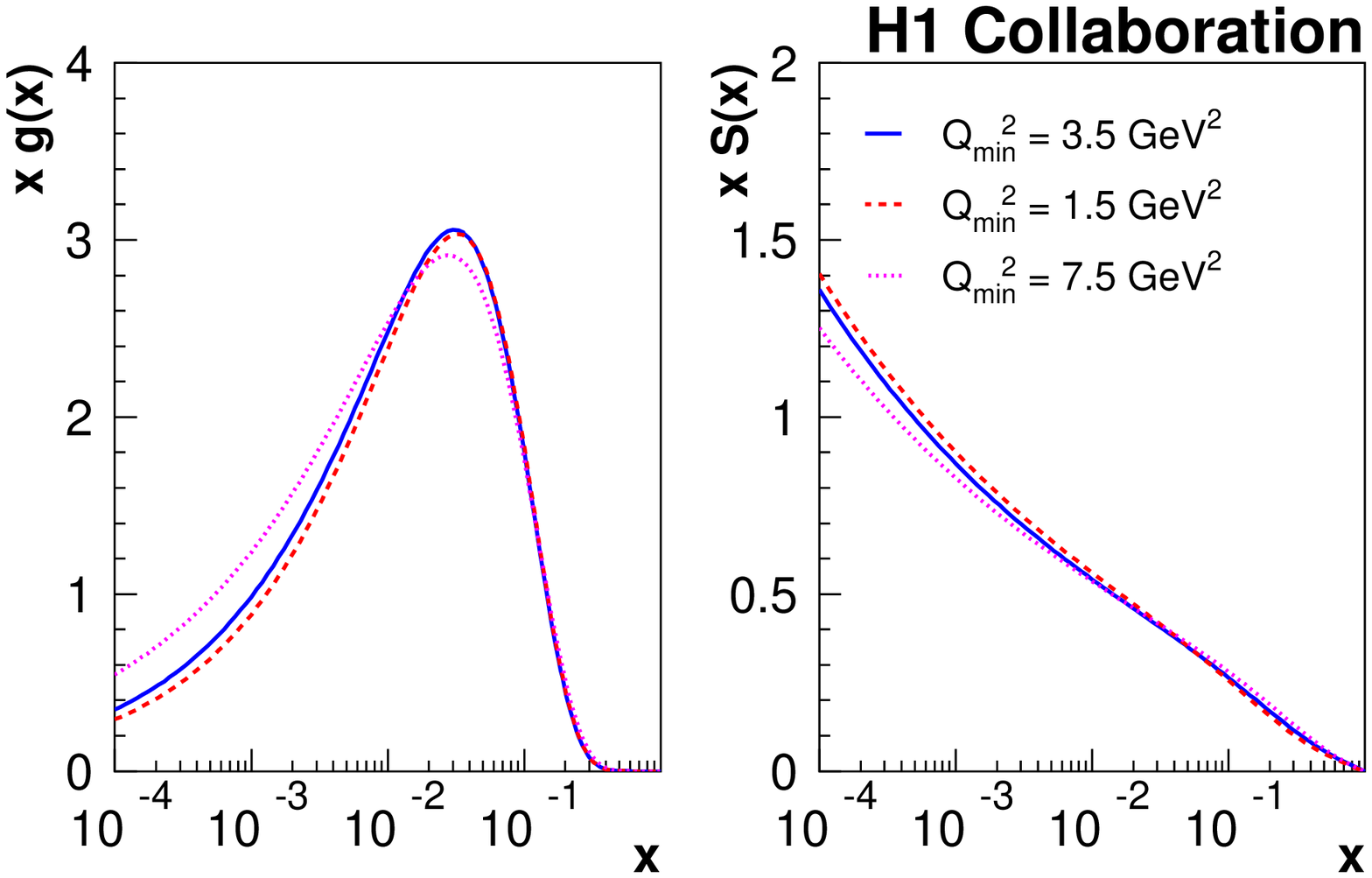,width=0.9\linewidth}}
\tablecaption{\label{fig:q2pdfs}Gluon and sea quark PDFs shown 
at the starting scale $Q^2_0=1.9$~GeV$^2$ for different values of $Q^2_{min}$.}
\end{figure}

The sensitivity of the fit to the inclusion of 
low $Q^2$ data is studied by varying the $Q^2_{min}$ cut. The variation
of the fit quality in terms of $\chi^2/\dof$ is summarised 
in \Tab~\ref{tab:q2scan}. Increasing the $Q^2_{min}$ cut leads to a steady
decrease in the $\chi^2/\dof$, suggesting that the fit has some
difficulties to describe the data at  low $Q^2$ values. 
\FFig~\ref{fig:acotf2} compares the structure function $F_2$ with the 
fits performed using different $Q^2_{min}$ cuts. 
At low $Q^2$ the shape of the measured structure function $F_2$ as a 
function of $x$ is somewhat different from those obtained
by the DGLAP fits based on the parameterisation described above.  
The fit obtained with a $Q^2_{min}$ cut of $7.5$~GeV$^2$ falls significantly 
below the data at small $x$ when 
extrapolating to the low $Q^2$ region.
\FFig~\ref{fig:q2pdfs} shows gluon and sea-quark distributions 
for different values of $Q^2_{min}$ at the
evolution starting scale $Q^2_0=1.9$~GeV$^2$. A change of $Q^2_{min}$
from $1.5$~GeV$^2$ to $7.5$~GeV$^2$ leads to an increase of the gluon
distribution while the sea-quark distribution becomes smaller at low $x$.
This suppression of the sea-quark contribution at small $x$ when using a
 $Q^2_{min}$ cut of $7.5$~GeV$^2$ 
is responsible for the smaller values of $F_2$ obtained by this fit at small 
$Q^2$ and small $x$. 

\begin{figure}
\centerline{\epsfig{file=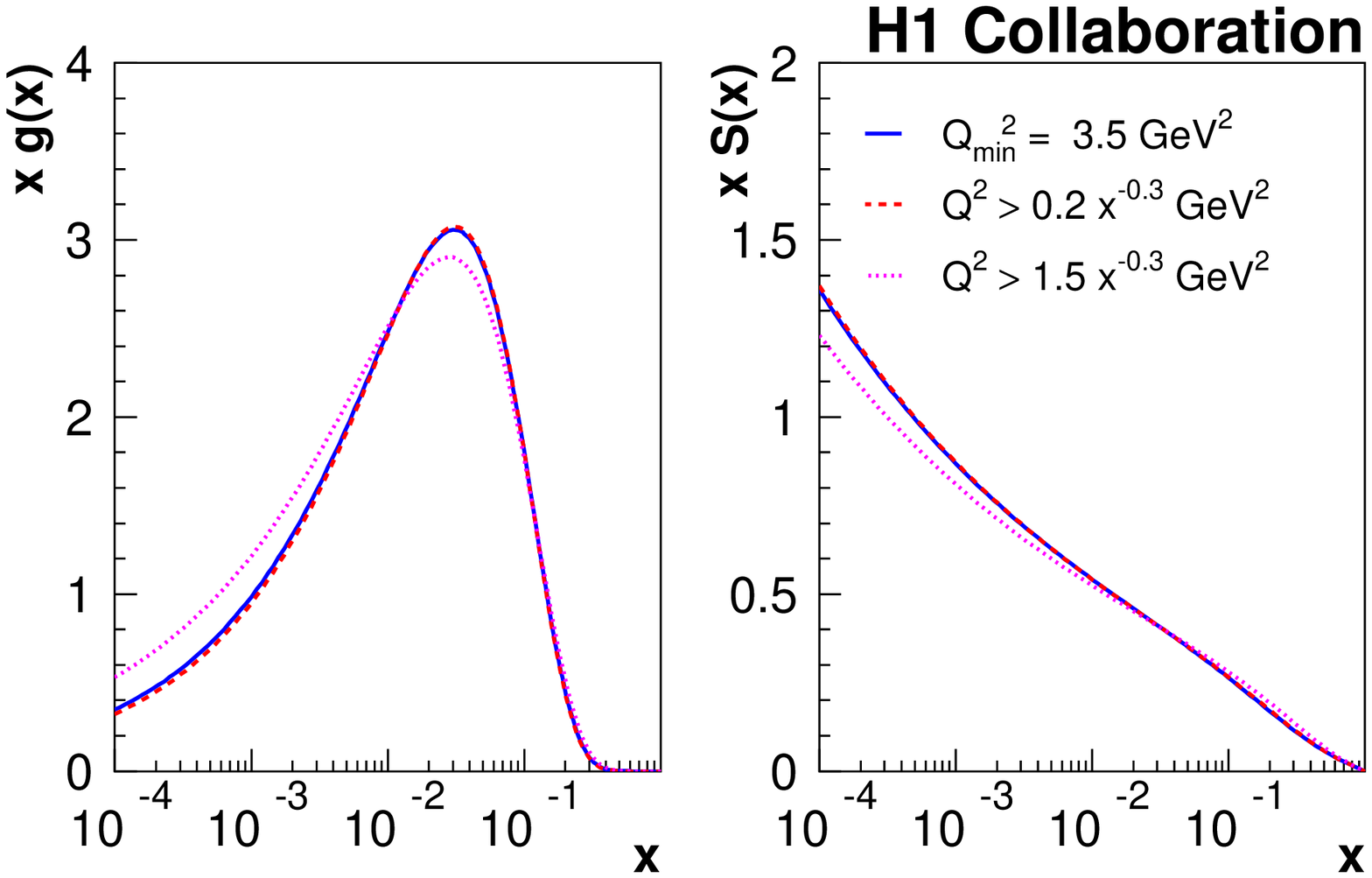,width=0.9\linewidth}}
\tablecaption{\label{fig:satpdfs}Gluon and sea quark PDFs shown 
at the starting scale $Q^2_0=1.9$~GeV$^2$ for 
the central fit with  $Q^2_{min}=3.5$~GeV$^2$ and for the fits with additional
cuts $Q^2>A_s x^{-0.3}$ where
 $A_{S}=0.2$ and $A_{s}=1.5$.}
\end{figure}

An alternative approach to the $Q^2_{min}$ variation is a saturation-inspired cut on the 
kinematic region depending on $x$ like
\begin{equation} \label{eq:asasas}
       Q^2 \ge A_S x^{-\lambda_S},
\end{equation}
with $\lambda_S = 0.3$ and different values of the parameter $A_S$, as 
suggested in~\cite{Caola:2009iy}.  The dependence of $\chi^2/\dof$ on $A_S$ is given in
\Tab~\ref{tab:satscan}. \FFig~\ref{fig:satpdfs} shows gluon and sea-quark distributions
for different values of $A_S$. The saturation-inspired cut has an
effect similar to the one of the  $Q^2_{min}$ variation. 
The fit quality improves
with increasing $A_S$ and the gluon becomes larger while the sea-quark 
density decreases at low $x$.

To facilitate the comparison of the data with dipole model predictions,  
DGLAP fits are also performed in the kinematic
domain $x<0.01$ and $Q^2\ge 3.5$~GeV$^2$, in which both the DGLAP theory and the 
dipole ansatz can be assumed to hold. The valence quark parameters cannot be determined
in this range. Therefore they are fixed to the values obtained by the full phase space fits.
There are six non-valence quark parameters,
$B_g$, $C_g$, $A_{\bar{D}}$, $B_{\bar{D}}$, $C_{\bar{D}}$ and
$C_{\bar{U}}$, as compared to three free parameters of the dipole models discussed below.
When restricted to this common kinematic domain, the ACOT fit is of very good quality, 
with  $\chi^2/\dof=\chacotqcut/\ndfdglapqcut$ while the RT fit yields
 $\chi^2/\dof=\chrtqcut/\ndfdglapqcut$.

\subsection{Dipole Model Fits}

\label{sec:dipole}
At low $x$ and low $Q^2$, virtual photon-proton scattering has been described
using the colour dipole model~\cite{Nikolaev:1990ja}. In this model, 
the scattering process is calculated as a fluctuation of the photon into a 
quark-antiquark pair (dipole), with a lifetime $\propto 1/x$, which interacts
with the proton.  

Several approaches were developed to phenomenologically describe the dipole-proton interaction 
cross section, three of which are subsequently  applied  to the data
of this paper. These are the original model
version (\gbw)\,\cite{Golec-Biernat:1998js},  
a model based on the colour glass condensate approach to the high parton density
regime
 (\iim)\,\cite{Iancu:2003ge}, and a model
with the generalised impact parameter dipole 
saturation (\bsat)\,\cite{Kowalski:2006hc}. 

In the GBW  model  the dipole-proton
cross section $\hat{\sigma}$ 
is given by
\begin{equation} \label{eq:gbw}
\hat{\sigma}(x,r) = \sigma_0
 \left\{1 - \exp \left[-r^2 / \left(4 r_0^2(x)\right) \right] \right\}\,,
\end{equation}
where $r$ corresponds to the transverse separation between the quark and 
the antiquark, and $r^2_0$ is an $x$ dependent scale parameter,
assumed to have the form
\begin{equation}
 r_0^2(x) \sim \left(x/x_0\right)^{\lambda} \label{eq:satrad}\,.
\end{equation}
The parameters of the fit are the cross-section normalisation 
$\sigma_0$ as well as $x_0$ and $\lambda$.
The IIM model has a modified expression for 
 $\hat{\sigma}$ using the parameter $R_{\rm IIM}$
instead of  $\sigma_0$. The B-SAT model modifies  \Eq~\ref{eq:gbw} by
adding effects of the DGLAP evolution. This model uses as an input a gluon density
\begin{equation}
xg (x, Q_0^2) = A_g x^{-\lambda_g} (1-x)^{5.6}
\end{equation}
with the starting scale $Q_0^2$, normalisation $A_g$ and low $x$ exponent $\lambda_g$ as free parameters
while the other dipole model parameters are kept fixed.

The dipole models are applicable
at low $x<0.01$ where the gluon and sea quark
densities dominate. 
The models are valid down to the 
photoproduction limit $Q^2 \approx 0$, therefore
no $Q^2$ cut is  applied for the central fits.
In DGLAP fits it is observed that the 
contribution of the valence quarks 
to the $ep$ scattering cross section is 
sizeable for the whole HERA kinematic range,
compared to the data precision. %For the measured data points 
This 
contribution varies between $5\%$ and $15\%$ for $x$ 
varying from $0.0001$ to $0.01$. Modified dipole fits are 
therefore performed, in which
the contribution from the valence quarks to the cross section,
as determined by the central DGLAP fit, is added to the dipole model 
prediction. These fits are performed requiring that
 $Q^2\ge 3.5$~GeV$^2$, to be consistent with DGLAP analyses.

It should be noted that the size of the valence-quark 
contribution  at low $x$ is rather uncertain because it is
only indirectly constrained by the HERA data. 
For $x>0.01$ it is 
determinedned by the combination of neutral and charged current scattering cross 
sections measurements. %but 
At lower $x$ however the valence quark contribution follows from the 
% by the 
parameterisation %shape 
and the fermion number sum rules. 
The uncertainty can reach
$\sim 30\%$ at $x=0.0001$~\cite{h1zeus:2009wt}.

\begin{table}
\begin{center}
\renewcommand{\arraystretch}{1.1}
\begin{small}
\begin{tabular}{l|cc|cc|cc}
\hline
{\bf Parameter} & {\bf Value} & {\bf Uncertainty} & {\bf Value} & {\bf Uncertainty} & {\bf Value} & {\bf Uncertainty} \\
\hline
         & \multicolumn{2}{c|}{\bf Nominal GBW} & \multicolumn{2}{c|}{\bf 
$\boldsymbol{Q^2\ge 3.5}$~GeV$^2$ \gbw} &  \multicolumn{2}{c}{\bf \gbwdglap} \\
\hline
$\sigma_{0}$ (mb) & $ 21.7$ & $  0.7$ & $ 18.4$ & $  0.7$ & $ 17.7$ & $  0.7$ \\
$\lambda$ & $0.287$ & $0.002$ & $ 0.296$ & $0.003$ & $ 0.336$ & $0.003$ \\
$x_{0}$ & $ 1.76 \times 10^{-4}$ & $ 0.25 \times 10^{-4}$ & $ 3.50 \times 10^{-4}$ & $ 0.60 \times 10^{-4}$ & $ 3.46 \times 10^{-4}$ & $ 0.57 \times 10^{-4}$ \\
\hline
         & \multicolumn{2}{c|}{\bf Nominal IIM} & \multicolumn{2}{c|}{\bf$\boldsymbol{Q^2\ge 3.5}$~GeV$^2$ \iim} &  \multicolumn{2}{c}{\bf \iimdglap} \\
\hline
$R_{\rm IIM}$ (fm) & $0.611$ & $0.007$ & $0.595$ & $0.007$ & $0.666$ & $0.009$ \\
$\lambda$ & $0.258$ & $0.004$ & $ 0.260$ & $0.004$ & $0.289$ & $0.005$ \\
$x_{0}$ & $ 0.48 \times 10^{-4}$ & $ 0.06 \times 10^{-4}$ & $ 0.61 \times 10^{-4}$ & $ 0.08 \times 10^{-4}$ & $ 0.15 \times 10^{-4}$ & $ 0.03 \times 10^{-4}$ \\
\hline
         & \multicolumn{2}{c|}{\bf Nominal B-SAT} & \multicolumn{2}{c|}{\bf $\boldsymbol{Q^2\ge 3.5}$~GeV$^2$ \bsat} &  \multicolumn{2}{c}{\bf \bsatdglap} \\
\hline
$A_g$ & $ 2.32$ & $ 0.06$ & $ 2.38$ & $ 0.09$ & $ 1.66$ & $ 0.05$ \\ 
$\lambda_g$  & $0.088$ & $0.010$ & $ 0.073$ & $0.014$ & $ 0.099$ & $0.011$ \\ 
$Q^2_0$ (GeV$^2$) & $2.25$ & $0.18$ & $ 2.04 $ & $ 0.20 $ & $ 1.51 $ & $ 0.11 $ \\  
\hline
\end{tabular}
\end{small}
\end{center}
\renewcommand{\arraystretch}{1.}
\tablecaption{\label{tab:dipole}
Parameters and total uncertainties  for 
GBW, IIM and B-SAT dipole model fits for various fit conditions described in 
the text. The $Q^2\ge 3.5$~GeV$^2$ and DGLAP$_{valence}$ fits are performed
in a kinematic phase spaces restricted to $Q^2\ge 3.5$~GeV$^2$.}
\end{table}

\begin{figure}
\centerline{\epsfig{file=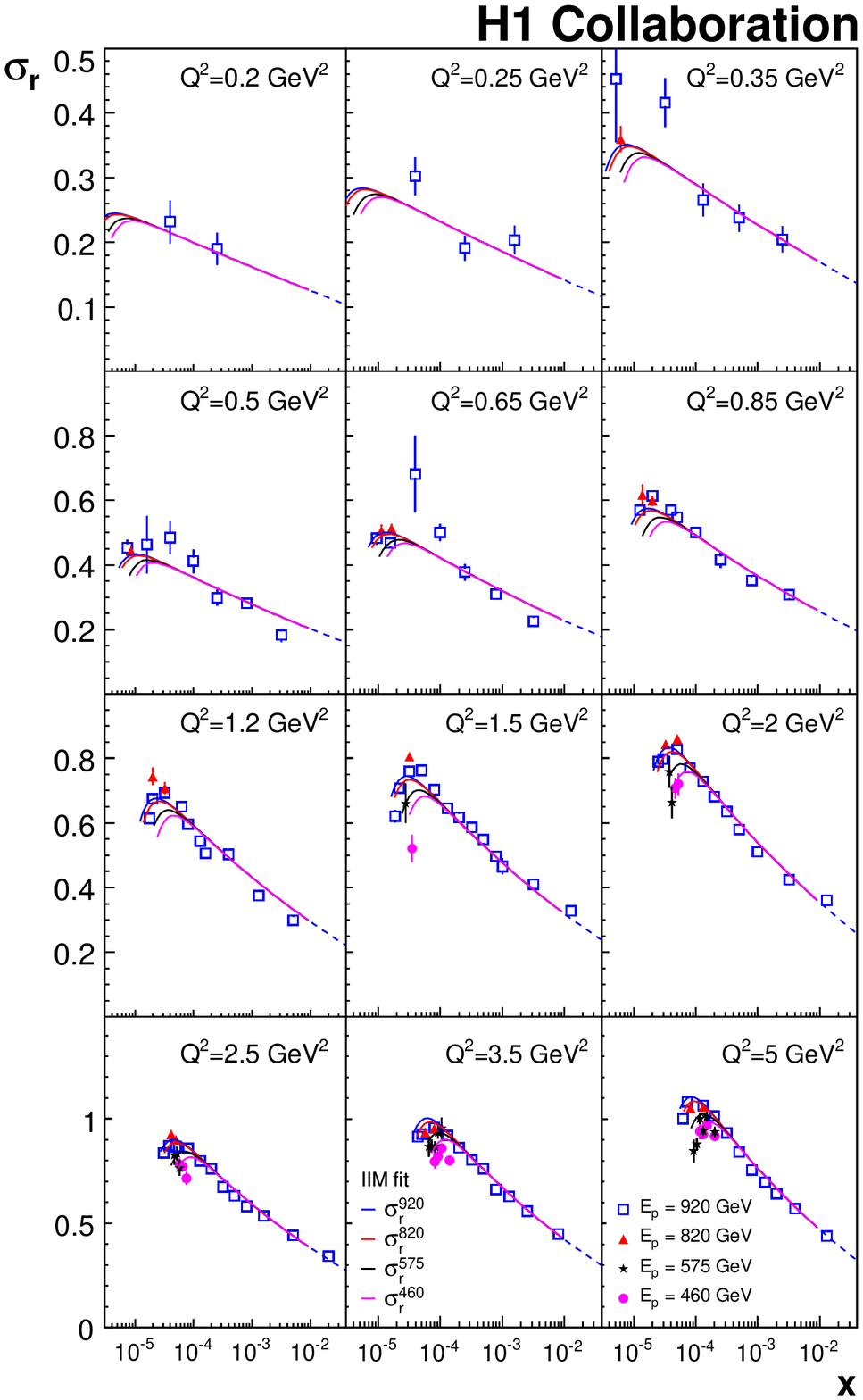,width=.9\linewidth}}
\caption{ \label{fig:iimfit1}
Reduced cross-section data taken at different proton beam energies $E_p$
compared to \iim\ fit results for $0.2 \le Q^2\le 5$~GeV$^2$.
}
\end{figure}

\begin{figure}
\centerline{\epsfig{file=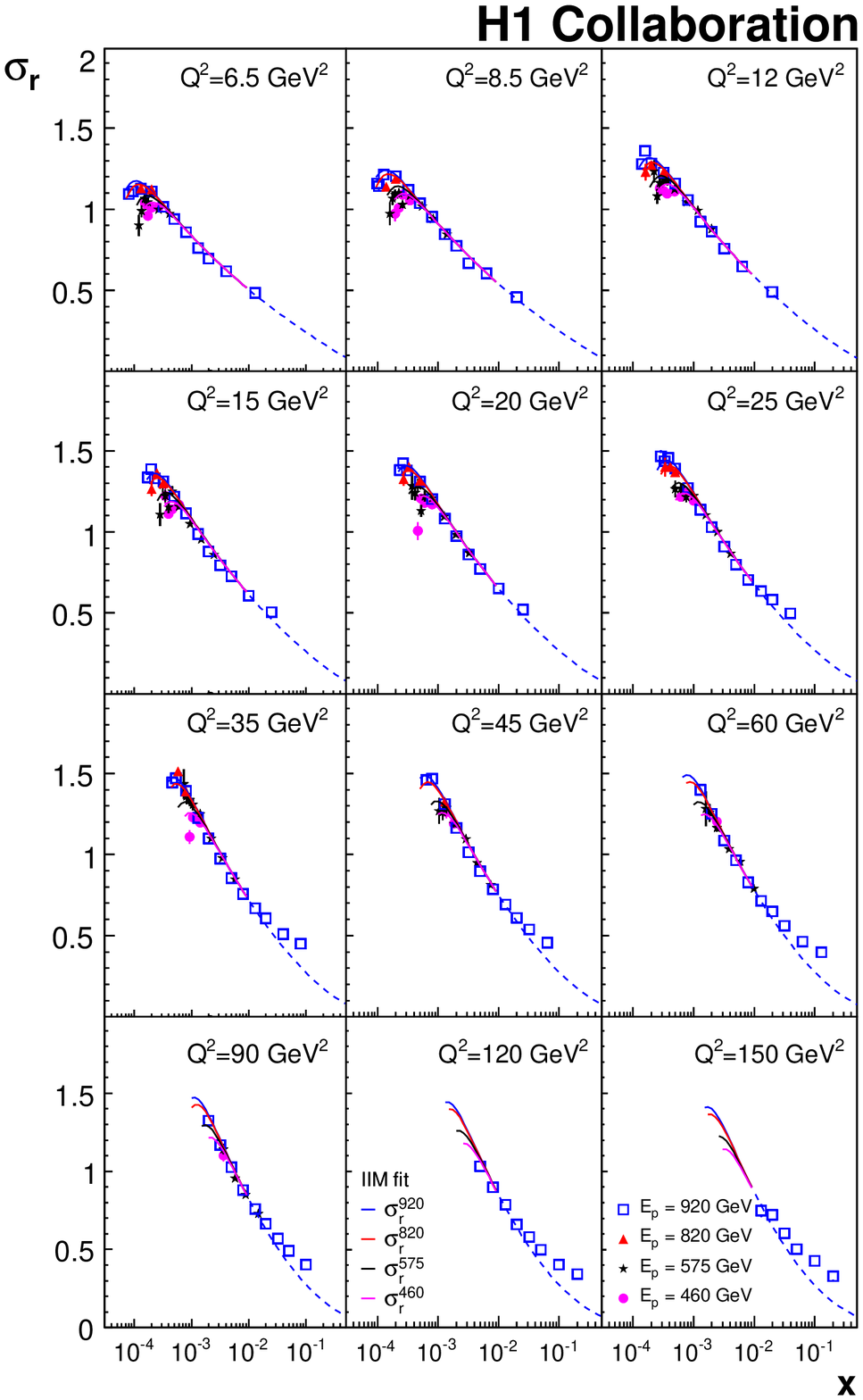,width=.9\linewidth}}
\caption{ \label{fig:iimfit2}
Reduced cross-section  data taken at different proton beam energies $E_p$
compared to \iim\ fit results for $6.5 \le Q^2\le 150$~GeV$^2$.
}
\end{figure}

\begin{figure}
\centerline{\epsfig{file=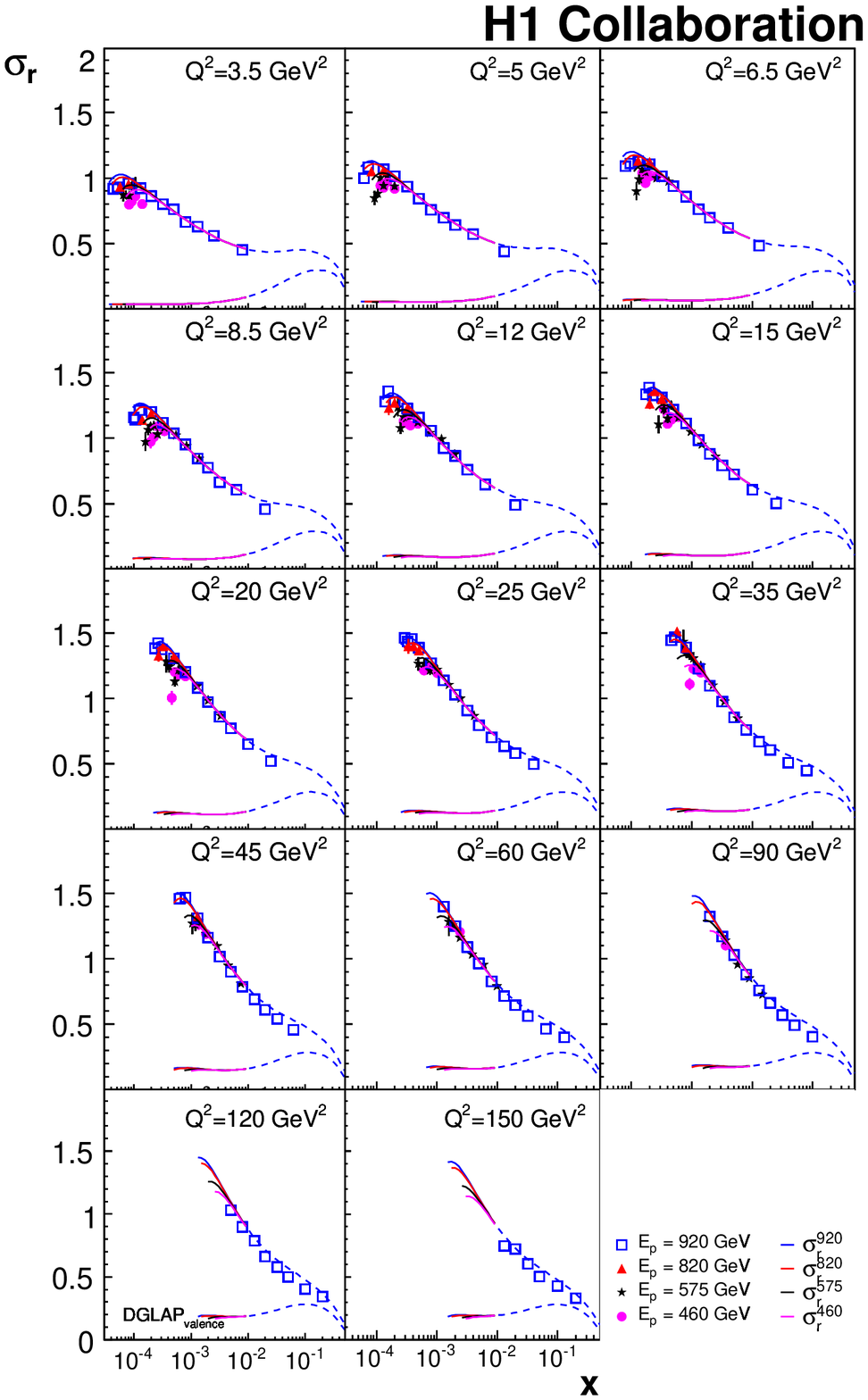,width=.9\linewidth}}
\caption{ \label{fig:iimdglap}
Reduced cross-section data taken at different proton beam energies $E_p$
compared to \iimdglap\ fit results for $3.5 \le Q^2\le 150$~GeV$^2$. The lower curves
show the contribution of DGLAP$_{\rm valence}$ calculation.
}
\end{figure}

Dipole models fits are performed using the same minimisation package  as  for the DGLAP fit. 
The values of parameters and their uncertainties are 
estimated using a Monte Carlo 
method \cite{Glazov:2010bw}\footnote{The Monte Carlo method  is preferred for the estimation of  
uncertainties compared to 
the MINUIT error estimation in order to avoid instabilities of numerical integration used in the Dipole codes.}.
The parameters of the  fits 
are given in \Tab~\ref{tab:dipole}.
The fit qualities are summarised together with the results obtained
for DGLAP fits in \Tab~\ref{tab:dipolchi}.
Among the dipole models, the \iim\ fit provides the best description of the data. It is shown in 
\Figs~\ref{fig:iimfit1} and \ref{fig:iimfit2}. 
The \bsat\ model, which includes some DGLAP evolution,
 provides a worse  fit to the data yet still with an acceptable $\chi^2/\dof$. 
The \gbw\ model, however, fails to describe the 
data. This fit agrees with the data well 
at low $Q^2$, but falls significantly below the data for $Q^2\ge 25$~GeV$^2$, where the DGLAP evolution, 
neglected in the model, plays an important role.
Fits with a DGLAP-based correction for the contribution of the valence 
quarks are  restricted to 
$Q^2\ge3.5$~GeV$^2$. In order to simplify comparisons, pure dipole model fits with the same $Q^2$ cut
are performed too. 
The addition of the valence-quark contribution allows an acceptable 
description of the data at 
high $x$, however, the overall fit quality is reduced. 
The fitted parameters of the models vary beyond their experimental uncertainties.
As an example, a comparison of the reduced cross-section data to the 
\iimdglap\ fit is given in \Fig~\ref{fig:iimdglap}.

Finally, the fits at $x<0.01$ and $Q^2\ge3.5$~GeV$^2$
are compared  in terms of $\chi^2/\dof$ for the
three dipole and two DGLAP models (\Tab~\ref{tab:dipolchi}). The best description
is obtained with the ACOT fit,  
followed by the pure dipole model
\iim\ and \bsat\ fits. The DGLAP RT fit
is of similar quality as the \iimdglap\ fit,
followed by the \bsatdglap\ fit. The \gbw\ fit
fails to describe the data in this kinematic domain.

\begin{figure}
\centerline{\epsfig{file=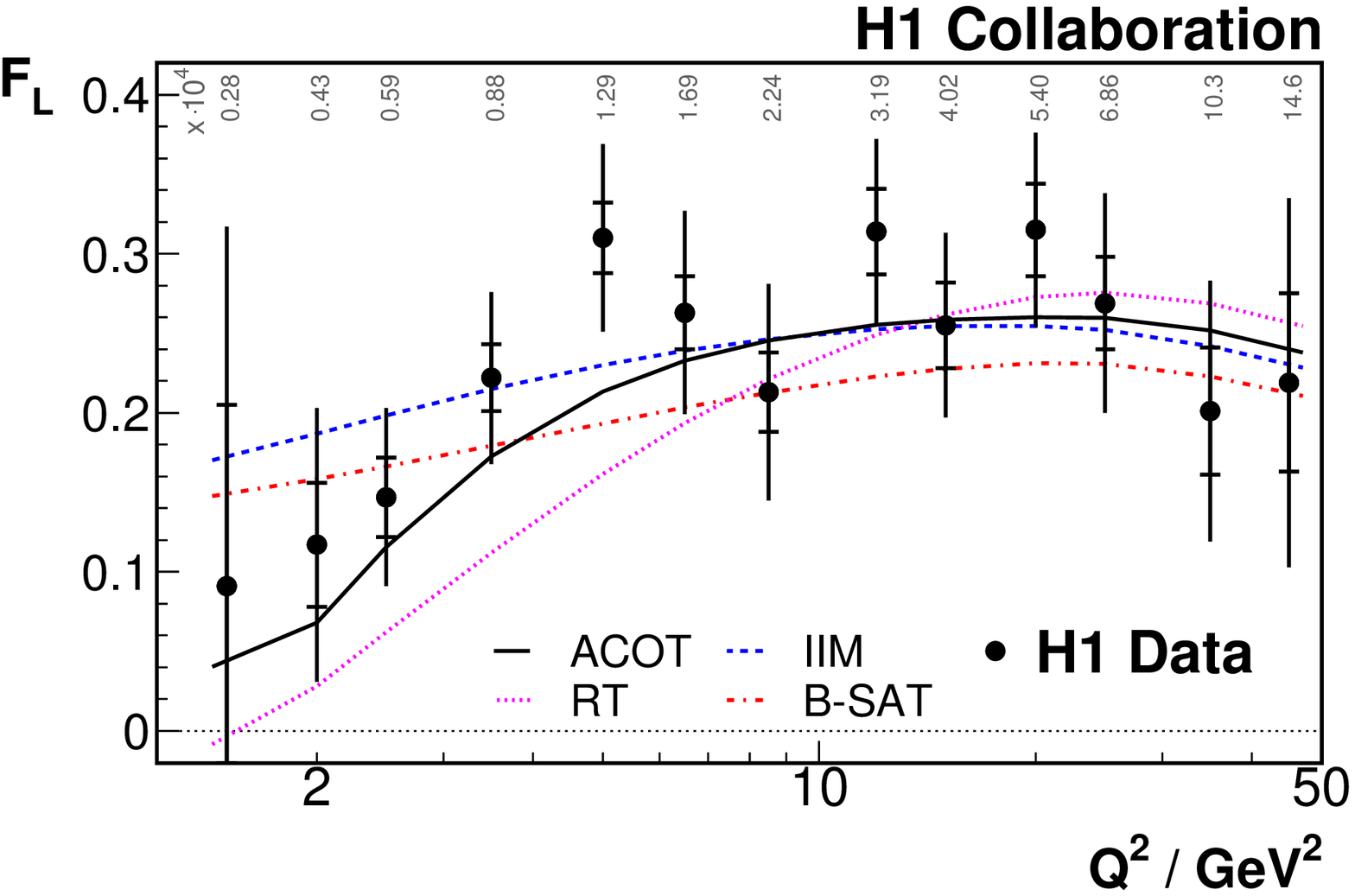,width=\linewidth}}
\caption{\label{fig:flave}The proton structure function $F_L$ 
shown as a function of $Q^2$. The average $x$ values for each $Q^2$  are indicated. 
The inner error bars
represent statistical error, the full error bars include the statistical and 
systematic uncertainties added in quadrature. The lines represent results of the  
DGLAP ACOT and RT as well as 
dipole \iim\ and \bsat\ model 
fits.
}
\end{figure}

In both the DGLAP and the dipole models the structure function $F_L$
can be calculated once the model parameters are fixed. It is thus
of interest to compare
the $F_L$ predictions of the different models with the data,
%structure function $F_L$, 
which is illustrated
in \Fig~\ref{fig:flave}. For high $Q^2>10$~GeV$^2$,
all models agree with the data and with each other  well. 
For lower $Q^2$ values, there is a 
significant difference between the predictions. The DGLAP fit in the RT scheme
predicts low values of $F_L$, while in the DGLAP fit in the ACOT scheme the decrease of $F_L$ 
occurs at lower values of $Q^2$.
%The dipole models considered vary less as a function of $Q^2$. 
The predictions of the dipol models considered here show only little variation with $Q^2$.
All predictions, except the DGLAP RT fit,  agree 
with the data well. The $Q^2$ dependence of $F_L$ is best reproduced by the DGLAP
ACOT fit.
%-- data mc comparisons -- better high x -- chi2 -- change of the parameters.
\begin{table}
\begin{center}
\begin{tabular}{l|ccc|cc}
\hline
                       & \multicolumn{5}{c}{\bf $\rm \boldsymbol{\chi^2/\dof}$ } \\
 {\bf Fit Conditions } &  {\bf \gbw }   & {\bf \iim } & {\bf \bsat } & {\bf ACOT} & {\bf RT}\\
\hline
 Nominal fit            & \gbwchsq    & \iimchsq  & \bsatchsq &
$\chacot/\ndfdef$ & $\chrt/\ndfdef$ \\
 $Q^2\ge 3.5$~GeV$^2$   & \gbwqchsq   & \iimqchsq  & \bsatqchsq\\
 DGLAP$_{\rm valence}$   & \gbwqdvchsq & \iimqdvchsq &\bsatqdvchsq & 
$\chacotqcut/\ndfdglapqcut$  & $\chrtqcut/\ndfdglapqcut$  \\
\hline
\end{tabular}
\end{center}
\tablecaption{\label{tab:dipolchi}Quality of fits in terms of $\chi^2/\dof$ 
for \gbw, \iim\ and \bsat\ dipole model as well as
ACOT and RT DGLAP models for various fit conditions described in the text. 
The $Q^2\ge 3.5$~GeV$^2$ and DGLAP$_{\rm valence}$ fits are performed in a 
kinematic phase space restricted to $Q^2\ge 3.5$~GeV$^2$ and $x<0.01$
which is valid for both dipole and DGLAP models.
}
\end{table}
% no discussion ...
% \subsection{Discussion of Fit Results}
% \input{discussion}
%\subsubsection{Framework and Parameterisations}
%\input{qcdset}
%\subsubsection{Results and Kinematic Cut Variations}
%\input{qcdresult}
%
%\subsection{Discussion}
%
\section{Summary}
\label{sec:conclusion}
A  measurement is presented of the inclusive double differential cross section for neutral current
deep inelastic $e^{\pm}p$ scattering at small Bjorken~$x$ and low 
absolute four-momentum 
transfers squared, $Q^2$. The measurement extends to high values of
inelasticity $y$. The data were collected with the H1 detector 
for
the proton beam energy of $E_p=920$~GeV, in the years $2003$ to $2006$, 
and for $E_p=575$~GeV and $E_p=460$~GeV,  in $2007$.
The integrated luminosities of the measurements are 
$103.5$\,\lunit, \lumimer\,\lunit\ and \lumiler\,\lunit\ 
for the $E_p=920$~GeV, $E_p=575$~GeV and $E_p=460$~GeV data samples, respectively.
The data at $E_p=920$~GeV significantly  
improve the accuracy of the cross-section measurements
at high $y$ when compared to the previous H1 data.
All data are combined with the HERA-I results to provide a new accurate
data sample covering 
 $0.2\le Q^2 \le 150$~GeV$^2$, $5\times10^{-6} < x < 0.15$
and $0.005<y\le 0.85$ which supersedes previous H1 measurements
of the DIS cross section and of $F_2$ in this kinematic domain.

The data at $E_p=460$~GeV and $E_p=575$~GeV, together with the measurements at
$E_p=920$~GeV
 are used to 
determine the structure function $F_L$.  This extraction applies a novel
method which takes into account the correlations of data points
due to systematic  uncertainties.  
This is the first measurement at low
$1.5\le Q^2 \le 45$~GeV$^2$ and $2.7 \times 10^{-5} < x < 2\times 10^{-3}$,
which became possible by employing a dedicated 
backward silicon tracker for the electron reconstruction. 
The data are reasonably well reproduced by the predictions based on NLO and NNLO QCD.

The measurements of $F_L$ are used to determine the ratio
 $R=F_L/(F_2-F_L)$. For $Q^2\ge 3.5$~GeV$^2$, the ratio $R$
%is found to exhibit
shows a constant behaviour with $R=\rval \pm \rerr$.

\begin{sloppypar}
The combined H1 data are subjected to phenomenological analyses. The rise
of the structure function $F_2$ towards low $x$ is examined using power-law
fits. As in  previous H1 analyses,
the power-law exponent $\lambda$ is found to be approximately constant 
for $Q^2\le 2$~GeV$^2$ but %exhibits a linear increase 
increases linearly with 
$\ln Q^2$ for higher $Q^2$ values. Closer inspection of the fits reveals,
however, a deterioration  of the fit quality for the $1\le Q^2\le10$~GeV$^2$ 
range. A parameterisation which allows for a $Q^2$ dependent $\ln x$ correction 
to a fixed power-law, for $\lambda=0.25$, provides an improved 
description of the data with the same number of parameters. This observation
suggests that the $x$ dependence of the structure function $F_2$ may
deviate from a simple power law 
at small $x$ and small $Q^2$  exhibiting a softer rise.
This confirms a QCD prediction of~\cite{DeRujula:1974rf}, according to
which the rise of
$F_2$  should be slower than any power 
of $1/x$ but faster than any power of $\ln 1/x$.
\end{sloppypar}

The data are found to be
 well described by an NLO DGLAP QCD analysis.
The ACOT
and the RT
schemes are used,  which differ in the treatment of the heavy-flavour 
and higher-order $F_L$ contributions
to the cross section. A comparison of ACOT and RT based fits to the data
reveals a
significant preference for the ACOT treatment.

The sensitivity of the DGLAP fits to low $Q^2$ and low $x$ effects
 is checked by varying the  
$Q^2_{min}$ of the data
and also by applying a saturation-model 
inspired~\cite{Caola:2009iy} selection of 
the data.
While for all variants of the cuts the fits provide a good
description of the data, the fit quality improves
as more data at low $x$
and low $Q^2$ are removed from the analysis.
This also leads to an increase of the gluon  and a decrease of the sea-quark 
densities at low $x$.

Dipole model based analyses are applied to
the  data at $x<0.01$  using three variants of them.
 The \gbw~model is unable to describe
the data at larger $Q^2$, while the \iim\ and \bsat\ models agreed with the data
generally well. The influence of valence quarks at low $x$ is investigated by 
adding their contribution as estimated from the ACOT fit. 
These models together
with the two DGLAP fits are compared to each other by
fitting the data in a common kinematic range. The DGLAP ACOT fit
provides the best description of the data, followed closely by the
dipole \iim\ and  \bsat\ models.
 All models agree well
with the $F_L$ measurement at $Q^2 > 10$~GeV$^2$. For lower $Q^2$, however,
the RT fit falls significantly below the data
while the other models describe the measured $F_L$ well.

The present measurement and the combination with all accurate
H1 DIS cross section data provide
a total cross section uncertainty of about $1$\% at low $Q^2$ and low $x$.
The structure function
$F_2(x,Q^2)$  rises towards low $x$.
The structure function $F_L(x,Q^2)$ is measured directly. Both structure
functions agree with pQCD expectations.
\section*{Acknowledgements}
\refstepcounter{pdfadd} \pdfbookmark[0]{Acknowledgements}{s:acknowledge}
%\begin{theacknowledgments}    
%
We are grateful to the HERA machine group whose outstanding
efforts have made this experiment possible.
We thank the engineers and technicians for their work in constructing
and maintaining the H1 detector, our funding agencies for
financial support, the DESY technical staff for continual assistance
and the DESY directorate for support and for the hospitality
which they extend to the non-DESY members of the collaboration.
%\end{theacknowledgments}

\bibliography{lowe.bib}  
\clearpage

\label{sec:tables}

\begin{table}
\begin{scriptsize}
\begin{center}
\begin{tabular}{cccccccccccc}
\hline
\\[-8pt]
 $\boldsymbol{Q^2}$ & $\boldsymbol{x}$ & $\boldsymbol{y}$ & $\boldsymbol{\sigma_r}$ & $\boldsymbol{\delta_{stat}}$ & $\boldsymbol{\delta_{unc}}$& $\boldsymbol{\delta_{tot}}$ & $\boldsymbol{\gamma_{\ee}}$ & $\boldsymbol{\gamma_{\thetae}}$ & $\boldsymbol{\gamma_{had}}$ & $\boldsymbol{\gamma_{noise}}$ & $\boldsymbol{\gamma_{acctag}}$ \\
{\bf GeV}$^2$ &   &     &             &  {\bf \% }         &  {\bf \% }        &  {\bf \% }   &         {\bf \% }        &    {\bf \%}           &   {\bf \%}          &   {\bf \%}      & {\bf \% }   \\
\hline
\\[-8pt]
$    2.5$ & $    2.900\times 10^{-5  } $ & $   0.85 $& $   0.828 $ &$    5.70$ &$    4.29 $& $   7.28 $ & $    1.26 $ & $   -0.15 $ & $    0.52 $ & $    0.53 $ & $    0.00 $ \\ 
 $    2.5$ & $    3.290\times 10^{-5  } $ & $   0.75 $& $   0.816 $ &$    3.99$ &$    3.27 $& $   5.36 $ & $    1.13 $ & $   -0.72 $ & $    0.35 $ & $    0.42 $ & $    0.00 $ \\ 
 $    2.5$ & $    3.790\times 10^{-5  } $ & $   0.65 $& $   0.896 $ &$    6.91$ &$    4.76 $& $   9.20 $ & $    3.70 $ & $   -0.76 $ & $   -0.05 $ & $   -0.05 $ & $    0.00 $ \\ 
 $    3.5$ & $    4.060\times 10^{-5  } $ & $   0.85 $& $   0.809 $ &$    6.13$ &$    4.17 $& $   7.57 $ & $    1.11 $ & $    0.96 $ & $    0.33 $ & $    0.35 $ & $    0.00 $ \\ 
 $    3.5$ & $    4.600\times 10^{-5  } $ & $   0.75 $& $   0.971 $ &$    3.04$ &$    2.99 $& $   4.35 $ & $    0.42 $ & $   -0.63 $ & $    0.28 $ & $    0.35 $ & $    0.00 $ \\ 
 $    3.5$ & $    5.310\times 10^{-5  } $ & $   0.65 $& $   0.887 $ &$    3.09$ &$    2.64 $& $   4.12 $ & $    0.53 $ & $   -0.17 $ & $    0.21 $ & $    0.20 $ & $    0.00 $ \\ 
 $    3.5$ & $    8.000\times 10^{-5  } $ & $   0.43 $& $   0.952 $ &$    5.62$ &$    3.66 $& $   7.28 $ & $    1.60 $ & $   -2.00 $ & $    0.31 $ & $    0.44 $ & $    1.07 $ \\ 
 $    5.0$ & $    5.800\times 10^{-5  } $ & $   0.85 $& $   0.939 $ &$    6.41$ &$    4.13 $& $   7.81 $ & $    1.46 $ & $    0.58 $ & $    0.35 $ & $    0.47 $ & $    0.00 $ \\ 
 $    5.0$ & $    6.580\times 10^{-5  } $ & $   0.75 $& $   1.002 $ &$    2.84$ &$    2.87 $& $   4.15 $ & $    0.74 $ & $   -0.39 $ & $    0.32 $ & $    0.36 $ & $    0.00 $ \\ 
 $    5.0$ & $    7.590\times 10^{-5  } $ & $   0.65 $& $   1.094 $ &$    1.71$ &$    2.29 $& $   3.08 $ & $    1.02 $ & $   -0.39 $ & $    0.24 $ & $    0.24 $ & $    0.00 $ \\ 
 $    5.0$ & $    1.300\times 10^{-4  } $ & $   0.38 $& $   1.050 $ &$    3.14$ &$    2.24 $& $   4.27 $ & $    1.40 $ & $   -1.00 $ & $    0.18 $ & $    0.26 $ & $    0.56 $ \\ 
 $    6.5$ & $    7.540\times 10^{-5  } $ & $   0.85 $& $   1.133 $ &$    6.23$ &$    4.15 $& $   7.55 $ & $    0.02 $ & $    0.89 $ & $    0.29 $ & $    0.34 $ & $    0.00 $ \\ 
 $    6.5$ & $    8.550\times 10^{-5  } $ & $   0.75 $& $   1.077 $ &$    2.83$ &$    2.82 $& $   4.25 $ & $    1.07 $ & $    0.85 $ & $    0.32 $ & $    0.34 $ & $    0.00 $ \\ 
 $    6.5$ & $    9.860\times 10^{-5  } $ & $   0.65 $& $   1.134 $ &$    2.09$ &$    2.31 $& $   3.29 $ & $    0.89 $ & $    0.45 $ & $    0.24 $ & $    0.23 $ & $    0.00 $ \\ 
 $    6.5$ & $    1.300\times 10^{-4  } $ & $   0.49 $& $   1.115 $ &$    1.51$ &$    1.86 $& $   2.55 $ & $    0.71 $ & $   -0.15 $ & $    0.17 $ & $    0.18 $ & $    0.40 $ \\ 
 $    6.5$ & $    2.000\times 10^{-4  } $ & $   0.32 $& $   1.082 $ &$    1.72$ &$    1.91 $& $   2.85 $ & $    0.85 $ & $   -0.89 $ & $    0.03 $ & $    0.05 $ & $    0.07 $ \\ 
 $    6.5$ & $    3.200\times 10^{-4  } $ & $   0.20 $& $   1.054 $ &$    3.17$ &$    2.25 $& $   4.11 $ & $    0.94 $ & $   -0.94 $ & $    0.00 $ & $    0.00 $ & $    0.16 $ \\ 
 $    8.5$ & $    9.860\times 10^{-5  } $ & $   0.85 $& $   1.146 $ &$    6.46$ &$    4.05 $& $   7.71 $ & $    1.07 $ & $   -0.04 $ & $    0.37 $ & $    0.35 $ & $    0.00 $ \\ 
 $    8.5$ & $    1.118\times 10^{-4  } $ & $   0.75 $& $   1.102 $ &$    3.31$ &$    2.82 $& $   4.46 $ & $    0.84 $ & $    0.43 $ & $    0.22 $ & $    0.22 $ & $    0.00 $ \\ 
 $    8.5$ & $    1.290\times 10^{-4  } $ & $   0.65 $& $   1.200 $ &$    1.69$ &$    2.22 $& $   3.04 $ & $    1.11 $ & $    0.38 $ & $    0.21 $ & $    0.21 $ & $    0.00 $ \\ 
 $    8.5$ & $    2.000\times 10^{-4  } $ & $   0.42 $& $   1.206 $ &$    1.24$ &$    1.80 $& $   2.29 $ & $    0.56 $ & $    0.27 $ & $    0.13 $ & $    0.17 $ & $    0.17 $ \\ 
 $    8.5$ & $    3.200\times 10^{-4  } $ & $   0.26 $& $   1.118 $ &$    1.38$ &$    1.82 $& $   2.42 $ & $    0.80 $ & $    0.03 $ & $    0.00 $ & $    0.00 $ & $    0.04 $ \\ 
 $   12.0$ & $    1.392\times 10^{-4  } $ & $   0.85 $& $   1.238 $ &$    5.34$ &$    3.97 $& $   6.80 $ & $    1.36 $ & $    0.10 $ & $    0.21 $ & $    0.25 $ & $    0.00 $ \\ 
 $   12.0$ & $    1.578\times 10^{-4  } $ & $   0.75 $& $   1.313 $ &$    3.41$ &$    2.82 $& $   4.47 $ & $    0.45 $ & $    0.29 $ & $    0.23 $ & $    0.26 $ & $    0.00 $ \\ 
 $   12.0$ & $    1.821\times 10^{-4  } $ & $   0.65 $& $   1.244 $ &$    2.51$ &$    2.30 $& $   3.54 $ & $    0.73 $ & $    0.54 $ & $    0.23 $ & $    0.20 $ & $    0.00 $ \\ 
 $   12.0$ & $    2.000\times 10^{-4  } $ & $   0.59 $& $   1.258 $ &$    1.67$ &$    2.20 $& $   2.98 $ & $    0.94 $ & $    0.52 $ & $    0.17 $ & $    0.19 $ & $    0.00 $ \\ 
 \hline
\end{tabular}
\end{center}
\end{scriptsize}
\tablecaption{\label{tab:bst920}
Reduced cross section $\sigma_{r}$, as measured with the
    {\lqbst} data sample.
The uncertainties are
    quoted in \% relative to $\sigma_r$.
$\delta_{\rm stat}$ is the statistical uncertainty,
    $\delta_{\rm unc}$ represents the uncorrelated systematic
    uncertainty and
  $\delta_{\rm tot}$ is the total uncertainty
    determined as the quadratic sum of systematic and statistical
    uncertainties. 
$\gamma_{E_e^{\prime}}$, $\gamma_{\theta_e}$,
    $\gamma_{had}$, $\gamma_{\rm noise}$ 
 and $\gamma_{\rm acctag}$
 are the bin-to-bin
    correlated systematic uncertainties in the cross section
    measurement due to uncertainties in the SpaCal electromagnetic
    energy scale, electron scattering angle, 
    calorimeter hadronic
    energy scale, LAr calorimeter noise and
electron tagger acceptance, respectively. 
   The global normalisation uncertainty of
    $3$\%  is not included in $\delta_{\rm tot}$.
}
\end{table}

\begin{table}
\begin{scriptsize}
\begin{center}
\begin{tabular}{cccccccccccc}
\hline
\\[-8pt]
 $\boldsymbol{Q^2}$ & $\boldsymbol{x}$ & $\boldsymbol{y}$ & $\boldsymbol{\sigma_r}$ & $\boldsymbol{\delta_{stat}}$ & $\boldsymbol{\delta_{unc}}$& $\boldsymbol{\delta_{tot}}$ & $\boldsymbol{\gamma_{\ee}}$ & $\boldsymbol{\gamma_{\thetae}}$ & $\boldsymbol{\gamma_{had}}$ & $\boldsymbol{\gamma_{noise}}$ & $\boldsymbol{\gamma_{acctag}}$ \\
{\bf GeV}$^2$ &   &     &             &  {\bf \% }         &  {\bf \% }        &  {\bf \% }   &         {\bf \% }        &    {\bf \%}           &   {\bf \%}          &   {\bf \%}      & {\bf \% }   \\
\hline
\\[-8pt]
$    8.5$ & $    9.860\times 10^{-5  } $ & $   0.85 $& $   1.172 $ &$    2.22$ &$    3.75 $& $   4.44 $ & $    0.76 $ & $    0.09 $ & $    0.29 $ & $    0.33 $ & $    0.00 $ \\ 
 $   12.0$ & $    1.392\times 10^{-4  } $ & $   0.85 $& $   1.304 $ &$    1.47$ &$    3.71 $& $   4.09 $ & $    0.77 $ & $    0.06 $ & $    0.28 $ & $    0.29 $ & $    0.00 $ \\ 
 $   12.0$ & $    1.578\times 10^{-4  } $ & $   0.75 $& $   1.394 $ &$    1.72$ &$    2.71 $& $   3.38 $ & $    0.98 $ & $    0.05 $ & $    0.28 $ & $    0.25 $ & $    0.00 $ \\ 
 $   15.0$ & $    1.741\times 10^{-4  } $ & $   0.85 $& $   1.349 $ &$    1.37$ &$    3.71 $& $   4.06 $ & $    0.72 $ & $    0.34 $ & $    0.31 $ & $    0.30 $ & $    0.00 $ \\ 
 $   15.0$ & $    1.973\times 10^{-4  } $ & $   0.75 $& $   1.400 $ &$    0.93$ &$    2.64 $& $   2.83 $ & $    0.26 $ & $    0.10 $ & $    0.25 $ & $    0.25 $ & $    0.00 $ \\ 
 $   15.0$ & $    2.276\times 10^{-4  } $ & $   0.65 $& $   1.342 $ &$    1.80$ &$    2.33 $& $   2.99 $ & $    0.37 $ & $   -0.12 $ & $    0.26 $ & $    0.20 $ & $    0.00 $ \\ 
 $   20.0$ & $    2.321\times 10^{-4  } $ & $   0.85 $& $   1.396 $ &$    1.52$ &$    3.73 $& $   4.12 $ & $    0.71 $ & $    0.32 $ & $    0.28 $ & $    0.25 $ & $    0.00 $ \\ 
 $   20.0$ & $    2.630\times 10^{-4  } $ & $   0.75 $& $   1.439 $ &$    0.83$ &$    2.63 $& $   2.92 $ & $    0.73 $ & $    0.50 $ & $    0.25 $ & $    0.21 $ & $    0.00 $ \\ 
 $   20.0$ & $    3.035\times 10^{-4  } $ & $   0.65 $& $   1.391 $ &$    0.76$ &$    2.16 $& $   2.35 $ & $    0.41 $ & $    0.13 $ & $    0.24 $ & $    0.19 $ & $    0.00 $ \\ 
 $   25.0$ & $    2.901\times 10^{-4  } $ & $   0.85 $& $   1.482 $ &$    2.46$ &$    3.86 $& $   4.64 $ & $    0.65 $ & $   -0.06 $ & $    0.32 $ & $    0.23 $ & $    0.00 $ \\ 
 $   25.0$ & $    3.288\times 10^{-4  } $ & $   0.75 $& $   1.450 $ &$    0.88$ &$    2.64 $& $   2.84 $ & $    0.33 $ & $    0.31 $ & $    0.24 $ & $    0.20 $ & $    0.00 $ \\ 
 $   25.0$ & $    3.794\times 10^{-4  } $ & $   0.65 $& $   1.469 $ &$    0.73$ &$    2.16 $& $   2.40 $ & $    0.55 $ & $    0.43 $ & $    0.24 $ & $    0.19 $ & $    0.00 $ \\ 
 $   25.0$ & $    5.000\times 10^{-4  } $ & $   0.49 $& $   1.403 $ &$    0.69$ &$    1.77 $& $   2.12 $ & $    0.74 $ & $    0.09 $ & $    0.22 $ & $    0.19 $ & $    0.51 $ \\ 
 $   25.0$ & $    8.000\times 10^{-4  } $ & $   0.31 $& $   1.281 $ &$    1.10$ &$    1.85 $& $   2.28 $ & $    0.70 $ & $    0.19 $ & $    0.10 $ & $    0.12 $ & $    0.08 $ \\ 
 $   35.0$ & $    4.603\times 10^{-4  } $ & $   0.75 $& $   1.457 $ &$    1.23$ &$    2.68 $& $   3.01 $ & $    0.33 $ & $    0.36 $ & $    0.29 $ & $    0.21 $ & $    0.00 $ \\ 
 $   35.0$ & $    5.311\times 10^{-4  } $ & $   0.65 $& $   1.483 $ &$    0.66$ &$    2.15 $& $   2.34 $ & $    0.31 $ & $    0.43 $ & $    0.26 $ & $    0.19 $ & $    0.00 $ \\ 
 $   35.0$ & $    8.000\times 10^{-4  } $ & $   0.43 $& $   1.404 $ &$    0.48$ &$    1.75 $& $   2.02 $ & $    0.78 $ & $    0.29 $ & $    0.18 $ & $    0.17 $ & $    0.16 $ \\ 
 $   35.0$ & $    1.300\times 10^{-3  } $ & $   0.27 $& $   1.236 $ &$    0.63$ &$    1.77 $& $   1.99 $ & $    0.64 $ & $    0.01 $ & $    0.01 $ & $    0.02 $ & $    0.01 $ \\ 
 $   35.0$ & $    2.000\times 10^{-3  } $ & $   0.17 $& $   1.112 $ &$    1.42$ &$    1.93 $& $   2.73 $ & $    1.25 $ & $   -0.35 $ & $    0.00 $ & $    0.00 $ & $    0.00 $ \\ 
 $   45.0$ & $    6.341\times 10^{-4  } $ & $   0.70 $& $   1.470 $ &$    1.00$ &$    2.21 $& $   2.58 $ & $    0.79 $ & $    0.28 $ & $    0.28 $ & $    0.18 $ & $    0.00 $ \\ 
 $   45.0$ & $    8.000\times 10^{-4  } $ & $   0.55 $& $   1.488 $ &$    0.69$ &$    2.16 $& $   2.47 $ & $    0.85 $ & $    0.41 $ & $    0.22 $ & $    0.19 $ & $    0.00 $ \\ 
 $   45.0$ & $    1.300\times 10^{-3  } $ & $   0.34 $& $   1.316 $ &$    0.45$ &$    1.75 $& $   1.94 $ & $    0.50 $ & $    0.45 $ & $    0.08 $ & $    0.09 $ & $    0.04 $ \\ 
 $   45.0$ & $    2.000\times 10^{-3  } $ & $   0.22 $& $   1.182 $ &$    0.52$ &$    1.76 $& $   1.94 $ & $    0.54 $ & $    0.30 $ & $    0.00 $ & $    0.00 $ & $    0.03 $ \\ 
 $   60.0$ & $    1.302\times 10^{-3  } $ & $   0.45 $& $   1.419 $ &$    0.61$ &$    1.77 $& $   1.97 $ & $    0.42 $ & $    0.35 $ & $    0.18 $ & $    0.14 $ & $    0.00 $ \\ 
 $   60.0$ & $    2.000\times 10^{-3  } $ & $   0.30 $& $   1.248 $ &$    0.50$ &$    1.76 $& $   1.95 $ & $    0.40 $ & $    0.56 $ & $    0.02 $ & $    0.03 $ & $    0.02 $ \\ 
 $   60.0$ & $    3.200\times 10^{-3  } $ & $   0.18 $& $   1.102 $ &$    0.71$ &$    1.78 $& $   2.12 $ & $    0.83 $ & $    0.33 $ & $    0.00 $ & $    0.00 $ & $    0.02 $ \\ 
 $   90.0$ & $    2.004\times 10^{-3  } $ & $   0.44 $& $   1.354 $ &$    0.79$ &$    1.80 $& $   2.09 $ & $    0.56 $ & $    0.42 $ & $    0.09 $ & $    0.08 $ & $    0.00 $ \\ 
 $   90.0$ & $    3.200\times 10^{-3  } $ & $   0.28 $& $   1.183 $ &$    0.60$ &$    1.77 $& $   1.95 $ & $    0.44 $ & $    0.32 $ & $    0.00 $ & $    0.00 $ & $    0.00 $ \\ 
 $   90.0$ & $    5.000\times 10^{-3  } $ & $   0.18 $& $   1.018 $ &$    1.50$ &$    1.96 $& $   2.50 $ & $    0.29 $ & $    0.33 $ & $    0.00 $ & $    0.00 $ & $    0.00 $ \\ 
 \hline
%\hline
\end{tabular}
\end{center}
\end{scriptsize}
\tablecaption{\label{tab:cjc920}
Reduced cross section $\sigma_{r}$, as measured with the
    {\hqcjc} data sample. 
Description of the columns is given in the caption of \Tab~\ref{tab:bst920}.
}
\end{table}

\begin{table}
\begin{scriptsize}
\begin{center}
\begin{tabular}{ccccccccccccc}
\hline
\\[-8pt]
 $\boldsymbol{Q^2}$ & $\boldsymbol{x}$ & $\boldsymbol{y}$ & $\boldsymbol{\sigma_r}$ & $\boldsymbol{\delta_{stat}}$ & $\boldsymbol{\delta_{unc}}$& $\boldsymbol{\delta_{tot}}$ & $\boldsymbol{\gamma_{\ee}}$ & $\boldsymbol{\gamma_{\thetae}}$ & $\boldsymbol{\gamma_{had}}$ & $\boldsymbol{\gamma_{noise}}$ & $\boldsymbol{\gamma_{\rm asym}}$
& $\boldsymbol{\gamma_{acctag}}$ \\
{\bf GeV}$^2$ &   &     &             &  {\bf \% }         &  {\bf \% }        &  {\bf \% }   &         {\bf \% }        &    {\bf \%}           &   {\bf \%}          &   {\bf \%}      & {\bf \% }  & {\bf \%}  \\
\hline
\\[-8pt]
$    1.5$ & $    3.480\times 10^{-5  } $ & $  0.850 $& $   0.520 $ &$    8.10$ &$    4.96 $& $  10.16 $ & $    0.69 $ & $    0.08 $ & $    0.45 $ & $    0.58 $ & $    3.47 $ & $    0.00 $ \\ 
 $    2.0$ & $    4.640\times 10^{-5  } $ & $  0.850 $& $   0.704 $ &$    4.57$ &$    4.31 $& $   6.89 $ & $    1.10 $ & $   -0.81 $ & $    0.49 $ & $    0.59 $ & $    2.34 $ & $    0.00 $ \\ 
 $    2.0$ & $    5.260\times 10^{-5  } $ & $  0.750 $& $   0.717 $ &$    4.59$ &$    3.93 $& $   6.38 $ & $    1.64 $ & $   -0.64 $ & $    0.32 $ & $    0.45 $ & $    0.86 $ & $    0.00 $ \\ 
 $    2.5$ & $    5.800\times 10^{-5  } $ & $  0.850 $& $   0.777 $ &$    4.18$ &$    4.16 $& $   6.73 $ & $    2.01 $ & $   -0.31 $ & $    0.60 $ & $    0.64 $ & $    2.36 $ & $    0.00 $ \\ 
 $    2.5$ & $    6.580\times 10^{-5  } $ & $  0.750 $& $   0.768 $ &$    2.64$ &$    3.03 $& $   4.24 $ & $    0.91 $ & $    0.42 $ & $    0.42 $ & $    0.37 $ & $    0.74 $ & $    0.00 $ \\ 
 $    2.5$ & $    7.590\times 10^{-5  } $ & $  0.650 $& $   0.711 $ &$    4.32$ &$    3.45 $& $   5.82 $ & $    1.33 $ & $   -1.12 $ & $    0.23 $ & $    0.33 $ & $    0.40 $ & $    0.00 $ \\ 
 $    3.5$ & $    8.120\times 10^{-5  } $ & $  0.850 $& $   0.794 $ &$    4.21$ &$    4.05 $& $   6.41 $ & $    0.82 $ & $    0.62 $ & $    0.38 $ & $    0.57 $ & $    2.32 $ & $    0.00 $ \\ 
 $    3.5$ & $    9.210\times 10^{-5  } $ & $  0.750 $& $   0.820 $ &$    2.22$ &$    2.86 $& $   4.09 $ & $    1.71 $ & $    0.30 $ & $    0.31 $ & $    0.38 $ & $    0.63 $ & $    0.00 $ \\ 
 $    3.5$ & $    1.062\times 10^{-4  } $ & $  0.650 $& $   0.857 $ &$    2.03$ &$    2.47 $& $   3.69 $ & $    1.73 $ & $   -0.43 $ & $    0.32 $ & $    0.32 $ & $    0.26 $ & $    0.00 $ \\ 
 $    3.5$ & $    1.409\times 10^{-4  } $ & $  0.490 $& $   0.797 $ &$    2.55$ &$    2.45 $& $   4.31 $ & $    1.32 $ & $   -1.66 $ & $    0.24 $ & $    0.25 $ & $    0.00 $ & $    1.19 $ \\ 
 $    5.0$ & $    1.160\times 10^{-4  } $ & $  0.850 $& $   0.939 $ &$    4.19$ &$    4.01 $& $   6.60 $ & $    1.92 $ & $    0.46 $ & $    0.45 $ & $    0.51 $ & $    2.37 $ & $    0.00 $ \\ 
 $    5.0$ & $    1.315\times 10^{-4  } $ & $  0.750 $& $   0.924 $ &$    2.07$ &$    2.80 $& $   3.82 $ & $    1.28 $ & $    0.58 $ & $    0.34 $ & $    0.37 $ & $    0.54 $ & $    0.00 $ \\ 
 $    5.0$ & $    1.517\times 10^{-4  } $ & $  0.650 $& $   0.966 $ &$    1.69$ &$    2.33 $& $   3.10 $ & $    1.02 $ & $   -0.25 $ & $    0.25 $ & $    0.30 $ & $    0.21 $ & $    0.00 $ \\ 
 $    5.0$ & $    2.013\times 10^{-4  } $ & $  0.490 $& $   0.914 $ &$    1.16$ &$    1.87 $& $   2.48 $ & $    0.66 $ & $   -0.59 $ & $    0.19 $ & $    0.20 $ & $    0.00 $ & $    0.65 $ \\ 
 $    6.5$ & $    1.509\times 10^{-4  } $ & $  0.850 $& $   1.034 $ &$    4.26$ &$    4.02 $& $   6.55 $ & $    0.94 $ & $    0.88 $ & $    0.41 $ & $    0.50 $ & $    2.57 $ & $    0.00 $ \\ 
 $    6.5$ & $    1.710\times 10^{-4  } $ & $  0.750 $& $   0.958 $ &$    2.15$ &$    2.80 $& $   3.79 $ & $    0.94 $ & $    0.65 $ & $    0.39 $ & $    0.43 $ & $    0.51 $ & $    0.00 $ \\ 
 $    6.5$ & $    1.973\times 10^{-4  } $ & $  0.650 $& $   1.007 $ &$    1.60$ &$    2.30 $& $   2.92 $ & $    0.64 $ & $    0.39 $ & $    0.23 $ & $    0.25 $ & $    0.17 $ & $    0.00 $ \\ 
 $    6.5$ & $    2.617\times 10^{-4  } $ & $  0.490 $& $   1.008 $ &$    0.90$ &$    1.82 $& $   2.26 $ & $    0.83 $ & $    0.20 $ & $    0.20 $ & $    0.22 $ & $    0.00 $ & $    0.40 $ \\ 
 $    6.5$ & $    4.136\times 10^{-4  } $ & $  0.310 $& $   0.962 $ &$    1.14$ &$    1.87 $& $   2.44 $ & $    1.00 $ & $   -0.39 $ & $    0.00 $ & $    0.01 $ & $    0.00 $ & $    0.00 $ \\ 
 $    8.5$ & $    1.973\times 10^{-4  } $ & $  0.850 $& $   0.969 $ &$    4.68$ &$    4.01 $& $   7.05 $ & $    1.28 $ & $    0.64 $ & $    0.40 $ & $    0.49 $ & $    3.05 $ & $    0.00 $ \\ 
 $    8.5$ & $    2.236\times 10^{-4  } $ & $  0.750 $& $   1.008 $ &$    2.38$ &$    2.84 $& $   3.95 $ & $    0.92 $ & $    0.79 $ & $    0.21 $ & $    0.26 $ & $    0.59 $ & $    0.00 $ \\ 
 $    8.5$ & $    2.580\times 10^{-4  } $ & $  0.650 $& $   1.087 $ &$    1.66$ &$    2.32 $& $   3.30 $ & $    1.40 $ & $    0.80 $ & $    0.22 $ & $    0.23 $ & $    0.15 $ & $    0.00 $ \\ 
 $    8.5$ & $    3.422\times 10^{-4  } $ & $  0.490 $& $   1.051 $ &$    0.85$ &$    1.81 $& $   2.25 $ & $    0.92 $ & $    0.18 $ & $    0.19 $ & $    0.21 $ & $    0.00 $ & $    0.31 $ \\ 
 $    8.5$ & $    5.409\times 10^{-4  } $ & $  0.310 $& $   1.016 $ &$    0.91$ &$    1.82 $& $   2.10 $ & $    0.44 $ & $    0.23 $ & $    0.01 $ & $    0.01 $ & $    0.00 $ & $    0.02 $ \\ 
 $    8.5$ & $    8.384\times 10^{-4  } $ & $  0.200 $& $   0.939 $ &$    1.01$ &$    1.84 $& $   2.29 $ & $    0.90 $ & $    0.14 $ & $    0.00 $ & $    0.00 $ & $    0.00 $ & $    0.00 $ \\ 
 $    8.5$ & $    1.397\times 10^{-3  } $ & $  0.120 $& $   0.857 $ &$    1.21$ &$    1.89 $& $   2.55 $ & $    1.11 $ & $   -0.48 $ & $    0.00 $ & $    0.00 $ & $    0.00 $ & $    0.00 $ \\ 
 $   12.0$ & $    2.785\times 10^{-4  } $ & $  0.850 $& $   1.127 $ &$    3.90$ &$    3.98 $& $   6.07 $ & $    0.97 $ & $    0.19 $ & $    0.34 $ & $    0.41 $ & $    2.13 $ & $    0.00 $ \\ 
 $   12.0$ & $    3.156\times 10^{-4  } $ & $  0.750 $& $   1.110 $ &$    2.38$ &$    2.85 $& $   3.92 $ & $    1.00 $ & $    0.10 $ & $    0.34 $ & $    0.33 $ & $    0.59 $ & $    0.00 $ \\ 
 $   12.0$ & $    3.642\times 10^{-4  } $ & $  0.650 $& $   1.095 $ &$    1.86$ &$    2.36 $& $   3.15 $ & $    0.58 $ & $    0.67 $ & $    0.22 $ & $    0.22 $ & $    0.18 $ & $    0.00 $ \\ 
 $   12.0$ & $    4.831\times 10^{-4  } $ & $  0.490 $& $   1.108 $ &$    0.89$ &$    1.81 $& $   2.25 $ & $    0.58 $ & $    0.76 $ & $    0.15 $ & $    0.16 $ & $    0.00 $ & $    0.17 $ \\ 
 $   12.0$ & $    7.636\times 10^{-4  } $ & $  0.310 $& $   1.028 $ &$    0.87$ &$    1.81 $& $   2.22 $ & $    0.79 $ & $    0.50 $ & $    0.01 $ & $    0.01 $ & $    0.00 $ & $    0.01 $ \\ 
 $   12.0$ & $    1.184\times 10^{-3  } $ & $  0.200 $& $   0.972 $ &$    0.93$ &$    1.83 $& $   2.22 $ & $    0.85 $ & $    0.06 $ & $    0.00 $ & $    0.00 $ & $    0.00 $ & $    0.00 $ \\ 
 $   12.0$ & $    1.973\times 10^{-3  } $ & $  0.120 $& $   0.880 $ &$    1.07$ &$    1.86 $& $   2.57 $ & $    1.37 $ & $    0.34 $ & $    0.00 $ & $    0.00 $ & $    0.00 $ & $    0.00 $ \\ 
 $   15.0$ & $    3.481\times 10^{-4  } $ & $  0.850 $& $   1.218 $ &$    3.90$ &$    4.06 $& $   5.91 $ & $    0.57 $ & $    0.33 $ & $    0.29 $ & $    0.30 $ & $    1.60 $ & $    0.00 $ \\ 
 $   15.0$ & $    3.945\times 10^{-4  } $ & $  0.750 $& $   1.109 $ &$    2.33$ &$    2.84 $& $   3.77 $ & $    0.30 $ & $    0.36 $ & $    0.30 $ & $    0.31 $ & $    0.53 $ & $    0.00 $ \\ 
 $   15.0$ & $    4.552\times 10^{-4  } $ & $  0.650 $& $   1.138 $ &$    2.05$ &$    2.40 $& $   3.44 $ & $    1.06 $ & $    0.78 $ & $    0.25 $ & $    0.24 $ & $    0.24 $ & $    0.00 $ \\ 
 $   15.0$ & $    6.039\times 10^{-4  } $ & $  0.490 $& $   1.161 $ &$    0.96$ &$    1.83 $& $   2.31 $ & $    0.87 $ & $    0.45 $ & $    0.20 $ & $    0.18 $ & $    0.00 $ & $    0.13 $ \\ 
 $   15.0$ & $    9.545\times 10^{-4  } $ & $  0.310 $& $   1.049 $ &$    0.91$ &$    1.82 $& $   2.21 $ & $    0.52 $ & $    0.68 $ & $    0.00 $ & $    0.01 $ & $    0.00 $ & $    0.04 $ \\ 
 $   15.0$ & $    1.479\times 10^{-3  } $ & $  0.200 $& $   0.939 $ &$    0.95$ &$    1.83 $& $   2.28 $ & $    0.81 $ & $    0.55 $ & $    0.00 $ & $    0.00 $ & $    0.00 $ & $    0.00 $ \\ 
 $   15.0$ & $    2.466\times 10^{-3  } $ & $  0.120 $& $   0.859 $ &$    1.06$ &$    1.85 $& $   2.68 $ & $    1.61 $ & $    0.17 $ & $    0.00 $ & $    0.00 $ & $    0.00 $ & $    0.00 $ \\ 
 \hline
%\hline
\end{tabular}
\end{center}
\end{scriptsize}
\tablecaption{\label{tab:ler1}
Reduced cross section $\sigma_{r}$, as measured with the
    {\ler} data sample. 
The uncertainties are
    quoted in \% relative to $\sigma_r$.
$\delta_{\rm stat}$ is the statistical uncertainty.
    $\delta_{\rm unc}$ represents the uncorrelated systematic
    uncertainty.  
  $\delta_{\rm tot}$ is the total uncertainty
    determined as the quadratic sum of systematic and statistical
    uncertainties. 
$\gamma_{E_e^{\prime}}$, $\gamma_{\theta_e}$,
    $\gamma_{had}$, $\gamma_{\rm noise}$, 
$\gamma_{asym}$ 
 and $\gamma_{\rm acctag}$ are the bin-to-bin
    correlated systematic uncertainties in the cross section
    measurement due to uncertainties in the 
SpaCal electromagnetic
    energy scale, electron scattering angle, 
    calorimeter hadronic
    energy scale, LAr calorimeter noise, background charge
   asymmetry and electron tagger acceptance, respectively. 
   The global normalisation uncertainty of
    $4$\%  is not included in $\delta_{\rm tot}$.
}
\end{table}

\begin{table}
\begin{scriptsize}
\begin{center}
\begin{tabular}{ccccccccccccc}
%\hline
\hline
\\[-8pt]
 $\boldsymbol{Q^2}$ & $\boldsymbol{x}$ & $\boldsymbol{y}$ & $\boldsymbol{\sigma_r}$ & $\boldsymbol{\delta_{stat}}$ & $\boldsymbol{\delta_{unc}}$& $\boldsymbol{\delta_{tot}}$ & $\boldsymbol{\gamma_{\ee}}$ & $\boldsymbol{\gamma_{\thetae}}$ & $\boldsymbol{\gamma_{had}}$ & $\boldsymbol{\gamma_{noise}}$ & $\boldsymbol{\gamma_{\rm asym}}$
& $\boldsymbol{\gamma_{acctag}}$ \\
{\bf GeV}$^2$ &   &     &             &  {\bf \% }         &  {\bf \% }        &  {\bf \% }   &         {\bf \% }        &    {\bf \%}           &   {\bf \%}          &   {\bf \%}      & {\bf \% }  & {\bf \%}  \\
\hline
\\[-8pt]
$   20.0$ & $    4.642\times 10^{-4  } $ & $  0.850 $& $   1.003 $ &$    5.34$ &$    4.21 $& $   7.12 $ & $    1.20 $ & $    0.59 $ & $    0.46 $ & $    0.46 $ & $    1.47 $ & $    0.00 $ \\ 
 $   20.0$ & $    5.261\times 10^{-4  } $ & $  0.750 $& $   1.200 $ &$    2.27$ &$    2.87 $& $   3.80 $ & $    0.88 $ & $    0.02 $ & $    0.28 $ & $    0.26 $ & $    0.36 $ & $    0.00 $ \\ 
 $   20.0$ & $    6.070\times 10^{-4  } $ & $  0.650 $& $   1.177 $ &$    1.97$ &$    2.39 $& $   3.15 $ & $    0.22 $ & $    0.36 $ & $    0.23 $ & $    0.20 $ & $    0.18 $ & $    0.00 $ \\ 
 $   20.0$ & $    8.052\times 10^{-4  } $ & $  0.490 $& $   1.166 $ &$    1.06$ &$    1.85 $& $   2.45 $ & $    0.93 $ & $    0.69 $ & $    0.13 $ & $    0.14 $ & $    0.00 $ & $    0.25 $ \\ 
 $   20.0$ & $    1.273\times 10^{-3  } $ & $  0.310 $& $   1.089 $ &$    1.00$ &$    1.84 $& $   2.31 $ & $    0.72 $ & $    0.65 $ & $    0.00 $ & $    0.00 $ & $    0.00 $ & $    0.01 $ \\ 
 $   20.0$ & $    1.973\times 10^{-3  } $ & $  0.200 $& $   0.987 $ &$    1.02$ &$    1.85 $& $   2.25 $ & $    0.56 $ & $    0.56 $ & $    0.00 $ & $    0.00 $ & $    0.00 $ & $    0.03 $ \\ 
 $   20.0$ & $    3.288\times 10^{-3  } $ & $  0.120 $& $   0.875 $ &$    1.14$ &$    1.87 $& $   2.49 $ & $    0.88 $ & $    0.78 $ & $    0.00 $ & $    0.00 $ & $    0.00 $ & $    0.00 $ \\ 
 $   25.0$ & $    6.165\times 10^{-4  } $ & $  0.800 $& $   1.207 $ &$    2.42$ &$    2.94 $& $   3.96 $ & $    0.23 $ & $    0.92 $ & $    0.33 $ & $    0.23 $ & $    0.36 $ & $    0.00 $ \\ 
 $   25.0$ & $    7.587\times 10^{-4  } $ & $  0.650 $& $   1.237 $ &$    2.00$ &$    2.43 $& $   3.24 $ & $   -0.17 $ & $    0.70 $ & $    0.19 $ & $    0.15 $ & $    0.11 $ & $    0.00 $ \\ 
 $   25.0$ & $    1.007\times 10^{-3  } $ & $  0.490 $& $   1.190 $ &$    1.10$ &$    1.86 $& $   2.42 $ & $    0.84 $ & $    0.61 $ & $    0.19 $ & $    0.18 $ & $    0.00 $ & $    0.26 $ \\ 
 $   25.0$ & $    1.591\times 10^{-3  } $ & $  0.310 $& $   1.079 $ &$    1.12$ &$    1.87 $& $   2.39 $ & $    0.74 $ & $    0.63 $ & $    0.00 $ & $    0.01 $ & $    0.00 $ & $    0.00 $ \\ 
 $   25.0$ & $    2.466\times 10^{-3  } $ & $  0.200 $& $   0.989 $ &$    1.15$ &$    1.88 $& $   2.47 $ & $    0.64 $ & $    0.91 $ & $    0.00 $ & $    0.00 $ & $    0.00 $ & $    0.02 $ \\ 
 $   25.0$ & $    4.110\times 10^{-3  } $ & $  0.120 $& $   0.854 $ &$    1.30$ &$    1.91 $& $   2.60 $ & $    0.75 $ & $    0.92 $ & $    0.00 $ & $    0.00 $ & $    0.00 $ & $    0.00 $ \\ 
 $   35.0$ & $    9.206\times 10^{-4  } $ & $  0.750 $& $   1.106 $ &$    3.84$ &$    3.32 $& $   5.32 $ & $    1.47 $ & $   -0.25 $ & $    0.46 $ & $    0.28 $ & $    0.19 $ & $    0.00 $ \\ 
 $   35.0$ & $    1.062\times 10^{-3  } $ & $  0.650 $& $   1.225 $ &$    2.23$ &$    2.50 $& $   3.57 $ & $    0.93 $ & $    0.77 $ & $    0.19 $ & $    0.13 $ & $    0.07 $ & $    0.00 $ \\ 
 $   35.0$ & $    1.409\times 10^{-3  } $ & $  0.490 $& $   1.195 $ &$    1.14$ &$    1.87 $& $   2.37 $ & $    0.60 $ & $    0.59 $ & $    0.20 $ & $    0.16 $ & $    0.00 $ & $    0.20 $ \\ 
 $   35.0$ & $    2.227\times 10^{-3  } $ & $  0.310 $& $   1.085 $ &$    1.16$ &$    1.88 $& $   2.31 $ & $    0.13 $ & $    0.67 $ & $    0.00 $ & $    0.01 $ & $    0.00 $ & $    0.02 $ \\ 
 $   35.0$ & $    3.452\times 10^{-3  } $ & $  0.200 $& $   0.984 $ &$    1.27$ &$    1.91 $& $   2.45 $ & $   -0.05 $ & $    0.85 $ & $    0.00 $ & $    0.00 $ & $    0.00 $ & $    0.00 $ \\ 
 $   35.0$ & $    5.754\times 10^{-3  } $ & $  0.120 $& $   0.847 $ &$    1.46$ &$    1.96 $& $   2.94 $ & $    1.29 $ & $    1.00 $ & $    0.00 $ & $    0.00 $ & $    0.00 $ & $    0.00 $ \\ 
 $   45.0$ & $    1.366\times 10^{-3  } $ & $  0.650 $& $   1.258 $ &$    3.13$ &$    2.84 $& $   4.27 $ & $   -0.15 $ & $    0.56 $ & $    0.19 $ & $    0.13 $ & $    0.06 $ & $    0.00 $ \\ 
 $   45.0$ & $    1.812\times 10^{-3  } $ & $  0.490 $& $   1.186 $ &$    1.29$ &$    1.91 $& $   2.50 $ & $    0.86 $ & $    0.38 $ & $    0.16 $ & $    0.16 $ & $    0.00 $ & $    0.05 $ \\ 
 $   45.0$ & $    2.864\times 10^{-3  } $ & $  0.310 $& $   1.081 $ &$    1.22$ &$    1.89 $& $   2.30 $ & $    0.23 $ & $    0.40 $ & $    0.00 $ & $    0.01 $ & $    0.00 $ & $    0.00 $ \\ 
 $   45.0$ & $    4.439\times 10^{-3  } $ & $  0.200 $& $   0.951 $ &$    1.29$ &$    1.91 $& $   2.47 $ & $    0.69 $ & $    0.52 $ & $    0.00 $ & $    0.00 $ & $    0.00 $ & $    0.00 $ \\ 
 $   45.0$ & $    7.398\times 10^{-3  } $ & $  0.120 $& $   0.806 $ &$    1.52$ &$    1.97 $& $   2.84 $ & $    1.28 $ & $    0.48 $ & $    0.00 $ & $    0.00 $ & $    0.00 $ & $    0.00 $ \\ 
 $   60.0$ & $    2.416\times 10^{-3  } $ & $  0.490 $& $   1.201 $ &$    1.63$ &$    2.01 $& $   2.77 $ & $    0.91 $ & $    0.24 $ & $    0.20 $ & $    0.14 $ & $    0.00 $ & $    0.08 $ \\ 
 $   60.0$ & $    3.818\times 10^{-3  } $ & $  0.310 $& $   1.038 $ &$    1.40$ &$    1.94 $& $   2.45 $ & $   -0.01 $ & $    0.55 $ & $    0.00 $ & $    0.00 $ & $    0.00 $ & $    0.00 $ \\ 
 $   60.0$ & $    5.918\times 10^{-3  } $ & $  0.200 $& $   0.929 $ &$    1.44$ &$    1.96 $& $   2.51 $ & $    0.06 $ & $    0.63 $ & $    0.00 $ & $    0.00 $ & $    0.00 $ & $    0.00 $ \\ 
 $   60.0$ & $    9.864\times 10^{-3  } $ & $  0.120 $& $   0.795 $ &$    1.63$ &$    2.02 $& $   2.85 $ & $    1.00 $ & $    0.63 $ & $    0.00 $ & $    0.00 $ & $    0.00 $ & $    0.00 $ \\ 
 $   90.0$ & $    3.623\times 10^{-3  } $ & $  0.490 $& $   1.097 $ &$    3.36$ &$    2.72 $& $   4.47 $ & $    1.03 $ & $    0.40 $ & $    0.20 $ & $    0.13 $ & $    0.00 $ & $    0.00 $ \\ 
 $   90.0$ & $    5.727\times 10^{-3  } $ & $  0.310 $& $   0.975 $ &$    1.76$ &$    2.05 $& $   2.73 $ & $    0.23 $ & $    0.33 $ & $    0.01 $ & $    0.01 $ & $    0.00 $ & $    0.00 $ \\ 
 $   90.0$ & $    8.877\times 10^{-3  } $ & $  0.200 $& $   0.835 $ &$    1.71$ &$    2.03 $& $   2.79 $ & $    0.73 $ & $    0.44 $ & $    0.00 $ & $    0.00 $ & $    0.00 $ & $    0.00 $ \\ 
 $   90.0$ & $    1.480\times 10^{-2  } $ & $  0.120 $& $   0.717 $ &$    1.90$ &$    2.10 $& $   2.96 $ & $    0.68 $ & $    0.54 $ & $    0.00 $ & $    0.00 $ & $    0.00 $ & $    0.00 $ \\ 
 \hline
%\hline
\end{tabular}
\end{center}
\end{scriptsize}
\tablecaption{\label{tab:460p2}
Continuation of \Tab~\ref{tab:ler1}
}
\end{table}

\begin{table}
\begin{scriptsize}
\begin{center}
\begin{tabular}{ccccccccccccc}
%\hline
\hline
\\[-8pt]
 $\boldsymbol{Q^2}$ & $\boldsymbol{x}$ & $\boldsymbol{y}$ & $\boldsymbol{\sigma_r}$ & $\boldsymbol{\delta_{stat}}$ & $\boldsymbol{\delta_{unc}}$& $\boldsymbol{\delta_{tot}}$ & $\boldsymbol{\gamma_{\ee}}$ & $\boldsymbol{\gamma_{\thetae}}$ & $\boldsymbol{\gamma_{had}}$ & $\boldsymbol{\gamma_{noise}}$ & $\boldsymbol{\gamma_{\rm asym}}$
& $\boldsymbol{\gamma_{acctag}}$ \\
{\bf GeV}$^2$ &   &     &             &  {\bf \% }         &  {\bf \% }        &  {\bf \% }   &         {\bf \% }        &    {\bf \%}           &   {\bf \%}          &   {\bf \%}      & {\bf \% }  & {\bf \%}  \\
\hline
\\[-8pt]
$    1.5$ & $    2.790\times 10^{-5  } $ & $  0.848 $& $   0.662 $ &$    9.36$ &$    4.94 $& $  11.36 $ & $    2.78 $ & $   -1.16 $ & $    0.36 $ & $    0.53 $ & $    2.73 $ & $    0.00 $ \\ 
 $    2.0$ & $    3.720\times 10^{-5  } $ & $  0.848 $& $   0.760 $ &$    6.37$ &$    4.34 $& $   8.32 $ & $    1.81 $ & $   -0.49 $ & $    0.56 $ & $    0.60 $ & $    2.38 $ & $    0.00 $ \\ 
 $    2.0$ & $    4.150\times 10^{-5  } $ & $  0.760 $& $   0.663 $ &$    7.43$ &$    3.97 $& $   8.82 $ & $    0.86 $ & $   -2.16 $ & $    0.26 $ & $    0.29 $ & $    1.10 $ & $    0.00 $ \\ 
 $    2.5$ & $    4.650\times 10^{-5  } $ & $  0.848 $& $   0.829 $ &$    5.43$ &$    4.12 $& $   7.51 $ & $    2.26 $ & $    0.15 $ & $    0.34 $ & $    0.40 $ & $    2.15 $ & $    0.00 $ \\ 
 $    2.5$ & $    5.190\times 10^{-5  } $ & $  0.760 $& $   0.837 $ &$    3.85$ &$    3.10 $& $   5.20 $ & $    1.19 $ & $    0.59 $ & $    0.38 $ & $    0.46 $ & $    0.69 $ & $    0.00 $ \\ 
 $    2.5$ & $    5.800\times 10^{-5  } $ & $  0.680 $& $   0.768 $ &$    4.98$ &$    3.14 $& $   5.92 $ & $    0.44 $ & $    0.15 $ & $    0.15 $ & $    0.10 $ & $    0.36 $ & $    0.00 $ \\ 
 $    3.5$ & $    6.510\times 10^{-5  } $ & $  0.848 $& $   0.871 $ &$    5.56$ &$    4.05 $& $   7.61 $ & $    2.33 $ & $   -0.19 $ & $    0.50 $ & $    0.55 $ & $    2.13 $ & $    0.00 $ \\ 
 $    3.5$ & $    7.270\times 10^{-5  } $ & $  0.760 $& $   0.873 $ &$    3.51$ &$    2.92 $& $   4.96 $ & $    1.74 $ & $    0.00 $ & $    0.30 $ & $    0.33 $ & $    0.69 $ & $    0.00 $ \\ 
 $    3.5$ & $    8.120\times 10^{-5  } $ & $  0.680 $& $   0.869 $ &$    3.15$ &$    2.49 $& $   4.54 $ & $    1.83 $ & $   -0.89 $ & $    0.27 $ & $    0.26 $ & $    0.36 $ & $    0.00 $ \\ 
 $    3.5$ & $    9.210\times 10^{-5  } $ & $  0.600 $& $   0.939 $ &$    3.36$ &$    2.66 $& $   4.34 $ & $    0.30 $ & $   -0.47 $ & $    0.30 $ & $    0.26 $ & $    0.12 $ & $    0.00 $ \\ 
 $    3.5$ & $    1.062\times 10^{-4  } $ & $  0.520 $& $   0.957 $ &$    5.39$ &$    3.27 $& $   6.67 $ & $    1.22 $ & $   -1.56 $ & $    0.07 $ & $    0.19 $ & $    0.00 $ & $    0.87 $ \\ 
 $    5.0$ & $    9.310\times 10^{-5  } $ & $  0.848 $& $   0.849 $ &$    6.48$ &$    3.99 $& $   8.24 $ & $    1.58 $ & $    0.08 $ & $    0.46 $ & $    0.55 $ & $    2.65 $ & $    0.00 $ \\ 
 $    5.0$ & $    1.038\times 10^{-4  } $ & $  0.760 $& $   0.883 $ &$    3.52$ &$    2.84 $& $   4.68 $ & $    0.83 $ & $    0.05 $ & $    0.32 $ & $    0.39 $ & $    0.68 $ & $    0.00 $ \\ 
 $    5.0$ & $    1.160\times 10^{-4  } $ & $  0.680 $& $   1.006 $ &$    2.66$ &$    2.37 $& $   3.77 $ & $    1.17 $ & $   -0.08 $ & $    0.19 $ & $    0.23 $ & $    0.28 $ & $    0.00 $ \\ 
 $    5.0$ & $    1.315\times 10^{-4  } $ & $  0.600 $& $   0.946 $ &$    2.60$ &$    2.37 $& $   3.78 $ & $    1.35 $ & $   -0.08 $ & $    0.20 $ & $    0.20 $ & $    0.14 $ & $    0.00 $ \\ 
 $    5.0$ & $    1.517\times 10^{-4  } $ & $  0.520 $& $   1.011 $ &$    2.43$ &$    2.04 $& $   3.34 $ & $    0.86 $ & $   -0.34 $ & $    0.18 $ & $    0.25 $ & $    0.00 $ & $    0.37 $ \\ 
 $    5.0$ & $    2.013\times 10^{-4  } $ & $  0.392 $& $   0.940 $ &$    2.58$ &$    2.06 $& $   3.50 $ & $    0.76 $ & $   -0.82 $ & $    0.17 $ & $    0.23 $ & $    0.00 $ & $    0.00 $ \\ 
 $    6.5$ & $    1.210\times 10^{-4  } $ & $  0.848 $& $   0.903 $ &$    7.18$ &$    4.03 $& $   8.91 $ & $    0.65 $ & $    0.62 $ & $    0.35 $ & $    0.40 $ & $    3.25 $ & $    0.00 $ \\ 
 $    6.5$ & $    1.350\times 10^{-4  } $ & $  0.760 $& $   0.992 $ &$    3.48$ &$    2.85 $& $   4.72 $ & $    0.98 $ & $    0.67 $ & $    0.29 $ & $    0.37 $ & $    0.56 $ & $    0.00 $ \\ 
 $    6.5$ & $    1.509\times 10^{-4  } $ & $  0.680 $& $   1.069 $ &$    2.56$ &$    2.35 $& $   3.70 $ & $    1.04 $ & $    0.57 $ & $    0.31 $ & $    0.35 $ & $    0.21 $ & $    0.00 $ \\ 
 $    6.5$ & $    1.710\times 10^{-4  } $ & $  0.600 $& $   1.048 $ &$    2.30$ &$    2.32 $& $   3.40 $ & $    0.75 $ & $   -0.46 $ & $    0.19 $ & $    0.19 $ & $    0.11 $ & $    0.00 $ \\ 
 $    6.5$ & $    1.973\times 10^{-4  } $ & $  0.520 $& $   1.106 $ &$    2.05$ &$    1.96 $& $   2.97 $ & $    0.81 $ & $    0.12 $ & $    0.19 $ & $    0.19 $ & $    0.00 $ & $    0.26 $ \\ 
 $    6.5$ & $    2.617\times 10^{-4  } $ & $  0.392 $& $   1.001 $ &$    1.38$ &$    1.83 $& $   2.50 $ & $    0.95 $ & $   -0.17 $ & $    0.10 $ & $    0.12 $ & $    0.00 $ & $    0.20 $ \\ 
 $    8.5$ & $    1.582\times 10^{-4  } $ & $  0.848 $& $   0.976 $ &$    7.10$ &$    4.00 $& $   9.20 $ & $    2.37 $ & $    0.47 $ & $    0.32 $ & $    0.34 $ & $    3.49 $ & $    0.00 $ \\ 
 $    8.5$ & $    1.765\times 10^{-4  } $ & $  0.760 $& $   1.072 $ &$    3.81$ &$    2.90 $& $   4.92 $ & $   -0.12 $ & $    0.87 $ & $    0.29 $ & $    0.28 $ & $    0.62 $ & $    0.00 $ \\ 
 $    8.5$ & $    1.973\times 10^{-4  } $ & $  0.680 $& $   1.092 $ &$    2.80$ &$    2.39 $& $   3.93 $ & $    1.27 $ & $    0.37 $ & $    0.26 $ & $    0.26 $ & $    0.19 $ & $    0.00 $ \\ 
 $    8.5$ & $    2.236\times 10^{-4  } $ & $  0.600 $& $   1.103 $ &$    2.32$ &$    2.33 $& $   3.41 $ & $    0.60 $ & $    0.65 $ & $    0.19 $ & $    0.18 $ & $    0.09 $ & $    0.00 $ \\ 
 $    8.5$ & $    2.580\times 10^{-4  } $ & $  0.520 $& $   1.033 $ &$    2.03$ &$    1.93 $& $   3.14 $ & $    1.08 $ & $    0.33 $ & $    0.19 $ & $    0.22 $ & $    0.00 $ & $    0.81 $ \\ 
 $    8.5$ & $    3.422\times 10^{-4  } $ & $  0.392 $& $   1.087 $ &$    1.20$ &$    1.81 $& $   2.29 $ & $    0.71 $ & $   -0.01 $ & $    0.09 $ & $    0.11 $ & $    0.00 $ & $    0.02 $ \\ 
 $    8.5$ & $    5.409\times 10^{-4  } $ & $  0.248 $& $   1.015 $ &$    1.32$ &$    1.83 $& $   2.35 $ & $    0.57 $ & $    0.35 $ & $    0.00 $ & $    0.00 $ & $    0.00 $ & $    0.00 $ \\ 
 $    8.5$ & $    8.384\times 10^{-4  } $ & $  0.160 $& $   0.941 $ &$    1.47$ &$    1.85 $& $   2.56 $ & $    0.92 $ & $   -0.33 $ & $    0.00 $ & $    0.00 $ & $    0.00 $ & $    0.00 $ \\ 
 $    8.5$ & $    1.397\times 10^{-3  } $ & $  0.096 $& $   0.818 $ &$    2.58$ &$    2.07 $& $   3.50 $ & $    1.11 $ & $   -0.16 $ & $    0.00 $ & $    0.00 $ & $    0.00 $ & $    0.10 $ \\ 
 $   12.0$ & $    2.233\times 10^{-4  } $ & $  0.848 $& $   1.238 $ &$    5.51$ &$    4.00 $& $   7.51 $ & $    2.15 $ & $    0.15 $ & $    0.47 $ & $    0.34 $ & $    2.24 $ & $    0.00 $ \\ 
 $   12.0$ & $    2.492\times 10^{-4  } $ & $  0.760 $& $   1.083 $ &$    4.06$ &$    2.91 $& $   5.19 $ & $    1.15 $ & $    0.15 $ & $    0.25 $ & $    0.24 $ & $    0.76 $ & $    0.00 $ \\ 
 $   12.0$ & $    2.785\times 10^{-4  } $ & $  0.680 $& $   1.167 $ &$    3.07$ &$    2.43 $& $   4.09 $ & $    0.29 $ & $    1.07 $ & $    0.26 $ & $    0.28 $ & $    0.24 $ & $    0.00 $ \\ 
 $   12.0$ & $    3.156\times 10^{-4  } $ & $  0.600 $& $   1.188 $ &$    2.55$ &$    2.37 $& $   3.70 $ & $    0.76 $ & $    0.97 $ & $    0.22 $ & $    0.19 $ & $    0.11 $ & $    0.00 $ \\ 
 $   12.0$ & $    3.642\times 10^{-4  } $ & $  0.520 $& $   1.175 $ &$    2.12$ &$    1.96 $& $   3.03 $ & $    0.43 $ & $    0.64 $ & $    0.21 $ & $    0.20 $ & $    0.00 $ & $    0.40 $ \\ 
 $   12.0$ & $    4.831\times 10^{-4  } $ & $  0.392 $& $   1.127 $ &$    1.19$ &$    1.81 $& $   2.33 $ & $    0.71 $ & $    0.50 $ & $    0.06 $ & $    0.08 $ & $    0.00 $ & $    0.04 $ \\ 
 $   12.0$ & $    7.636\times 10^{-4  } $ & $  0.248 $& $   1.053 $ &$    1.23$ &$    1.82 $& $   2.31 $ & $    0.64 $ & $    0.31 $ & $    0.00 $ & $    0.00 $ & $    0.00 $ & $    0.05 $ \\ 
 $   12.0$ & $    1.184\times 10^{-3  } $ & $  0.160 $& $   1.017 $ &$    1.29$ &$    1.83 $& $   2.39 $ & $    0.77 $ & $    0.30 $ & $    0.00 $ & $    0.00 $ & $    0.00 $ & $    0.00 $ \\ 
 $   12.0$ & $    1.973\times 10^{-3  } $ & $  0.096 $& $   0.871 $ &$    2.25$ &$    2.00 $& $   3.36 $ & $    1.48 $ & $   -0.04 $ & $    0.00 $ & $    0.00 $ & $    0.00 $ & $    0.00 $ \\ 
 $   15.0$ & $    2.792\times 10^{-4  } $ & $  0.848 $& $   1.110 $ &$    6.31$ &$    4.04 $& $   7.92 $ & $    1.47 $ & $    0.34 $ & $    0.37 $ & $    0.36 $ & $    2.01 $ & $    0.00 $ \\ 
 $   15.0$ & $    3.115\times 10^{-4  } $ & $  0.760 $& $   1.294 $ &$    3.40$ &$    2.90 $& $   4.56 $ & $    0.47 $ & $    0.38 $ & $    0.33 $ & $    0.31 $ & $    0.48 $ & $    0.00 $ \\ 
 $   15.0$ & $    3.481\times 10^{-4  } $ & $  0.680 $& $   1.226 $ &$    3.18$ &$    2.47 $& $   4.05 $ & $    0.24 $ & $    0.28 $ & $    0.16 $ & $    0.17 $ & $    0.24 $ & $    0.00 $ \\ 
 $   15.0$ & $    3.945\times 10^{-4  } $ & $  0.600 $& $   1.156 $ &$    2.97$ &$    2.43 $& $   3.96 $ & $    0.39 $ & $    0.81 $ & $    0.25 $ & $    0.21 $ & $    0.15 $ & $    0.00 $ \\ 
 $   15.0$ & $    4.552\times 10^{-4  } $ & $  0.520 $& $   1.255 $ &$    2.33$ &$    2.01 $& $   3.23 $ & $    0.69 $ & $    0.60 $ & $    0.20 $ & $    0.18 $ & $    0.00 $ & $    0.26 $ \\ 
 $   15.0$ & $    6.039\times 10^{-4  } $ & $  0.392 $& $   1.162 $ &$    1.29$ &$    1.82 $& $   2.53 $ & $    0.95 $ & $    0.72 $ & $    0.08 $ & $    0.11 $ & $    0.00 $ & $    0.03 $ \\ 
 $   15.0$ & $    9.545\times 10^{-4  } $ & $  0.248 $& $   1.044 $ &$    1.28$ &$    1.82 $& $   2.30 $ & $    0.37 $ & $    0.43 $ & $    0.00 $ & $    0.00 $ & $    0.00 $ & $    0.00 $ \\ 
 $   15.0$ & $    1.479\times 10^{-3  } $ & $  0.160 $& $   0.973 $ &$    1.32$ &$    1.83 $& $   2.58 $ & $    0.99 $ & $    0.78 $ & $    0.00 $ & $    0.00 $ & $    0.00 $ & $    0.00 $ \\ 
 $   15.0$ & $    2.466\times 10^{-3  } $ & $  0.096 $& $   0.856 $ &$    2.17$ &$    1.98 $& $   3.36 $ & $    1.31 $ & $    0.94 $ & $    0.00 $ & $    0.00 $ & $    0.00 $ & $    0.07 $ \\ 
 \hline
%\hline
\end{tabular}
\end{center}
\end{scriptsize}
\tablecaption{\label{tab:mer}
Reduced cross section $\sigma_{r}$, as measured with the
    {\mer} data sample. 
Description of the columns is given in the caption of 
\Tab~\ref{tab:ler1}.
}
\end{table}

\begin{table}
\begin{scriptsize}
\begin{center}
\begin{tabular}{ccccccccccccc}
%\hline
\hline
\\[-8pt]
 $\boldsymbol{Q^2}$ & $\boldsymbol{x}$ & $\boldsymbol{y}$ & $\boldsymbol{\sigma_r}$ & $\boldsymbol{\delta_{stat}}$ & $\boldsymbol{\delta_{unc}}$& $\boldsymbol{\delta_{tot}}$ & $\boldsymbol{\gamma_{\ee}}$ & $\boldsymbol{\gamma_{\thetae}}$ & $\boldsymbol{\gamma_{had}}$ & $\boldsymbol{\gamma_{noise}}$ & $\boldsymbol{\gamma_{\rm asym}}$
& $\boldsymbol{\gamma_{acctag}}$ \\
{\bf GeV}$^2$ &   &     &             &  {\bf \% }         &  {\bf \% }        &  {\bf \% }   &         {\bf \% }        &    {\bf \%}           &   {\bf \%}          &   {\bf \%}      & {\bf \% }  & {\bf \%}  \\
\hline
\\[-8pt]
$   20.0$ & $    3.722\times 10^{-4  } $ & $  0.848 $& $   1.287 $ &$    6.59$ &$    4.25 $& $   8.28 $ & $    2.28 $ & $   -0.09 $ & $    0.32 $ & $    0.35 $ & $    1.27 $ & $    0.00 $ \\ 
 $   20.0$ & $    4.153\times 10^{-4  } $ & $  0.760 $& $   1.246 $ &$    3.65$ &$    2.93 $& $   4.80 $ & $    0.89 $ & $    0.17 $ & $    0.28 $ & $    0.24 $ & $    0.42 $ & $    0.00 $ \\ 
 $   20.0$ & $    4.642\times 10^{-4  } $ & $  0.680 $& $   1.293 $ &$    3.04$ &$    2.46 $& $   3.95 $ & $    0.20 $ & $    0.47 $ & $    0.19 $ & $    0.17 $ & $    0.18 $ & $    0.00 $ \\ 
 $   20.0$ & $    5.261\times 10^{-4  } $ & $  0.600 $& $   1.135 $ &$    3.17$ &$    2.46 $& $   4.16 $ & $    0.73 $ & $    0.67 $ & $    0.27 $ & $    0.24 $ & $    0.17 $ & $    0.00 $ \\ 
 $   20.0$ & $    6.070\times 10^{-4  } $ & $  0.520 $& $   1.224 $ &$    2.57$ &$    2.06 $& $   3.64 $ & $    0.65 $ & $    1.27 $ & $    0.21 $ & $    0.19 $ & $    0.00 $ & $    0.52 $ \\ 
 $   20.0$ & $    8.052\times 10^{-4  } $ & $  0.392 $& $   1.212 $ &$    1.41$ &$    1.84 $& $   2.50 $ & $    0.44 $ & $    0.80 $ & $    0.10 $ & $    0.10 $ & $    0.00 $ & $    0.03 $ \\ 
 $   20.0$ & $    1.273\times 10^{-3  } $ & $  0.248 $& $   1.093 $ &$    1.39$ &$    1.84 $& $   2.53 $ & $    0.53 $ & $    0.89 $ & $    0.00 $ & $    0.00 $ & $    0.00 $ & $    0.00 $ \\ 
 $   20.0$ & $    1.973\times 10^{-3  } $ & $  0.160 $& $   0.981 $ &$    1.44$ &$    1.85 $& $   2.63 $ & $    0.87 $ & $    0.84 $ & $    0.00 $ & $    0.00 $ & $    0.00 $ & $    0.00 $ \\ 
 $   20.0$ & $    3.288\times 10^{-3  } $ & $  0.096 $& $   0.853 $ &$    2.39$ &$    2.03 $& $   3.41 $ & $    1.28 $ & $    0.43 $ & $    0.00 $ & $    0.00 $ & $    0.00 $ & $    0.00 $ \\ 
 $   25.0$ & $    4.932\times 10^{-4  } $ & $  0.800 $& $   1.264 $ &$    3.94$ &$    3.02 $& $   5.02 $ & $    0.44 $ & $    0.12 $ & $    0.30 $ & $    0.27 $ & $    0.47 $ & $    0.00 $ \\ 
 $   25.0$ & $    6.165\times 10^{-4  } $ & $  0.640 $& $   1.272 $ &$    2.19$ &$    2.46 $& $   3.35 $ & $    0.41 $ & $    0.37 $ & $    0.27 $ & $    0.18 $ & $    0.12 $ & $    0.00 $ \\ 
 $   25.0$ & $    7.587\times 10^{-4  } $ & $  0.520 $& $   1.220 $ &$    2.61$ &$    2.06 $& $   3.49 $ & $    0.92 $ & $    0.33 $ & $    0.20 $ & $    0.15 $ & $    0.00 $ & $    0.31 $ \\ 
 $   25.0$ & $    1.007\times 10^{-3  } $ & $  0.392 $& $   1.228 $ &$    1.54$ &$    1.86 $& $   2.58 $ & $    0.48 $ & $    0.74 $ & $    0.07 $ & $    0.10 $ & $    0.00 $ & $    0.11 $ \\ 
 $   25.0$ & $    1.591\times 10^{-3  } $ & $  0.248 $& $   1.131 $ &$    1.55$ &$    1.87 $& $   2.51 $ & $    0.39 $ & $    0.52 $ & $    0.00 $ & $    0.00 $ & $    0.00 $ & $    0.04 $ \\ 
 $   25.0$ & $    2.466\times 10^{-3  } $ & $  0.160 $& $   1.015 $ &$    1.58$ &$    1.88 $& $   2.93 $ & $    1.22 $ & $    1.04 $ & $    0.00 $ & $    0.00 $ & $    0.00 $ & $    0.00 $ \\ 
 $   25.0$ & $    4.110\times 10^{-3  } $ & $  0.096 $& $   0.896 $ &$    2.63$ &$    2.11 $& $   3.57 $ & $    0.34 $ & $    1.11 $ & $    0.00 $ & $    0.00 $ & $    0.00 $ & $    0.00 $ \\ 
 $   35.0$ & $    7.268\times 10^{-4  } $ & $  0.760 $& $   1.438 $ &$    6.27$ &$    3.72 $& $   7.52 $ & $   -1.72 $ & $   -0.48 $ & $    0.28 $ & $    0.19 $ & $    0.22 $ & $    0.00 $ \\ 
 $   35.0$ & $    8.123\times 10^{-4  } $ & $  0.680 $& $   1.364 $ &$    3.51$ &$    2.60 $& $   4.54 $ & $    1.03 $ & $    0.56 $ & $    0.29 $ & $    0.24 $ & $    0.08 $ & $    0.00 $ \\ 
 $   35.0$ & $    9.206\times 10^{-4  } $ & $  0.600 $& $   1.343 $ &$    3.12$ &$    2.52 $& $   4.09 $ & $    0.31 $ & $    0.68 $ & $    0.26 $ & $    0.18 $ & $    0.06 $ & $    0.00 $ \\ 
 $   35.0$ & $    1.062\times 10^{-3  } $ & $  0.520 $& $   1.314 $ &$    2.73$ &$    2.11 $& $   3.62 $ & $    0.86 $ & $    0.65 $ & $    0.16 $ & $    0.15 $ & $    0.00 $ & $    0.11 $ \\ 
 $   35.0$ & $    1.409\times 10^{-3  } $ & $  0.392 $& $   1.254 $ &$    1.55$ &$    1.87 $& $   2.55 $ & $    0.54 $ & $    0.54 $ & $    0.08 $ & $    0.11 $ & $    0.00 $ & $    0.00 $ \\ 
 $   35.0$ & $    2.227\times 10^{-3  } $ & $  0.248 $& $   1.111 $ &$    1.68$ &$    1.89 $& $   2.68 $ & $    0.62 $ & $    0.63 $ & $    0.00 $ & $    0.00 $ & $    0.00 $ & $    0.00 $ \\ 
 $   35.0$ & $    3.452\times 10^{-3  } $ & $  0.160 $& $   0.967 $ &$    1.83$ &$    1.91 $& $   2.77 $ & $    0.32 $ & $    0.75 $ & $    0.00 $ & $    0.00 $ & $    0.00 $ & $    0.00 $ \\ 
 $   35.0$ & $    5.754\times 10^{-3  } $ & $  0.096 $& $   0.851 $ &$    3.01$ &$    2.20 $& $   4.23 $ & $    1.90 $ & $    0.60 $ & $    0.00 $ & $    0.00 $ & $    0.00 $ & $    0.00 $ \\ 
 $   45.0$ & $    1.044\times 10^{-3  } $ & $  0.680 $& $   1.274 $ &$    6.36$ &$    3.42 $& $   7.23 $ & $    0.24 $ & $    0.30 $ & $    0.17 $ & $    0.09 $ & $    0.06 $ & $    0.00 $ \\ 
 $   45.0$ & $    1.184\times 10^{-3  } $ & $  0.600 $& $   1.266 $ &$    3.80$ &$    2.68 $& $   4.71 $ & $    0.32 $ & $    0.54 $ & $    0.31 $ & $    0.22 $ & $    0.04 $ & $    0.00 $ \\ 
 $   45.0$ & $    1.366\times 10^{-3  } $ & $  0.520 $& $   1.291 $ &$    3.08$ &$    2.20 $& $   3.81 $ & $    0.29 $ & $    0.11 $ & $    0.20 $ & $    0.13 $ & $    0.00 $ & $    0.14 $ \\ 
 $   45.0$ & $    1.812\times 10^{-3  } $ & $  0.392 $& $   1.193 $ &$    1.70$ &$    1.89 $& $   2.64 $ & $    0.40 $ & $    0.56 $ & $    0.09 $ & $    0.10 $ & $    0.00 $ & $    0.04 $ \\ 
 $   45.0$ & $    2.864\times 10^{-3  } $ & $  0.248 $& $   1.106 $ &$    1.71$ &$    1.90 $& $   2.64 $ & $    0.29 $ & $    0.60 $ & $    0.00 $ & $    0.00 $ & $    0.00 $ & $    0.00 $ \\ 
 $   45.0$ & $    4.439\times 10^{-3  } $ & $  0.160 $& $   0.942 $ &$    1.87$ &$    1.92 $& $   2.86 $ & $    0.48 $ & $    0.89 $ & $    0.00 $ & $    0.00 $ & $    0.00 $ & $    0.00 $ \\ 
 $   45.0$ & $    7.398\times 10^{-3  } $ & $  0.096 $& $   0.822 $ &$    3.15$ &$    2.23 $& $   4.33 $ & $    1.63 $ & $    1.11 $ & $    0.00 $ & $    0.00 $ & $    0.00 $ & $    0.00 $ \\ 
 $   60.0$ & $    1.578\times 10^{-3  } $ & $  0.600 $& $   1.286 $ &$    8.41$ &$    4.22 $& $   9.57 $ & $    1.69 $ & $    0.28 $ & $    0.00 $ & $    0.07 $ & $    0.04 $ & $    0.00 $ \\ 
 $   60.0$ & $    1.821\times 10^{-3  } $ & $  0.520 $& $   1.263 $ &$    4.23$ &$    2.54 $& $   4.98 $ & $    0.23 $ & $    0.63 $ & $    0.10 $ & $    0.12 $ & $    0.00 $ & $    0.00 $ \\ 
 $   60.0$ & $    2.416\times 10^{-3  } $ & $  0.392 $& $   1.167 $ &$    1.96$ &$    1.94 $& $   2.78 $ & $    0.28 $ & $    0.18 $ & $    0.06 $ & $    0.06 $ & $    0.00 $ & $    0.00 $ \\ 
 $   60.0$ & $    3.818\times 10^{-3  } $ & $  0.248 $& $   1.018 $ &$    1.98$ &$    1.94 $& $   2.84 $ & $    0.35 $ & $    0.50 $ & $    0.00 $ & $    0.00 $ & $    0.00 $ & $    0.06 $ \\ 
 $   60.0$ & $    5.918\times 10^{-3  } $ & $  0.160 $& $   0.992 $ &$    1.97$ &$    1.96 $& $   2.93 $ & $    0.75 $ & $    0.53 $ & $    0.00 $ & $    0.00 $ & $    0.00 $ & $    0.00 $ \\ 
 $   60.0$ & $    9.864\times 10^{-3  } $ & $  0.096 $& $   0.777 $ &$    3.38$ &$    2.30 $& $   4.23 $ & $    0.14 $ & $    1.07 $ & $    0.00 $ & $    0.00 $ & $    0.00 $ & $    0.00 $ \\ 
 $   90.0$ & $    3.623\times 10^{-3  } $ & $  0.392 $& $   1.147 $ &$    2.91$ &$    2.17 $& $   3.67 $ & $   -0.05 $ & $    0.51 $ & $    0.07 $ & $    0.09 $ & $    0.00 $ & $    0.00 $ \\ 
 $   90.0$ & $    5.727\times 10^{-3  } $ & $  0.248 $& $   0.921 $ &$    2.37$ &$    2.01 $& $   3.16 $ & $    0.55 $ & $    0.18 $ & $    0.00 $ & $    0.00 $ & $    0.00 $ & $    0.00 $ \\ 
 $   90.0$ & $    8.877\times 10^{-3  } $ & $  0.160 $& $   0.869 $ &$    2.31$ &$    2.02 $& $   3.20 $ & $    0.78 $ & $    0.49 $ & $    0.00 $ & $    0.00 $ & $    0.00 $ & $    0.00 $ \\ 
 $   90.0$ & $    1.480\times 10^{-2  } $ & $  0.096 $& $   0.755 $ &$    3.84$ &$    2.47 $& $   4.63 $ & $    0.68 $ & $    0.34 $ & $    0.00 $ & $    0.00 $ & $    0.00 $ & $    0.00 $ \\ 
 \hline
%\hline
\end{tabular}
\end{center}
\end{scriptsize}
\tablecaption{
Continuation of \Tab~\ref{tab:mer}
\label{tab:h1575last}}
\end{table}

\begin{table}
\begin{tiny}
\begin{center}
\begin{tabular}{l|cccccccccc}
\hline
\\[-6pt]
{\bf Bin} &  $\boldsymbol{Q^2}$  & $\boldsymbol{x}$ & 
$\boldsymbol{y}$ &  $\boldsymbol{\sigma^{\rm ave}_{r}}$ & $\boldsymbol{F_2^{\rm ave}}$ 
 & $\boldsymbol{\delta_{\rm ave,stat}}$ & $\boldsymbol{\delta_{\rm ave,uncor}}$ & 
$\boldsymbol{\delta_{\rm ave,cor}}$ 
 &  $\boldsymbol{\delta_{\rm ave,tot}}$ 
 & $\boldsymbol{\sqrt{s}}$\\
$\boldsymbol{\#}$     & {\bf GeV}$\boldsymbol{^2}$ &   &       &       &         &       {\bf \%}               & {\bf \%}  &    
\%              &   {\bf \%}  & {\bf GeV} \\
\hline
\\[-7pt]
$    1 $ &$    0.2 $ &$ 0.398 \times 10^{-4}$ & $ 0.050 $ &$ 0.232 $ &$ 0.232 $ &$ 14.27 $ &$ 11.96 $ &$  6.97 $ &$19.88$ & $   319$ \\
$    2 $ &$    0.2 $ &$ 0.251 \times 10^{-3}$ & $ 0.008 $ &$ 0.190 $ &$ 0.190 $ &$ 13.12 $ &$  6.18 $ &$  3.89 $ &$15.01$ & $   319$ \\
$    3 $ &$   0.25 $ &$ 0.398 \times 10^{-4}$ & $ 0.062 $ &$ 0.302 $ &$ 0.302 $ &$  9.79 $ &$ 11.26 $ &$  7.48 $ &$16.69$ & $   319$ \\
$    4 $ &$   0.25 $ &$ 0.251 \times 10^{-3}$ & $ 0.010 $ &$ 0.191 $ &$ 0.191 $ &$ 10.00 $ &$  4.70 $ &$  4.78 $ &$12.03$ & $   319$ \\
$    5 $ &$   0.25 $ &$ 0.158 \times 10^{-2}$ & $ 0.002 $ &$ 0.204 $ &$ 0.204 $ &$ 10.84 $ &$  5.29 $ &$  2.69 $ &$12.35$ & $   319$ \\
$    6 $ &$   0.35 $ &$ 0.511 \times 10^{-5}$ & $ 0.675 $ &$ 0.452 $ &$ 0.494 $ &$ 21.67 $ &$ 12.79 $ &$  2.28 $ &$25.27$ & $   319$ \\
$    7 $ &$   0.35 $ &$ 0.611 \times 10^{-5}$ & $ 0.634 $ &$ 0.359 $ &$ 0.387 $ &$  5.73 $ &$ 11.03 $ &$  5.25 $ &$13.49$ & $   301$ \\
$    8 $ &$   0.35 $ &$ 0.320 \times 10^{-4}$ & $ 0.108 $ &$ 0.416 $ &$ 0.416 $ &$  9.06 $ &$ 11.10 $ &$ 14.08 $ &$20.09$ & $   319$ \\
$    9 $ &$   0.35 $ &$ 0.130 \times 10^{-3}$ & $ 0.027 $ &$ 0.266 $ &$ 0.266 $ &$  9.59 $ &$  4.38 $ &$  2.94 $ &$10.95$ & $   319$ \\
$   10 $ &$   0.35 $ &$ 0.500 \times 10^{-3}$ & $ 0.007 $ &$ 0.237 $ &$ 0.237 $ &$  8.80 $ &$  4.19 $ &$  2.51 $ &$10.06$ & $   319$ \\
$   11 $ &$   0.35 $ &$ 0.251 \times 10^{-2}$ & $ 0.001 $ &$ 0.205 $ &$ 0.205 $ &$  9.91 $ &$  4.55 $ &$  1.82 $ &$11.06$ & $   319$ \\
$   12 $ &$    0.5 $ &$ 0.731 \times 10^{-5}$ & $ 0.675 $ &$ 0.453 $ &$ 0.495 $ &$  5.40 $ &$  5.74 $ &$  4.97 $ &$ 9.32$ & $   319$ \\
$   13 $ &$    0.5 $ &$ 0.860 \times 10^{-5}$ & $ 0.650 $ &$ 0.444 $ &$ 0.481 $ &$  3.74 $ &$  9.17 $ &$  4.01 $ &$10.68$ & $   301$ \\
$   14 $ &$    0.5 $ &$ 0.158 \times 10^{-4}$ & $ 0.312 $ &$ 0.463 $ &$ 0.470 $ &$ 18.93 $ &$  9.84 $ &$  3.07 $ &$21.56$ & $   319$ \\
$   15 $ &$    0.5 $ &$ 0.398 \times 10^{-4}$ & $ 0.124 $ &$ 0.484 $ &$ 0.485 $ &$ 10.08 $ &$  6.07 $ &$ 10.88 $ &$16.03$ & $   319$ \\
$   16 $ &$    0.5 $ &$ 0.100 \times 10^{-3}$ & $ 0.049 $ &$ 0.412 $ &$ 0.412 $ &$  8.83 $ &$  4.87 $ &$  3.09 $ &$10.55$ & $   319$ \\
$   17 $ &$    0.5 $ &$ 0.251 \times 10^{-3}$ & $ 0.020 $ &$ 0.297 $ &$ 0.297 $ &$  8.33 $ &$  4.25 $ &$  2.61 $ &$ 9.71$ & $   319$ \\
$   18 $ &$    0.5 $ &$ 0.800 \times 10^{-3}$ & $ 0.006 $ &$ 0.281 $ &$ 0.281 $ &$  5.88 $ &$  3.49 $ &$  1.69 $ &$ 7.04$ & $   319$ \\
$   19 $ &$    0.5 $ &$ 0.320 \times 10^{-2}$ & $ 0.002 $ &$ 0.183 $ &$ 0.183 $ &$ 11.37 $ &$  6.39 $ &$  1.37 $ &$13.11$ & $   319$ \\
$   20 $ &$   0.65 $ &$ 0.950 \times 10^{-5}$ & $ 0.675 $ &$ 0.482 $ &$ 0.527 $ &$  3.95 $ &$  2.90 $ &$  3.02 $ &$ 5.76$ & $   319$ \\
$   21 $ &$   0.65 $ &$ 0.112 \times 10^{-4}$ & $ 0.650 $ &$ 0.506 $ &$ 0.549 $ &$  3.73 $ &$  8.21 $ &$  4.04 $ &$ 9.88$ & $   301$ \\
$   22 $ &$   0.65 $ &$ 0.158 \times 10^{-4}$ & $ 0.406 $ &$ 0.468 $ &$ 0.480 $ &$  3.08 $ &$  5.44 $ &$  1.70 $ &$ 6.48$ & $   319$ \\
$   23 $ &$   0.65 $ &$ 0.164 \times 10^{-4}$ & $ 0.439 $ &$ 0.512 $ &$ 0.528 $ &$  3.01 $ &$  7.28 $ &$  2.66 $ &$ 8.31$ & $   301$ \\
$   24 $ &$   0.65 $ &$ 0.398 \times 10^{-4}$ & $ 0.161 $ &$ 0.681 $ &$ 0.683 $ &$ 17.43 $ &$ 11.16 $ &$  4.24 $ &$21.13$ & $   319$ \\
$   25 $ &$   0.65 $ &$ 0.100 \times 10^{-3}$ & $ 0.064 $ &$ 0.501 $ &$ 0.501 $ &$  5.14 $ &$  5.84 $ &$  7.32 $ &$10.68$ & $   319$ \\
$   26 $ &$   0.65 $ &$ 0.251 \times 10^{-3}$ & $ 0.026 $ &$ 0.378 $ &$ 0.378 $ &$  6.78 $ &$  3.48 $ &$  2.25 $ &$ 7.94$ & $   319$ \\
$   27 $ &$   0.65 $ &$ 0.800 \times 10^{-3}$ & $ 0.008 $ &$ 0.309 $ &$ 0.309 $ &$  4.91 $ &$  3.06 $ &$  2.09 $ &$ 6.15$ & $   319$ \\
$   28 $ &$   0.65 $ &$ 0.320 \times 10^{-2}$ & $ 0.002 $ &$ 0.226 $ &$ 0.226 $ &$  5.79 $ &$  3.19 $ &$  1.35 $ &$ 6.75$ & $   319$ \\
$   29 $ &$   0.85 $ &$ 0.124 \times 10^{-4}$ & $ 0.675 $ &$ 0.569 $ &$ 0.621 $ &$  2.54 $ &$  2.52 $ &$  2.56 $ &$ 4.40$ & $   319$ \\
$   30 $ &$   0.85 $ &$ 0.138 \times 10^{-4}$ & $ 0.675 $ &$ 0.617 $ &$ 0.675 $ &$  5.19 $ &$  9.45 $ &$  5.62 $ &$12.16$ & $   301$ \\
$   31 $ &$   0.85 $ &$ 0.200 \times 10^{-4}$ & $ 0.470 $ &$ 0.598 $ &$ 0.620 $ &$  2.64 $ &$  4.98 $ &$  2.72 $ &$ 6.26$ & $   301$ \\
$   32 $ &$   0.85 $ &$ 0.200 \times 10^{-4}$ & $ 0.419 $ &$ 0.614 $ &$ 0.631 $ &$  1.95 $ &$  5.36 $ &$  1.75 $ &$ 5.97$ & $   319$ \\
$   33 $ &$   0.85 $ &$ 0.398 \times 10^{-4}$ & $ 0.211 $ &$ 0.569 $ &$ 0.572 $ &$  1.58 $ &$  3.49 $ &$  1.61 $ &$ 4.15$ & $   319$ \\
$   34 $ &$   0.85 $ &$ 0.500 \times 10^{-4}$ & $ 0.168 $ &$ 0.548 $ &$ 0.550 $ &$  2.91 $ &$  4.52 $ &$  2.54 $ &$ 5.95$ & $   319$ \\
$   35 $ &$   0.85 $ &$ 0.100 \times 10^{-3}$ & $ 0.084 $ &$ 0.501 $ &$ 0.502 $ &$  2.65 $ &$  3.78 $ &$  3.72 $ &$ 5.93$ & $   319$ \\
$   36 $ &$   0.85 $ &$ 0.251 \times 10^{-3}$ & $ 0.033 $ &$ 0.415 $ &$ 0.415 $ &$  5.88 $ &$  2.98 $ &$  3.08 $ &$ 7.28$ & $   319$ \\
$   37 $ &$   0.85 $ &$ 0.800 \times 10^{-3}$ & $ 0.010 $ &$ 0.352 $ &$ 0.352 $ &$  4.59 $ &$  2.67 $ &$  1.66 $ &$ 5.56$ & $   319$ \\
$   38 $ &$   0.85 $ &$ 0.320 \times 10^{-2}$ & $ 0.003 $ &$ 0.308 $ &$ 0.308 $ &$  4.54 $ &$  2.83 $ &$  1.14 $ &$ 5.47$ & $   301$ \\
$   39 $ &$    1.2 $ &$ 0.176 \times 10^{-4}$ & $ 0.675 $ &$ 0.613 $ &$ 0.670 $ &$  2.51 $ &$  2.16 $ &$  2.90 $ &$ 4.40$ & $   319$ \\
$   40 $ &$    1.2 $ &$ 0.200 \times 10^{-4}$ & $ 0.675 $ &$ 0.744 $ &$ 0.813 $ &$  3.59 $ &$  8.36 $ &$  4.07 $ &$ 9.97$ & $   301$ \\
$   41 $ &$    1.2 $ &$ 0.200 \times 10^{-4}$ & $ 0.592 $ &$ 0.675 $ &$ 0.719 $ &$  2.61 $ &$  2.51 $ &$  1.45 $ &$ 3.90$ & $   319$ \\
$   42 $ &$    1.2 $ &$ 0.320 \times 10^{-4}$ & $ 0.415 $ &$ 0.708 $ &$ 0.727 $ &$  2.67 $ &$  4.55 $ &$  2.44 $ &$ 5.81$ & $   301$ \\
$   43 $ &$    1.2 $ &$ 0.320 \times 10^{-4}$ & $ 0.370 $ &$ 0.692 $ &$ 0.706 $ &$  1.67 $ &$  2.73 $ &$  1.45 $ &$ 3.51$ & $   319$ \\
$   44 $ &$    1.2 $ &$ 0.631 \times 10^{-4}$ & $ 0.188 $ &$ 0.649 $ &$ 0.652 $ &$  1.18 $ &$  2.27 $ &$  1.71 $ &$ 3.08$ & $   319$ \\
$   45 $ &$    1.2 $ &$ 0.800 \times 10^{-4}$ & $ 0.148 $ &$ 0.596 $ &$ 0.598 $ &$  2.18 $ &$  4.03 $ &$  2.52 $ &$ 5.22$ & $   319$ \\
$   46 $ &$    1.2 $ &$ 0.130 \times 10^{-3}$ & $ 0.091 $ &$ 0.544 $ &$ 0.545 $ &$  2.42 $ &$  4.97 $ &$  1.64 $ &$ 5.76$ & $   319$ \\
$   47 $ &$    1.2 $ &$ 0.158 \times 10^{-3}$ & $ 0.075 $ &$ 0.506 $ &$ 0.506 $ &$  1.54 $ &$  2.35 $ &$  1.46 $ &$ 3.17$ & $   319$ \\
$   48 $ &$    1.2 $ &$ 0.398 \times 10^{-3}$ & $ 0.030 $ &$ 0.503 $ &$ 0.503 $ &$  2.09 $ &$  3.37 $ &$  1.63 $ &$ 4.29$ & $   319$ \\
$   49 $ &$    1.2 $ &$ 0.130 \times 10^{-2}$ & $ 0.009 $ &$ 0.375 $ &$ 0.375 $ &$  3.54 $ &$  2.67 $ &$  1.61 $ &$ 4.72$ & $   319$ \\
$   50 $ &$    1.2 $ &$ 0.500 \times 10^{-2}$ & $ 0.002 $ &$ 0.298 $ &$ 0.298 $ &$  4.50 $ &$  2.61 $ &$  1.64 $ &$ 5.46$ & $   319$ \\
$   51 $ &$    1.5 $ &$ 0.185 \times 10^{-4}$ & $ 0.800 $ &$ 0.621 $ &$ 0.711 $ &$  3.14 $ &$  3.48 $ &$  4.33 $ &$ 6.38$ & $   319$ \\
$   52 $ &$    1.5 $ &$ 0.219 \times 10^{-4}$ & $ 0.675 $ &$ 0.707 $ &$ 0.773 $ &$  1.93 $ &$  1.79 $ &$  1.60 $ &$ 3.08$ & $   319$ \\
$   53 $ &$    1.5 $ &$ 0.320 \times 10^{-4}$ & $ 0.519 $ &$ 0.805 $ &$ 0.843 $ &$  1.19 $ &$  3.20 $ &$  2.85 $ &$ 4.45$ & $   301$ \\
$   54 $ &$    1.5 $ &$ 0.320 \times 10^{-4}$ & $ 0.462 $ &$ 0.759 $ &$ 0.786 $ &$  1.73 $ &$  2.15 $ &$  1.24 $ &$ 3.02$ & $   319$ \\
$   55 $ &$    1.5 $ &$ 0.500 \times 10^{-4}$ & $ 0.296 $ &$ 0.762 $ &$ 0.772 $ &$  1.04 $ &$  1.98 $ &$  1.29 $ &$ 2.58$ & $   319$ \\
$   56 $ &$    1.5 $ &$ 0.800 \times 10^{-4}$ & $ 0.185 $ &$ 0.702 $ &$ 0.705 $ &$  1.26 $ &$  2.15 $ &$  1.52 $ &$ 2.92$ & $   319$ \\
$   57 $ &$    1.5 $ &$ 0.130 \times 10^{-3}$ & $ 0.114 $ &$ 0.645 $ &$ 0.646 $ &$  1.46 $ &$  2.43 $ &$  1.68 $ &$ 3.29$ & $   319$ \\
$   58 $ &$    1.5 $ &$ 0.200 \times 10^{-3}$ & $ 0.074 $ &$ 0.617 $ &$ 0.618 $ &$  2.07 $ &$  2.86 $ &$  1.78 $ &$ 3.95$ & $   319$ \\
$   59 $ &$    1.5 $ &$ 0.320 \times 10^{-3}$ & $ 0.046 $ &$ 0.586 $ &$ 0.586 $ &$  1.48 $ &$  2.26 $ &$  1.85 $ &$ 3.27$ & $   319$ \\
$   60 $ &$    1.5 $ &$ 0.500 \times 10^{-3}$ & $ 0.030 $ &$ 0.550 $ &$ 0.550 $ &$  2.51 $ &$  7.05 $ &$  1.94 $ &$ 7.73$ & $   319$ \\
$   61 $ &$    1.5 $ &$ 0.800 \times 10^{-3}$ & $ 0.018 $ &$ 0.497 $ &$ 0.497 $ &$  2.35 $ &$  2.47 $ &$  1.63 $ &$ 3.78$ & $   319$ \\
$   62 $ &$    1.5 $ &$ 0.100 \times 10^{-2}$ & $ 0.015 $ &$ 0.465 $ &$ 0.465 $ &$  5.21 $ &$  3.74 $ &$  1.50 $ &$ 6.58$ & $   319$ \\
$   63 $ &$    1.5 $ &$ 0.320 \times 10^{-2}$ & $ 0.005 $ &$ 0.410 $ &$ 0.410 $ &$  2.31 $ &$  2.04 $ &$  1.59 $ &$ 3.47$ & $   301$ \\
$   64 $ &$    1.5 $ &$ 0.130 \times 10^{-1}$ & $ 0.001 $ &$ 0.327 $ &$ 0.327 $ &$  3.99 $ &$  2.49 $ &$  5.17 $ &$ 6.99$ & $   319$ \\
\hline
\end{tabular}
\end{center}
\end{tiny}
\tablecaption{\label{tab615a1}
%\small 
Combined reduced cross section $\sigma_{r}$  for $E_p=920$~GeV and $E_p=820$~GeV. 
$F_2^{\rm ave}$ represents the structure function $F_2$ calculated
from $\sigma^{\rm ave}_{r}$ by using 
$R=0.26$.
$\delta_{\rm ave,stat}$, $\delta_{\rm ave,uncor}$, $\delta_{\rm ave,cor}$ and
$\delta_{\rm ave,tot}$ represent the statistical, uncorrelated
systematic, correlated systematic and total experimental uncertainty, respectively. 
The uncertainties are quoted in percent relative to $\sigma^{\rm ave}_{r}$. 
The overall normalisation uncertainty of $0.5\%$ is not included.
}
\end{table}

\begin{table}
\begin{tiny}
\begin{center}
\begin{tabular}{l|cccccccccc}
\hline
\\[-6pt]
{\bf Bin} &  $\boldsymbol{Q^2}$  & $\boldsymbol{x}$ & 
$\boldsymbol{y}$ &  $\boldsymbol{\sigma^{\rm ave}_{r}}$ & $\boldsymbol{F_2^{\rm ave}}$ 
 & $\boldsymbol{\delta_{\rm ave,stat}}$ & $\boldsymbol{\delta_{\rm ave,uncor}}$ & 
$\boldsymbol{\delta_{\rm ave,cor}}$ 
 &  $\boldsymbol{\delta_{\rm ave,tot}}$ 
 & $\boldsymbol{\sqrt{s}}$\\
$\boldsymbol{\#}$     & {\bf GeV}$\boldsymbol{^2}$ &   &       &       &         &       {\bf \%}               & {\bf \%}  &    
\%              &   {\bf \%}  & {\bf GeV} \\
\hline
\\[-7pt]
$   65 $ &$     2. $ &$ 0.247 \times 10^{-4}$ & $ 0.800 $ &$ 0.788 $ &$ 0.903 $ &$  2.30 $ &$  2.70 $ &$  3.33 $ &$ 4.86$ & $   319$ \\
$   66 $ &$     2. $ &$ 0.295 \times 10^{-4}$ & $ 0.675 $ &$ 0.798 $ &$ 0.872 $ &$  1.48 $ &$  1.65 $ &$  1.43 $ &$ 2.64$ & $   319$ \\
$   67 $ &$     2. $ &$ 0.328 \times 10^{-4}$ & $ 0.675 $ &$ 0.843 $ &$ 0.922 $ &$  1.40 $ &$  5.32 $ &$  3.10 $ &$ 6.32$ & $   301$ \\
$   68 $ &$     2. $ &$ 0.500 \times 10^{-4}$ & $ 0.443 $ &$ 0.860 $ &$ 0.887 $ &$  0.80 $ &$  2.49 $ &$  1.41 $ &$ 2.97$ & $   301$ \\
\hline
\end{tabular}
\end{center}
\end{tiny}
\tablecaption{\label{tab615a2}Continuation of \Tab~\ref{tab615a1}}
%\small 
\end{table}

\begin{table}
\begin{tiny}
\begin{center}
\begin{tabular}{l|cccccccccc}
\hline
\\[-6pt]
{\bf Bin} &  $\boldsymbol{Q^2}$  & $\boldsymbol{x}$ & 
$\boldsymbol{y}$ &  $\boldsymbol{\sigma^{\rm ave}_{r}}$ & $\boldsymbol{F_2^{\rm ave}}$ 
 & $\boldsymbol{\delta_{\rm ave,stat}}$ & $\boldsymbol{\delta_{\rm ave,uncor}}$ & 
$\boldsymbol{\delta_{\rm ave,cor}}$ 
 &  $\boldsymbol{\delta_{\rm ave,tot}}$ 
 & $\boldsymbol{\sqrt{s}}$\\
$\boldsymbol{\#}$     & {\bf GeV}$\boldsymbol{^2}$ &   &       &       &         &       {\bf \%}               & {\bf \%}  &    
\%              &   {\bf \%}  & {\bf GeV} \\
\hline
\\[-7pt]
$  131 $ &$    8.5 $ &$ 0.986 \times 10^{-4}$ & $ 0.850 $ &$ 1.158 $ &$ 1.355 $ &$  2.32 $ &$  2.99 $ &$  0.96 $ &$ 3.90$ & $   319$ \\
$  132 $ &$    8.5 $ &$ 0.105 \times 10^{-3}$ & $ 0.800 $ &$ 1.143 $ &$ 1.310 $ &$  2.47 $ &$  2.00 $ &$  1.27 $ &$ 3.42$ & $   319$ \\
$  133 $ &$    8.5 $ &$ 0.124 \times 10^{-3}$ & $ 0.675 $ &$ 1.214 $ &$ 1.326 $ &$  1.36 $ &$  1.60 $ &$  1.07 $ &$ 2.36$ & $   319$ \\
$  134 $ &$    8.5 $ &$ 0.139 \times 10^{-3}$ & $ 0.675 $ &$ 1.140 $ &$ 1.246 $ &$  2.06 $ &$  1.82 $ &$  3.58 $ &$ 4.51$ & $   301$ \\
\hline
\end{tabular}
\end{center}
\end{tiny}
\tablecaption{\label{tab615a3}Continuation of \Tab~\ref{tab615a1}.}
%\small 
\end{table}

\begin{table}
\begin{tiny}
\begin{center}
\begin{tabular}{l|cccccccccc}
\hline
\\[-6pt]
{\bf Bin} &  $\boldsymbol{Q^2}$  & $\boldsymbol{x}$ & 
$\boldsymbol{y}$ &  $\boldsymbol{\sigma^{\rm ave}_{r}}$ & $\boldsymbol{F_2^{\rm ave}}$ 
 & $\boldsymbol{\delta_{\rm ave,stat}}$ & $\boldsymbol{\delta_{\rm ave,uncor}}$ & 
$\boldsymbol{\delta_{\rm ave,cor}}$ 
 &  $\boldsymbol{\delta_{\rm ave,tot}}$ 
 & $\boldsymbol{\sqrt{s}}$\\
$\boldsymbol{\#}$     & {\bf GeV}$\boldsymbol{^2}$ &   &       &       &         &       {\bf \%}               & {\bf \%}  &    
\%              &   {\bf \%}  & {\bf GeV} \\
\hline
\\[-7pt]
$  204 $ &$    35. $ &$ 0.460 \times 10^{-3}$ & $ 0.750 $ &$ 1.444 $ &$ 1.621 $ &$  1.24 $ &$  2.68 $ &$  1.05 $ &$ 3.13$ & $   319$ \\
$  205 $ &$    35. $ &$ 0.531 \times 10^{-3}$ & $ 0.650 $ &$ 1.469 $ &$ 1.593 $ &$  0.66 $ &$  2.15 $ &$  1.06 $ &$ 2.49$ & $   319$ \\
$  206 $ &$    35. $ &$ 0.575 \times 10^{-3}$ & $ 0.675 $ &$ 1.512 $ &$ 1.652 $ &$  1.36 $ &$  2.01 $ &$  1.15 $ &$ 2.69$ & $   301$ \\
$  207 $ &$    35. $ &$ 0.800 \times 10^{-3}$ & $ 0.484 $ &$ 1.389 $ &$ 1.444 $ &$  0.88 $ &$  1.75 $ &$  1.02 $ &$ 2.20$ & $   301$ \\
\hline
\end{tabular}
\end{center}
\end{tiny}
\tablecaption{\label{tab615a4}Continuation of \Tab~\ref{tab615a1}.}
%\small 
\end{table}

\begin{table}
\begin{tiny}
\begin{center}
\begin{tabular}{l|cccccccccc}
\hline
\\[-6pt]
{\bf Bin} &  $\boldsymbol{Q^2}$  & $\boldsymbol{x}$ & 
$\boldsymbol{y}$ &  $\boldsymbol{\sigma^{\rm ave}_{r}}$ & $\boldsymbol{F_2^{\rm ave}}$ 
 & $\boldsymbol{\delta_{\rm ave,stat}}$ & $\boldsymbol{\delta_{\rm ave,uncor}}$ & 
$\boldsymbol{\delta_{\rm ave,cor}}$ 
 &  $\boldsymbol{\delta_{\rm ave,tot}}$ 
 & $\boldsymbol{\sqrt{s}}$\\
$\boldsymbol{\#}$     & {\bf GeV}$\boldsymbol{^2}$ &   &       &       &         &       {\bf \%}               & {\bf \%}  &    
\%              &   {\bf \%}  & {\bf GeV} \\
\hline
\\[-7pt]
$    1 $ &$    1.5 $ &$ 0.279 \times 10^{-4}$ & $ 0.848 $ &$ 0.661 $ &$ 0.773 $ &$  9.37 $ &$  4.94 $ &$  4.80 $ &$11.63$ & $   252$ \\
$    2 $ &$    1.5 $ &$ 0.348 \times 10^{-4}$ & $ 0.850 $ &$ 0.521 $ &$ 0.610 $ &$  8.09 $ &$  4.96 $ &$  4.60 $ &$10.54$ & $   225$ \\
$    3 $ &$     2. $ &$ 0.372 \times 10^{-4}$ & $ 0.848 $ &$ 0.758 $ &$ 0.887 $ &$  6.38 $ &$  4.34 $ &$  4.14 $ &$ 8.76$ & $   252$ \\
$    4 $ &$     2. $ &$ 0.415 \times 10^{-4}$ & $ 0.761 $ &$ 0.663 $ &$ 0.747 $ &$  7.43 $ &$  3.97 $ &$  3.73 $ &$ 9.21$ & $   252$ \\
\hline
\end{tabular}
\end{center}
\end{tiny}
\tablecaption{\label{tablowea1}
%\small 
Combined reduced cross section $\sigma_{r}$  for $E_p=460$~GeV and $E_p=575$~GeV. 
Description of the columns is given in the caption of \Tab~\ref{tab615a1}.
}
\end{table}

\begin{table}
\begin{tiny}
\begin{center}
\begin{tabular}{l|cccccccccc}
\hline
\\[-6pt]
{\bf Bin} &  $\boldsymbol{Q^2}$  & $\boldsymbol{x}$ & 
$\boldsymbol{y}$ &  $\boldsymbol{\sigma^{\rm ave}_{r}}$ & $\boldsymbol{F_2^{\rm ave}}$ 
 & $\boldsymbol{\delta_{\rm ave,stat}}$ & $\boldsymbol{\delta_{\rm ave,uncor}}$ & 
$\boldsymbol{\delta_{\rm ave,cor}}$ 
 &  $\boldsymbol{\delta_{\rm ave,tot}}$ 
 & $\boldsymbol{\sqrt{s}}$\\
$\boldsymbol{\#}$     & {\bf GeV}$\boldsymbol{^2}$ &   &       &       &         &       {\bf \%}               & {\bf \%}  &    
\%              &   {\bf \%}  & {\bf GeV} \\
\hline
\\[-7pt]
$   69 $ &$    15. $ &$ 0.279 \times 10^{-3}$ & $ 0.848 $ &$ 1.106 $ &$ 1.294 $ &$  6.32 $ &$  4.04 $ &$  3.77 $ &$ 8.39$ & $   252$ \\
$   70 $ &$    15. $ &$ 0.312 \times 10^{-3}$ & $ 0.760 $ &$ 1.290 $ &$ 1.453 $ &$  3.41 $ &$  2.90 $ &$  2.97 $ &$ 5.37$ & $   252$ \\
$   71 $ &$    15. $ &$ 0.348 \times 10^{-3}$ & $ 0.850 $ &$ 1.222 $ &$ 1.430 $ &$  3.89 $ &$  4.06 $ &$  3.36 $ &$ 6.55$ & $   225$ \\
$   72 $ &$    15. $ &$ 0.348 \times 10^{-3}$ & $ 0.680 $ &$ 1.222 $ &$ 1.338 $ &$  3.19 $ &$  2.47 $ &$  2.88 $ &$ 4.95$ & $   252$ \\
\hline
\end{tabular}
\end{center}
\end{tiny}
\tablecaption{\label{tablowea2}Continuation of \Tab~\ref{tablowea1}}
%\small 
\end{table}

\begin{table}
\begin{center}
\begin{tiny}
\begin{tabular}{cc|ccccc|ccccc|c}
\hline
\\[-6pt]
 $\boldsymbol{Q^2}$ & $\boldsymbol{x}$ & 
$\boldsymbol{F_L}$ & $\boldsymbol{\Delta_{\rm stat}F_L}$ & $\boldsymbol{\Delta_{\rm uncor}F_L}$ & $\boldsymbol{\Delta_{\rm cor}F_L}$ & $\boldsymbol{\Delta_{\rm tot}F_L}$
&
$\boldsymbol{F_2}$ & $\boldsymbol{\Delta_{\rm stat} F_2}$ & $\boldsymbol{\Delta_{\rm uncor} F_2}$ & $\boldsymbol{\Delta_{\rm cor} F_2}$ & $\boldsymbol{\Delta_{\rm tot} F_2}$ 
& $\boldsymbol{\rho}$ 
\\
\hline
\\[-6pt]
$   1.5 $ &$ 0.279 \times 10^{-4}$ & $ 0.091 $ &$ 0.114 $ &$ 0.188 $ &$ 0.054 $ &$ 0.226 $ &$ 0.740 $ &$ 0.066 $ &$0.097$ & $0.029$ & $0.121$ & $0.882$  \\
$    2. $ &$ 0.372 \times 10^{-4}$ & $ 0.102 $ &$ 0.069 $ &$ 0.131 $ &$ 0.062 $ &$ 0.160 $ &$ 0.843 $ &$ 0.028 $ &$0.051$ & $0.032$ & $0.066$ & $0.855$  \\
$    2. $ &$ 0.415 \times 10^{-4}$ & $ 0.419 $ &$ 0.110 $ &$ 0.181 $ &$ 0.071 $ &$ 0.223 $ &$ 0.903 $ &$ 0.039 $ &$0.060$ & $0.030$ & $0.078$ & $0.852$  \\
$    2. $ &$ 0.464 \times 10^{-4}$ & $ 0.035 $ &$ 0.052 $ &$ 0.104 $ &$ 0.033 $ &$ 0.121 $ &$ 0.741 $ &$ 0.033 $ &$0.052$ & $0.009$ & $0.062$ & $0.821$  \\
$   2.5 $ &$ 0.465 \times 10^{-4}$ & $ 0.007 $ &$ 0.057 $ &$ 0.120 $ &$ 0.047 $ &$ 0.141 $ &$ 0.847 $ &$ 0.022 $ &$0.045$ & $0.016$ & $0.053$ & $0.856$  \\
\hline
\end{tabular}
\end{tiny}
\end{center}
\tablecaption{\label{tab:flf2xq}The proton structure 
functions $F_L$ and $F_2$ measured
at the given values of $Q^2$ and $x$  without model assumptions.
$\Delta_{\rm stat}F_L$, $\Delta_{\rm uncor}F_L$, $\Delta_{\rm cor}F_L$ and
$\Delta_{\rm tot}F_L$ are 
the statistical, uncorrelated systematic, correlated systematic
and total uncertainty on $F_L$, respectively. 
$\Delta_{\rm stat}F_2$, $\Delta_{\rm uncor}F_2$, $\Delta_{\rm cor}F_2$ and
$\Delta_{\rm tot}F_2$ are 
the statistical, uncorrelated systematic, correlated systematic
and total uncertainty on $F_2$, respectively. 
$\rho$ is the correlation 
coefficient between the $F_L$ and $F_2$ values.
 }
\end{table}

\begin{table}
\begin{scriptsize}
\begin{center}
\begin{tabular}{cccccccc}
\hline
\\[-8pt]
$\boldsymbol{Q^2}$ & $\boldsymbol{x}$ & $\boldsymbol{F_L}$ & 
$\boldsymbol{\Delta_{\rm stat}}$ & $\boldsymbol{\Delta_{\rm uncor}}$ & 
$\boldsymbol{\Delta_{\rm cor}}$ & $\boldsymbol{\Delta_{\rm tot}}$\\
    {\bf GeV}$\boldsymbol{^2}$& \\
\hline
\\[-8pt]
$   1.5 $ &$ 0.279 \times 10^{-4}$ & $ 0.091 $ &$ 0.114 $ &$ 0.188 $ &$ 0.054 $ &$0.226$  \\
$    2. $ &$ 0.427 \times 10^{-4}$ & $ 0.117 $ &$ 0.039 $ &$ 0.074 $ &$ 0.017 $ &$0.086$  \\
$   2.5 $ &$ 0.588 \times 10^{-4}$ & $ 0.147 $ &$ 0.025 $ &$ 0.050 $ &$ 0.008 $ &$0.056$  \\
$   3.5 $ &$ 0.877 \times 10^{-4}$ & $ 0.222 $ &$ 0.021 $ &$ 0.049 $ &$ 0.006 $ &$0.054$  \\
$    5. $ &$ 0.129 \times 10^{-3}$ & $ 0.310 $ &$ 0.022 $ &$ 0.055 $ &$ 0.007 $ &$0.059$  \\
$   6.5 $ &$ 0.169 \times 10^{-3}$ & $ 0.263 $ &$ 0.023 $ &$ 0.059 $ &$ 0.008 $ &$0.064$  \\
$   8.5 $ &$ 0.224 \times 10^{-3}$ & $ 0.213 $ &$ 0.025 $ &$ 0.063 $ &$ 0.009 $ &$0.068$  \\
$   12. $ &$ 0.319 \times 10^{-3}$ & $ 0.314 $ &$ 0.027 $ &$ 0.051 $ &$ 0.008 $ &$0.058$  \\
$   15. $ &$ 0.402 \times 10^{-3}$ & $ 0.255 $ &$ 0.027 $ &$ 0.051 $ &$ 0.008 $ &$0.058$  \\
$   20. $ &$ 0.540 \times 10^{-3}$ & $ 0.315 $ &$ 0.029 $ &$ 0.053 $ &$ 0.009 $ &$0.061$  \\
$   25. $ &$ 0.686 \times 10^{-3}$ & $ 0.269 $ &$ 0.029 $ &$ 0.061 $ &$ 0.010 $ &$0.069$  \\
$   35. $ &$ 0.103 \times 10^{-2}$ & $ 0.201 $ &$ 0.040 $ &$ 0.070 $ &$ 0.014 $ &$0.082$  \\
$   45. $ &$ 0.146 \times 10^{-2}$ & $ 0.219 $ &$ 0.056 $ &$ 0.098 $ &$ 0.024 $ &$0.116$  \\
\hline
\end{tabular}
\end{center}
\end{scriptsize}
\tablecaption{\label{tab:flave}
The proton structure function $F_L(x,Q^2)$ obtained by averaging
$F_L$ data from \Tab~\ref{tab:flf2xq} for each $Q^2$ bin at the given values
of $Q^2$ and $x$. 
$\Delta_{\rm stat}$, $\Delta_{\rm uncor}$, $\Delta_{\rm cor}$ and
$\Delta_{\rm tot}$ are 
the statistical, uncorrelated systematic, correlated systematic
and total uncertainty on $F_L$, respectively. 
}
\end{table}

\clearpage

%\begin{figure}
%\centerline{\epsfig{file=figs/control06pm_v1001.eps,width=0.9\linewidth}}
%\caption{Control plots BST \label{fig:cont920c}}
%\end{figure}

%\begin{figure}
%\centerline{\epsfig{file=figs/e_peak_371_5.eps,width=0.9\linewidth}}
%\caption{Kin. peak 2006 comb.}
%\end{figure}

%\begin{figure}
%\centerline{\epsfig{file=figs/allavelow.eps,width=0.9\linewidth}}
%\caption{Combined}
%\end{figure}

%\begin{figure}
%\centerline{\epsfig{file=figs/allave.eps,width=0.9\linewidth}}
%\caption{Combined}
%\end{figure}

\clearpage

%\begin{figure}[t]
%\begin{center}
%\begin{minipage}{0.9\linewidth}
%\epsfig{file=figs/all_q21f2.eps,width=\linewidth}
%\put(-22,3){\large\bf x}
%\put(-68,3){\large\bf x}
%\put(-110,3){\large\bf x}
%\put(-157,55){\begin{sideways}\large\bf $\sigma_{r}$\end{sideways}}
%\put(-157,105){\begin{sideways}\large\bf $\sigma_{r}$\end{sideways}}
%\put(-157,155){\begin{sideways}\large\bf $\sigma_{r}$\end{sideways}}
%\put(-157,205){\begin{sideways}\large\bf $\sigma_{r}$\end{sideways}}
%\end{minipage}
%\end{center}
%\caption{
%\label{fig:xlamf2a}}
%\end{figure}

%\begin{figure}[t]
%\begin{center}
%\begin{minipage}{0.9\linewidth}
%\epsfig{file=figs/all_q22f2.eps,width=\linewidth}
%\put(-22,3){\large\bf x}
%\put(-68,3){\large\bf x}
%\put(-110,3){\large\bf x}
%\put(-157,55){\begin{sideways}\large\bf $\sigma_{r}$\end{sideways}}
%\put(-157,105){\begin{sideways}\large\bf $\sigma_{r}$\end{sideways}}
%\put(-157,155){\begin{sideways}\large\bf $\sigma_{r}$\end{sideways}}
%\put(-157,205){\begin{sideways}\large\bf $\sigma_{r}$\end{sideways}}
%\end{minipage}
%\end{center}
%\caption{\FFig~\ref{fig:xlamf2a} continued. \label{fig:xlamf2b}}
%\end{figure}

%\begin{figure}
%\centerline{\epsfig{file=figs/2_q2ll.eps,width=\linewidth}}
%\caption{
%$\lambda$ and $\lambda'$ fit results for the structure function 
%$F_2(x,Q^2)$  as a function of $x$  
%at $Q^2=5$~GeV$^2$ and $Q^2=90$~GeV$^2$. 
%\label{fig:2f2}}
%\end{figure}

%\begin{figure}
%\epsfig{file=figs/r_vs_q2.eps,width=\linewidth}
%\put(-50,-3){\large\bf $Q^{2}/GeV^{2}$}
%\put(-153,75){\begin{sideways}\large\bf R\end{sideways}}
%\caption{$R$ vs $Q^2$ }
%\end{figure}

\end{document}